\documentclass[12pt,english,reqno]{smfart}
\usepackage{mathptmx}
\theoremstyle{plain}
\newtheorem{thm}{Theorem}[section]
\newtheorem{lem}[thm]{Lemma}
\newtheorem{prop}[thm]{Proposition}
\newtheorem{cor}[thm]{Corollary}
\newtheorem{defn}[thm]{Definition}
\theoremstyle{definition}
\newtheorem*{rem} {Remark}
\newtheorem*{ex} {Example}
\renewcommand{\theequation}{\arabic{section}.\arabic{equation}}
\title[Howe pairs, supersymmetry, and ratios]
{Howe pairs, supersymmetry, and ratios \\
of random characteristic polynomials \\
for the unitary groups $\mathrm{U}_N$}
\author{J.B.\ Conrey \and D.W.\ Farmer}
\address{American Institute of Mathematics, Palo Alto, USA}
\author{M.R.\ Zirnbauer}
\address{Institut f\"ur Theoretische Physik, Universit\"at zu
K\"oln, Germany}
\email{conrey@aimath.org, farmer@aimath.org,
zirn@thp.uni-koeln.de}
\date{August 7, 2007}
\begin{document}
\begin{abstract}
  For the classical compact Lie groups $K \equiv \mathrm{U}_N$ the
  autocorrelation functions of ratios of characteristic polynomials
  $(z\, ,w) \mapsto \mathrm{Det}(z-k) / \mathrm{Det}(w - k)$ are
  studied with $k \in K$ as random variable. Basic to our treatment is
  a property shared by the spinor representation of the spin group
  with the Shale-Weil representation of the metaplectic group: in both
  cases the character is the analytic square root of a determinant or
  the reciprocal thereof. By combining this fact with Howe's theory of
  supersymmetric dual pairs $(\mathfrak{g},K)$, we express the
  $K$-Haar average product of $p$ ratios of characteristic polynomials
  and $q$ conjugate ratios as a character $\chi$ which is associated
  with an irreducible representation of the Lie superalgebra
  $\mathfrak{g} = \mathfrak{gl}_{n|n}$ for $n = p + q\,$. The
  primitive character $\chi$ is shown to extend to an analytic radial
  section of a real supermanifold related to $\mathfrak{gl}_{n|n}\,$,
  and is computed by invoking Berezin's description of the radial
  parts of Laplace-Casimir operators for $\mathfrak{gl}_{n|n} \,$. The
  final result for $\chi$ looks like a natural transcription of the
  Weyl character formula to the context of highest-weight
  representations of Lie supergroups.

  While several other works have recently reproduced our results in
  the stable range $N\ge \mathrm{max}(p,q)$, the present approach
  covers the \emph{full} range of matrix dimensions $N \in
  \mathbb{N}\,$.

  To make this paper accessible to the non-expert reader, we have
  included a chapter containing the required background material from
  superanalysis.
\end{abstract}
\maketitle
\tableofcontents

\section{Introduction}
\label{sect:intro}

Let $K$ be one of the compact Lie groups $\mathrm{U}_N \,$,
$\mathrm{O}_N\,$, or $\mathrm{USp}_N\,$, and let $K$ be equipped
with Haar measure $dk$ normalized by $\int_K dk = 1$. We are
interested in products of ratios of characteristic polynomials
$\mathrm{Det}(z - k)$, averaged with respect to the probability
measure $dk\,$. Our goal is to derive exact expressions for such
autocorrelation functions.

Quantities of a similar kind, for Gaussian distributed Hermitian
matrices, were first computed some time ago in the large-$N$ limit by
Andreev and Simons \cite{as}, using an elaboration of the
supersymmetry method pioneered by Wegner and Efetov.  Then, in
1996-97, one of us developed a variant \cite{circular} of the
Wegner-Efetov method to handle the case of Haar-ensemble averages for
the compact Lie groups above. With guidance from a perceptive remark
of Sarnak \cite{sarnak}, it soon transpired that the mathematical
foundation of that variant had existed in classical invariant theory
for decades, and is known there as \emph{Howe duality}. The main
ideas -- namely, Howe dual pairs acting in the
\emph{spinor-oscillator} module, the formula for the
spinor-oscillator character, and the fact that the $K$-Haar
autocorrelation can be regarded as the character associated with a
\emph{highest-weight representation} of some \emph{Lie superalgebra}
$\mathfrak{g}$ -- were presented at the MSRI workshop on "Random
matrices and their Applications" (Berkeley, June 1999).

The second step of that original approach was to express the
primitive $\mathfrak{g}$-character as a co-adjoint orbit integral
(actually, a Berezin superintegral) \'a la Kirillov, and finally to
compute this integral by a super version of the localization
principle for equivariantly closed differential forms \cite{BGV}.
While this last step works in the so-called \emph{stable range},
i.e., for large enough $N$, the superintegral becomes too subtle an
object to handle for small values of $N$ and, in any case, this type
of analysis has not yet been perfected to the stage of yielding
mathematical theorems. In the present paper, we replace it by a
combination of algebraic and analytical techniques \emph{avoiding}
superintegration.

While this research was initially motivated by quantum chaos and
the physics of disordered systems, the main motivation for
bringing it to completion comes from analytic number theory. In
that area, random matrices have been a recurrent and growing theme
ever since Montgomery conjectured the pair correlation function of
the non-trivial zeroes of the Riemann zeta function to be
asymptotic to Dyson's pair correlation function for the Gaussian
Unitary Ensemble (GUE) and since Odlyzko's high-precision
numerical study demonstrated asymptotic agreement of the
distribution of spacings between the Riemann zeroes with the
spacing distribution predicted from GUE.

The conjectured relation between the zeroes of zeta functions and
eigenvalues of random matrices was much clarified by the work of
Katz and Sarnak \cite{KS}, which shifted the emphasis from Riemann
zeta to \emph{families} of $L$-functions.  Carried out in the
setting of $L$-functions over finite fields, this work explained
how a limit procedure for families leads to equidistribution on
one of the classical compact Lie groups $\mathrm{SU}_N\,$,
$\mathrm{U}_N\,$, $\mathrm{SO}_{2N+1}\,$, $\mathrm{USp}_N\,$, or
$\mathrm{SO}_{2N}\,$, the group being determined by the symmetry
of the family.

Keating and Snaith made the proposal that the connection between
$L$-functions and random matrices should carry over to the case of
global $L$-functions.  In particular, they conjectured \cite{ks} that
the critical value distribution of the Riemann zeta function (or,
more precisely, a certain universal part thereof, which is hard to
access by number theory methods) can be modelled by the
characteristic polynomial of a random matrix $u \in \mathrm{U}_N \,$.
A sizable number of precise generalizations of this proposal have
since been made, although most of the theory to date remains at the
conjectural level.

%
In a companion paper \cite{cfz-2} we present various conjectures
that are suggested for $L$-functions by the Keating-Snaith
conjecture and the results of our work. The present paper is
solely concerned with the random-matrix aspect of this unfolding
story; the reader is referred to our second paper for the related
details concerning $L$-functions.

\subsection{Statement of result}\label{sect:statement}

We now define the quantity to be studied in this paper. Postponing
the cases of the compact Lie groups $K = \mathrm{O}_N$ and $K =
\mathrm{USp}_{2N}$ to another publication we focus on the case of $K
= \mathrm{U}_N$ and, fixing a pair of positive integers $p$ and
$q\,$, consider for $u \in \mathrm{U}_N$ the product of ratios of
characteristic polynomials
\begin{equation}\label{R for U(N)}
    Z(t,u) = \mathrm{e}^{\lambda_N}\,
    \prod_{j=1}^p \frac{\mathrm{Det}(\mathrm{Id}_N -
    \mathrm{e}^{\mathrm{i}\psi_j}\,u)}{\mathrm{Det}
    (\mathrm{Id}_N - \mathrm{e}^{\phi_j}\,u)}
    \prod_{l=p+1}^{p+q}  \frac{\mathrm{Det}(\mathrm{Id}_N -
    \mathrm{e}^{-\mathrm{i}\psi_l}\,\bar{u})}{\mathrm{Det}
    (\mathrm{Id}_N-\mathrm{e}^{-\phi_l}\,\bar{u})}\;,
\end{equation}
where $\bar{u}$ is the complex conjugate of the unitary matrix $u\,$.
The parameters $\psi_1, \ldots, \phi_{p+q}$ are complex, with the
range of the $\phi_k$ restricted by $\mathfrak{Re}\, \phi_j < 0 <
\mathfrak{Re}\, \phi_l$ for $j = 1, \ldots, p$ and $l = p+1, \ldots,
p+q\,$. It will be convenient to think of these parameters as
defining a diagonal matrix $t \in \mathrm{End}(\mathbb{C}^{2(p+q)})$
by
\begin{equation}\label{t for U(N)}
    t = \mathrm{diag}\, ( \mathrm{e}^{\mathrm{i}\psi_1}, \ldots,
    \mathrm{e}^{\mathrm{i}\psi_{p+q}}, \mathrm{e}^{\phi_1},
    \ldots, \mathrm{e}^{\phi_{p+q}} ) \;.
\end{equation}
The multiplier $\mathrm{e}^{\lambda_N}$ is given by the linear
combination of parameters
\begin{equation}\label{HW U(N)}
    \lambda_N = N \sum\nolimits_{l=p+1}^{p+q}
    (\mathrm{i}\psi_l - \phi_l) \;.
\end{equation}
Notice that a more concise form of (\ref{R for U(N)}) is
\begin{equation}\label{concise R}
    Z(t,u) = \prod_{k=1}^n \frac{\mathrm{Det}(\mathrm{Id}_N
    - \mathrm{e}^{\mathrm{i}\psi_k} u)}{\mathrm{Det}
    (\mathrm{Id}_N - \mathrm{e}^{\phi_k}\,u)}\qquad (n = p+q)\;.
\end{equation}

Why do we exponentiate our parameters as $\mathrm{e}^{\mathrm{i}
\psi_k}$ in the numerator as opposed to $\mathrm{e}^{\phi_k}$ in the
denominator? The answer is that at some point we will take advantage
of a certain Riemannian structure, by restricting the complex
parameters $\psi_k$ and $\phi_k$ to $\mathbb{R}$.

If $du$ denotes the $\mathrm{U}_N$-Haar measure normalized by
$\int_{\mathrm{U}_N} du = 1\,$, our goal is to establish an exact
expression for the ensemble average
\begin{equation}\label{def chi}
    \chi(t) := \int_{\mathrm{U}_N} Z(t,u)\, du
\end{equation}
for all matrix dimensions $N\,$. To state our result for $\chi\,$,
let $\mathrm{S}_{p+q}$ be the group of permutations of the $p+q$
objects
\begin{displaymath}
    \psi_1\, , \ldots, \psi_p\, , \psi_{p+1} \, , \ldots,
    \psi_{p+q} \;.
\end{displaymath}
\begin{thm}\label{thm0}
For all $N \in \mathbb{N}$ the Haar-ensemble average $\chi(t) =
\int_{\mathrm{U}_N} Z(t,u) \, du$ is expressed by
\begin{equation}\label{eq thm0}
    \chi(t) = \frac{1}{p!\, q!} \sum_{w \in \mathrm{S}_{p+q}}
    \,\, \prod_{l=p+1}^{p+q} \frac{\mathrm{e}^{N\,\mathrm{i}
    w(\psi_l)}} {\mathrm{e}^{N\phi_l}} \prod_{j = 1}^p
    \frac{(1 - \mathrm{e}^{\phi_j - \mathrm{i} w(\psi_l)})
    \, (1 - \mathrm{e}^{\mathrm{i} w(\psi_j) - \phi_l})}
    {(1-\mathrm{e}^{\mathrm{i} w(\psi_j)-\mathrm{i}w(\psi_l)})
    \, (1 - \mathrm{e}^{\phi_j - \phi_l})} \;.
\end{equation}
\end{thm}
\begin{rem}
Let $\mathrm{S}_p \times \mathrm{S}_q$ be the subgroup of
$\mathrm{S}_{p+q}$ which permutes the first $p$ parameters $\psi_1\,
, \ldots, \psi_p$ separately from the last $q$ parameters
$\psi_{p+1}\, , \ldots, \psi_{p+q}\,$.  It is easy to see that the
terms in the sum for $\chi$ do not change under the substitution $w
\to w w^\prime$ if $w^\prime \in \mathrm{S}_p \times \mathrm{S}_q\,$.
Therefore, without changing the result, one may replace the sum over
$w \in \mathrm{S}_{p+q}$ by a sum over cosets $[w] \in \mathrm{S}_{p
+q} / (\mathrm{S}_p \times \mathrm{S}_q)$ and drop the factor of
$(p!\, q!)^{-1}$.
\end{rem}
To give a smooth statement without restrictions on the values of
$N\,$, we have taken the number of characteristic polynomials in the
numerator and denominator of (\ref{R for U(N)}) to be equal. This, of
course, is not a serious limitation; from formula (\ref{eq thm0}) one
immediately produces answers for the case of unequal numbers by
sending one or several of the parameters $\mathrm{e}^{\pm \mathrm{i}
\psi_k}$ or $\mathrm{e}^{\pm\phi_k}$ to zero on both sides of the
equation.  In fact, an easy induction (spelled out in Sect.\
\ref{sect:cor1.2}) yields the following statement.
\begin{cor}\label{cor 1.2}
If $p^\prime \le p + N$ and $q^\prime \le q + N$ then
\begin{eqnarray*}
    &&\int\limits_{\mathrm{U}_N} \frac{\prod_{j=1}^p
    \mathrm{Det}(\mathrm{Id}_N - \mathrm{e}^{\mathrm{i}
    \psi_j}\,u)\, \prod_{l=p+1}^{p+q} \mathrm{Det}
    (\mathrm{Id}_N - \mathrm{e}^{-\mathrm{i}\psi_l}\,
    \bar{u})}{\prod_{j^\prime = 1}^{p^\prime}\mathrm{Det}
    (\mathrm{Id}_N - \mathrm{e}^{\phi_{j^\prime}} u)\,
    \prod_{l^\prime = p^\prime +1}^{p^\prime + q^\prime}
    \mathrm{Det}(\mathrm{Id}_N-\mathrm{e}^{-\phi_{l^\prime}}
    \,\bar{u})} \, du \\ &=& \sum_{[w] \in \mathrm{S}_{p+q} /
    \mathrm{S}_p \times \mathrm{S}_q}\,\, \prod_{k=p+1}^{p+q}
    \frac{\mathrm{e}^{N\,\mathrm{i} w(\psi_k)}}
    {\mathrm{e}^{N\,\mathrm{i} \psi_k}} \times
    \, \frac{\prod_{j^\prime,l}(1 - \mathrm{e}^{\phi_{j^\prime}
    - \mathrm{i} w(\psi_l)}) \, \prod_{j,l^\prime}
    (1 - \mathrm{e}^{\mathrm{i} w(\psi_j) - \phi_{l^\prime}})}
    {\prod_{j,l} (1-\mathrm{e}^{\mathrm{i} w(\psi_j) -
    \mathrm{i}w(\psi_l)}) \, \prod_{j^\prime , l^\prime}
    (1 - \mathrm{e}^{\phi_{j^\prime} - \phi_{l^\prime}})} \;.
\end{eqnarray*}
\end{cor}
When the matrix dimension $N$ is sufficiently large, formula (\ref{eq
thm0}) can be obtained from a variety of different approaches. The
present paper develops the point of view that (\ref{eq thm0}) is best
seen as a result in the theory of Lie superalgebras and supergroups:
it is the latter framework that delivers the answer for \emph{all}
values of $N \ge 1\,$.

\subsubsection{Howe duality}

The method we will use to prove Thm.\ \ref{thm0} rests on the fact
that $\mathrm{U}_N$ is in Howe duality with a Lie superalgebra
$\mathfrak{gl}_{n|n}$ for $n = p + q\,$.  The details are briefly
speaking as follows. Let $V = V_0 \oplus V_1$ with $V_0 \simeq V_1$
be a $\mathbb{Z}_2$-graded vector space, where $V_0$ is the direct
sum of $p$ fundamental modules and $q$ co-fundamental modules for
$\mathrm{U}_N\,$. Let $\mathcal{A}_V$ be the spinor-oscillator module
of $V$, i.e., the tensor product of the exterior algebra of $V_1$
with the symmetric algebra of $V_0 :$
\begin{equation}
    \mathcal{A}_V = \wedge (V_1) \otimes \mathrm{S} (V_0) \;.
\end{equation}
$\mathrm{U}_N$ acts on $\mathcal{A}_V\,$, and so does
$\mathfrak{gl}_{n|n} \,$, and these two actions commute.

Now recall the definition of the diagonal parameter matrix $t$ by
(\ref{t for U(N)}), and think of its logarithm as lying in a Cartan
subalgebra $\mathfrak{h} \subset \mathfrak{gl}_{n|n}\,$. Then $t =
\mathrm{e}^{\ln t}$ operates on $\mathcal{A}_V$ by the exponential of
the action of $\mathfrak{gl}_{n|n}$ and, as will be explained in some
pedagogical detail in Sect.\ \ref{sect:2}, the supertrace of the
operator representing the pair $(t,u)$ on $\mathcal{A}_V$ equals the
product of ratios (\ref{R for U(N)}). Moreover, the joint action of
the Howe dual pair $(\mathfrak{gl}_{n|n}\, , \mathrm{U}_N)$ on
$\mathcal{A}_V$ is \emph{multiplicity-free}, which leads to the
following key observation.
\begin{prop}\label{prop 1.2}
The function $\chi: \, t \mapsto \int_{\mathrm{U}_N} Z(t,u)\, du$ is
a primitive $\mathfrak{gl}_{n|n}$-character, and is given by the
$\mathfrak{gl}_{n|n}$-representation on the space of
$\mathrm{U}_N$-invariants in $\mathcal{A}_V\,$.
\end{prop}
{}From Prop.\ \ref{prop 1.2} (which summarizes Prop.\
\ref{howe-char}), our task is to compute the primitive character $t
\mapsto \chi(t)$. This is not entirely straightforward as the
representation at hand is \emph{atypical} and lies outside the range
of applicability of known character formulas. Some answer for $\chi$
(in the form of a Littlewood-Richardson sum over products of
$\mathfrak{gl}_{p|p}$- and $\mathfrak{gl}_{q|q}$-characters, using
the $\mathfrak{gl}_{p+q|p+q} \to \mathfrak{gl}_{p|p} \times
\mathfrak{gl}_{q|q}$ branching rules) has recently been given in
\cite{cheng}; that answer, however, is not of the explicit form we
are looking for.

In this paper $\chi(t)$ will be computed explicitly by drawing on
some of Berezin's results for the radial parts of the Laplace-Casimir
operators for the classical \emph{Lie supergroup} $\mathrm{U}_
{n|n}\,$.  The outcome is the formula stated in Thm.\ \ref{thm0}. Let
us now rewrite that formula using the standard language of
representation theory.

\subsubsection{Representation-theoretic formulation}

It is appropriate to regard the $2n$ parameters $\psi_1 \, , \ldots,
\psi_n \, , \phi_1 \, , \ldots, \phi_n$ as linear functions on a
Cartan subalgebra $\mathfrak{h} \simeq \mathbb{C}^n \oplus
\mathbb{C}^n$ of diagonal transformations in the Lie superalgebra
$\mathfrak{g} \equiv \mathfrak{gl}_{n|n}\,$; by linearly combining
them one can then form the roots of $\mathfrak{g}$ and the weights of
its representations.

Roots $\alpha: \, \mathfrak{h} \to \mathbb{C}$ are defined as the
eigenvalues of the adjoint action of $\mathfrak{h}$ on $\mathfrak
{g}$ as usual: $[H , X] = \alpha(H) X\,$.  A root $\alpha$ is
called \emph{even} or \emph{odd} depending on whether the
corresponding eigenvector $X$ is an even or odd element of the Lie
superalgebra $\mathfrak{g}\,$.

The representation of $\mathfrak{g} = \mathfrak{gl}_{n|n}$ on the
subspace of $\mathrm{U}_N$-invariants in $\mathcal{A}_V$ will be seen
to be a highest-weight representation, with the role of highest
weight being played by the linear function $\lambda_N : \,
\mathfrak{h} \to \mathbb{C}$ of (\ref{HW U(N)}). Our result for
$\chi$ is to be expressed in terms of the $\lambda_N$-positive roots,
i.e., the sets of even and odd roots of ($\mathfrak{h}$ acting on)
$\mathfrak{g}$ which are positive with respect to the highest weight
$\lambda_N\, $. These are the sets
\begin{eqnarray}
    \Delta_{\lambda,0}^+ &:& \mathrm{i}\psi_l - \mathrm{i}\psi_j
    \;, \quad \phi_l -\phi_j \;, \nonumber \\
    \Delta_{\lambda,1}^+ &:& \phi_l -\mathrm{i}\psi_j \;;
    \quad \mathrm{i}\psi_l - \phi_j \quad
    (1 \le j \le p < l \le n) \;. \label{roots U(N)}
\end{eqnarray}

The primitive characters of a compact Lie group $K$ are expressed by
the Weyl character formula (see, e.g., \cite{knapp}),
%
%
which says that if $\rho : \, K \to \mathrm{U}(V_\lambda)$ is an
irreducible representation with regular highest weight $\lambda$,
then
\begin{displaymath}
    \mathrm{Tr}_{\, V_\lambda}\, \rho(t) = \sum_{w \in W} \frac{
    \mathrm{e}^{w(\lambda)}} {\prod_{\alpha \in \Delta_\lambda^+}
    (1 - \mathrm{e}^{-w(\alpha)})} (\ln t) \;,
\end{displaymath}
where $W$ is the Weyl group of $K$, and $\Delta_\lambda^+$ denotes
the system of $\lambda$-positive roots.

Our answer for $\chi$ will look like a natural generalization of this
formula to the case of Lie supergroups. The Weyl group $W$ of our
problem is the same as the symmetric group $\mathrm{S}_n$ of
permutations of the $n$ basis vectors of the first summand in
$\mathfrak{h} \simeq \mathbb{C}^n \oplus \mathbb{C}^n$ (which are
dual to the coordinate functions $\psi_1 \, , \ldots, \psi_n$). While
the Weyl group $W$ is defined by its action on the Cartan subalgebra
$\mathfrak{h}\,$, what is needed here is the induced action on linear
functions $f \in \mathfrak{h}^\ast$ by $w(f) := f \circ w^{-1}$.

Because the highest weight $\lambda_N$ of the $\mathfrak
{g}$-representation at hand turns out to be non-regular, it is not
$W$ that matters but its quotient by the subgroup $W_\lambda$ that
fixes $\lambda_N \,$. This is the subgroup $W_\lambda \simeq
\mathrm{S}_p \times \mathrm{S}_q$ stabilizing the decomposition
$\mathbb{C}^n = \mathbb{C}^p \oplus \mathbb{C}^q$.

Of course every coset $[w] \in W/W_\lambda$ uniquely defines a
transformed highest weight $w(\lambda_N)\,$. It also determines
transformed sets of $\lambda_N$-positive roots $w(\Delta_{\lambda
,0}^+)$ and $w(\Delta_{\lambda ,1}^+)$ which are independent of
the choice of representative $w$.

Given these definitions and the insight from Prop.\ \ref{prop 1.2},
the statement of Thm.\ \ref{thm0} can be rephrased as follows.
\begin{prop}\label{thm1}
For all $N \in \mathbb{N}$ the character $t \mapsto \chi(t)$ is given
by a generalization of the Weyl character formula:
\begin{equation}\label{eq thm1}
    \chi(t) = \sum_{[w]\in W/W_\lambda}\mathrm{e}^{w(\lambda_N)}
    \, \frac{\prod_{\beta \in \Delta_{\lambda,1}^+} (1 -
    \mathrm{e}^{-w(\beta)})}{\prod_{\alpha\in\Delta_{\lambda,0}^+}
    (1 - \mathrm{e}^{-w(\alpha)})} \, (\ln t)\;.
\end{equation}
\end{prop}
\begin{rem}
In a companion paper \cite{HPZ} it is proved that a completely
analogous result holds for the case of ratios of random
characteristic polynomials for the compact Lie groups $K =
\mathrm{USp}_{2N}\,$, $K = \mathrm{O}_N\,$, and $K = \mathrm{SO}_N
\,$.
\end{rem}
A number of further remarks on the scope of Thm.\ \ref{thm0} are
called for.

\subsubsection{Compact sector}

Consider now removing all of the $\phi$-parameters, i.e., send
$\phi_j \to -\infty$ in $\mathrm{Det}(\mathrm{Id}_N -
\mathrm{e}^{\phi_j}\, u)^{-1}$ and $\phi_l \to +\infty$ in
$\mathrm{Det}(\mathrm{Id}_N - \mathrm{e}^{-\phi_l}\, \bar{u})^{-1}$.
Then the formula of Thm.\ \ref{thm0} reduces to
\begin{eqnarray}
    &&\int\limits_{\mathrm{U}_N} \prod_{j=1}^p \mathrm{Det}
    (\mathrm{Id}_N - \mathrm{e}^{\mathrm{i}\psi_j}\,u)
    \prod_{l=p+1}^{p+q}\mathrm{Det}(\mathrm{e}^{\mathrm{i}
    \psi_l}\mathrm{Id}_N - \bar{u})\, du \nonumber \\ &=&
    \sum_{[w] \in \mathrm{S}_{p+q} / \mathrm{S}_p \times
    \mathrm{S}_q}\,\, \prod_{l = p+1}^{p+q} \mathrm{e}^{N\,
    \mathrm{i}w(\psi_l)} \prod_{j = 1}^p \, \frac{1}
    {1-\mathrm{e}^{\mathrm{i} w(\psi_j) -
    \mathrm{i}w(\psi_l)}}\;, \label{eq 1.9}
\end{eqnarray}
which is a result that can also be extracted from \cite{CFKRS}. The
mathematical situation in this limit is ruled by a Howe dual pair
\cite{howe1} of compact Lie groups
\begin{displaymath}
    K^\prime \times K = \mathrm{U}_{p+q} \times \mathrm{U}_N\;;
\end{displaymath}
which is to say that $K^\prime \times K$ acts without multiplicity on
an exterior algebra $\wedge(V_1)$, and the left-hand side of (\ref{eq
1.9}) is the character of the irreducible representation of
$K^\prime$ on the subalgebra of $K$-invariants in $\wedge(V_1)$. The
right-hand side of (\ref{eq 1.9}) is nothing but the classical Weyl
character formula evaluated for the irreducible
$\mathrm{U}_{p+q}$-representation with (non-regular) highest weight
\begin{equation}
    \lambda_N^\prime = N\sum\nolimits_{l=p+1}^{p+q}
    \mathrm{i}\psi_l \;.
\end{equation}

\subsubsection{Non-compact sector}

When all of the $\psi$-parameters are removed, Thm.\ \ref{thm0}
reduces to a statement about the autocorrelations of
\emph{reciprocals} of characteristic polynomials. This particular
limit is ruled by the Howe dual pair
\begin{displaymath}
    G \times K = \mathrm{U}_{p,\,q} \times \mathrm{U}_N \;,
\end{displaymath}
where $\mathrm{U}_{p,\,q}$ is the non-compact group of isometries of
the pseudo-unitary vector space $\mathbb{C}^p \oplus \mathbb{C}^q$.
The pair $G \times K$ acts without multiplicity on a symmetric
algebra $\mathrm{S}(V_0)$ (generated by a direct sum of $p$ copies of
$\mathbb{C}^N$ and $q$ copies of the dual vector space of
$\mathbb{C}^N$) by the \emph{Shale-Weil} or \emph{oscillator
representation}.

Taking a product of $p + q$ reciprocals of characteristic polynomials
and their conjugates and averaging with Haar measure of $K\,$, one
again gets a primitive character:
\begin{equation}
    \int_{\mathrm{U}_N} \prod_{j=1}^p \mathrm{Det}(\mathrm{Id}_N -
    \mathrm{e}^{\phi_j}\,u)^{-1} \prod_{l=p+1}^{p+q} \mathrm{Det}
    (\mathrm{e}^{\phi_l} \mathrm{Id}_N - \bar{u})^{-1} \, du =:
    \chi^{\prime\prime}(t_0) \;;
\end{equation}
which is the character $\chi^{\prime\prime}(t_0) \equiv
\chi^{\prime\prime} (\mathrm{e}^{\phi_1} , \ldots,
\mathrm{e}^{\phi_{p+q}})$ of the representation of $G$ on the
subalgebra of $K$-invariants in $\mathrm{S}(V_0)$.  For all positive
integers $p$, $q$, and $N$, these $G$-repre\-sentations are
irreducible, unitary, and with highest weight
\begin{equation}\label{HW dprime}
    \lambda_N^{\prime\prime} =
    - N \sum\nolimits_{l=p+1}^{p+q} \phi_l \;.
\end{equation}

\subsubsection{Stable versus non-stable range}

In the range $N \ge p + q\,$, the representations with highest weight
(\ref{HW dprime}) belong to the \emph{principal discrete series} of
$G = \mathrm{U}_{p,\,q}\,$, which is to say that they can be realized
by Hilbert spaces of $L^2$-functions on $G$ and are covered by an
analog of the Borel-Weil correspondence known for the compact case.
The characters in that range turn out to have a particularly simple
expression:
\begin{equation}\label{stable U(N)}
    \chi^{\prime\prime}(\mathrm{e}^{\phi_1},\ldots, \mathrm{e}^{
    \phi_{p+q}}) = \mathrm{e}^{\lambda_N^{\prime\prime}}
    \prod_{j=1}^p \prod_{l=p+1}^{p+q} \frac{1}{1 -
    \mathrm{e}^{\phi_j - \phi_l}} \;.
\end{equation}
This formula actually continues to hold as long as $N \ge
\mathrm{max}(p,q)$. As a matter of fact, in that range the algebra of
$K$-invariants in the oscillator module $\mathrm{S}(V_0)$ is
\emph{freely generated} by the quadratic invariants that correspond
to the system of $\lambda_N^{\prime \prime}$-positive roots.

The range $N \ge \mathrm{max}(p,q)$ where (\ref{stable U(N)}) holds
is called the \emph{stable} range. From the vantage point of (\ref{eq
thm0}) the simplicity of the result (\ref{stable U(N)}) comes about
because throughout the stable range, only the term from the identity
coset $[e]$ in the sum over $[w] \in \mathrm{S}_{p+q} / (\mathrm{S}_p
\times \mathrm{S}_q)$ survives the limit of removing the parameters
$\psi_k \,$.

On the other hand, the said characters \emph{outside} the stable
range are of a different nature. Their representations have no
realization by $L^2$-functions on $G$ and are called \emph{singular}.
The algebra of $K$-invariants outside the stable range is \emph{not}
freely generated; there exist relations and the characters therefore
do not have any expression as simple as (\ref{stable U(N)}).  In
fact, when these characters are extracted from our supersymmetric
master formula (\ref{eq thm0}), there is a build up of combinatorial
complexity because the product over odd-root factors $(1 -
\mathrm{e}^{\phi_j - \mathrm{i} w(\psi_l)}) (1 - \mathrm{e}^{
\mathrm{i}w(\psi_j) - \phi_l})$ starts competing with the highest
weight factor $\mathrm{e}^{N\sum \mathrm{i}w(\psi_l)}$ so that an
increasing number of terms survives the process of removing the
$\psi$-parameters.

\subsubsection{Other approaches}

Given the existence of a stable and a "non-stable" range for the
non-compact sector, one should expect the same distinction to be
visible also in the full, supersymmetric situation.  This is indeed
the case.

Our answer (\ref{eq thm0}) has recently been reproduced partially by
two independent approaches: by the work of Conrey, Forrester and
Snaith \cite{cfs} which is based on the results for orthogonal
polynomials by Fyodorov and Strahov \cite{fs} and Baik, Deift and
Strahov \cite{bds}, and in a preprint by Bump and Gamburd \cite{bg}
using symmetric function theory. Both make a statement only in the
stable range for $N\,$. A fourth viable approach in that range is our
original 1999 method of computation (i.e., expressing the character
in question as a co-adjoint orbit superintegral and then calculating
it by an equivariant localization technique) which is how the
formulas (\ref{eq thm1}) and (\ref{eq thm0}) were first obtained.

From our experience, proving Thm.\ \ref{thm0} \emph{outside} the
stable range for $N$ is qualitatively more difficult!  In fact, in
view of the singular nature of the representations in the underlying
non-compact situation, we consider it somewhat of a miracle that a
formula as universal and simple as (\ref{eq thm0}) happens to be true
for all $N \in \mathbb{N}$. The present paper owes its length to our
desire to give a transparent and self-contained proof of it.

\subsection{Some basic definitions of superanalysis}
\label{sect:basics}

Let $V = V_1 \oplus V_0$ be a $\mathbb{Z}_2$-graded vector space over
$\mathbb{K} = \mathbb{R}$ or $\mathbb{K} = \mathbb{C}\,$. An element
$v \in V$ is called \emph{odd} if $v \in V_1 \setminus \{ 0 \}$ and
\emph{even} if $v \in V_0 \setminus \{ 0 \}\,$. Elements which are
either even or odd are called \emph{homogeneous}. $V$ comes with a
\emph{parity function} $| \cdot |$ which is defined on homogeneous
elements and takes the value $|v| = 1$ for $v$ odd and $|v|=0$ for
$v$ even. If $V_1 = \mathbb{K}^p$ and $V_0 = \mathbb{K}^q$, one also
writes $V = \mathbb{K}^{p|q}$.  Depending on the context we will
sometimes write $V = V_0 \oplus V_1\,$, and sometimes $V = V_1 \oplus
V_0\,$, but we always mean the same thing.

Given the $\mathbb{Z}_2$-graded vector space $V$, consider the
$\mathbb{K}$-linear transformations $\mathrm{End}(V)$ of $V = V_1
\oplus V_0\,$. The elements $X \in \mathrm{End}(V_1 \oplus V_0)$
decompose into blocks:
\begin{equation}\label{M-blocks}
    X = \begin{pmatrix} \mathsf{A} &\mathsf{B} \\
    \mathsf{C} &\mathsf{D} \end{pmatrix} \;,
\end{equation}
where $\mathsf{A} \in \mathrm{End}(V_1)$, $\mathsf{B} \in
\mathrm{Hom}(V_0\, , V_1)$, $\mathsf{C} \in \mathrm{Hom}(V_1, V_0)$,
and $\mathsf{D} \in \mathrm{End}(V_0)$. The vector space
$\mathrm{End}(V)$ inherits from $V$ a natural $\mathbb{Z}_2 $-grading
$\mathrm{End}(V) = \mathrm{End}(V)_0 \oplus \mathrm{End}(V)_1$ by
\begin{equation}\label{grading}
    \begin{pmatrix} \mathsf{A} &\mathsf{B} \\
    \mathsf{C} &\mathsf{D} \end{pmatrix} =
    \begin{pmatrix} \mathsf{A} &0 \\ 0 &\mathsf{D}
    \end{pmatrix} + \begin{pmatrix} 0 &\mathsf{B} \\
    \mathsf{C} &0 \end{pmatrix} \;,
\end{equation}
i.e., $\mathrm{End}(V)_0 \simeq \mathrm{End}(V_0) \oplus \mathrm{End}
(V_1)$ and $\mathrm{End}(V)_1 \simeq \mathrm{Hom}(V_0\, , V_1) \oplus
\mathrm{Hom}(V_1 , V_0)$.  The \emph{supertrace} on $V$ is the linear
function $\mathrm{STr}_V : \, \mathrm{End}(V) \to \mathbb{K}$ defined
by
\begin{equation}\label{STrace}
    \mathrm{STr}_V \, X = \mathrm{Tr}\, X\vert_{V_0 \to V_0} -
    \mathrm{Tr}\, X\vert_{V_1 \to V_1} = \mathrm{Tr}\, \mathsf{D}
    - \mathrm{Tr} \, \mathsf{A} \;.
\end{equation}
It has cyclic invariance in the $\mathbb{Z}_2$-graded sense: $\mathrm
{STr}\, (XY) = (-1)^{|X| \, |Y|} \mathrm{STr}\, (YX)$.

As a $\mathbb{Z}_2$-graded vector space, $\mathrm{End}(V)$ carries
a natural bracket operation,
\begin{displaymath}
    [ \,\, , \,\, ] : \quad \mathrm{End}(V) \times \mathrm{End}(V)
    \to \mathrm{End}(V) \;,
\end{displaymath}
which for any two homogeneous elements $X, Y \in \mathrm{End}(V)$
is defined by
\begin{equation}\label{sbracket}
    [X,Y] = XY - (-1)^{|X| \, |Y|} YX \;.
\end{equation}
Thus the bracket of two odd elements is the anti-commutator, while in
the other cases it is the commutator.  This definition is extended to
all of $\mathrm{End}(V)$ by linearity. The algebra $\mathrm{End} (V)$
equipped with the bracket $[ \, , \, ]$ is denoted by $\mathfrak{gl}
(V)$, and is an example of a \emph{Lie superalgebra}. If $V =
\mathbb{K}^{p|q}$, one writes $\mathfrak{gl}(V) \equiv
\mathfrak{gl}_{p|q}(\mathbb{K})\,$.

More generally a Lie superalgebra $\mathfrak{g}$ is a $\mathbb{Z}
_2$-graded vector space $\mathfrak{g} = \mathfrak{g}_0 \oplus
\mathfrak{g}_1$ over $\mathbb{K}$ equipped with a bilinear bracket
operation that has the following three properties:
\begin{eqnarray*}
    &\mathrm{(i)}& | [X,Y] | = |X| + |Y| \quad (\text{mod } 2) \;,
    \\ &\mathrm{(ii)}& [X , Y] = - (-1)^{|X| \, |Y|} [Y , X] \;,
    \\ &\mathrm{(iii)}& [X,[Y,Z]] = [[X,Y],Z] + (-1)^{|X| \, |Y|}
    [Y,[X,Z]]\;,
\end{eqnarray*}
for all $X, Y, Z \in \mathfrak{g}$ of definite parity. Property
(iii) is called the super-Jacobi identity.
\begin{ex}
Let each of $V_1$ and $V_0$ carry a non-degenerate
$\mathbb{K}$-bilinear form,
\begin{displaymath}
    S : \,\, V_1 \times V_1 \to \mathbb{K} \;, \quad
    A : \,\, V_0 \times V_0 \to \mathbb{K} \;,
\end{displaymath}
and let $S$ be symmetric and $A$ alternating (which of course implies
that the dimension of $V_0$ is even). On $V = V_1 \oplus V_0$
introduce a $\mathbb{K}$-bilinear form $Q : \, V \times V \to
\mathbb{K}$ by the requirements that $V_1$ be $Q$-orthogonal to
$V_0\,$, and that $Q \vert_{V_1 \times V_1} = S$ and $Q \vert_{V_0
\times V_0} = A\,$. A $\mathbb{Z}_2$-graded vector space $V$ with
such a form $Q$ is called \emph{orthosymplectic}.

Consider then the set of $\mathbb{K}$-linear transformations of $V$
that are skew with respect $Q = S + A\,$. This set is a vector space,
commonly denoted by $\mathfrak{osp}(V)$, which inherits from
$\mathrm{End}(V)$ the $\mathbb{Z}_2$-grading (\ref{grading}), and
skewness of a homogeneous element $X \in \mathrm{End}(V)$ here means
that the relation
\begin{displaymath}
    Q(X v , v^\prime) + (-1)^{|X| \, |v|} Q(v, X v^\prime) = 0
\end{displaymath}
holds for all $v, v^\prime \in V$ with homogeneous $v\,$.

It is not difficult to check that $\mathfrak{osp}(V)$ closes under
the bracket operation (\ref{sbracket}) in $\mathfrak{gl}(V)$ and thus
is a Lie superalgebra; one calls it the \emph{orthosymplectic} Lie
superalgebra of $V$. If $V_1 = \mathbb {K}^p$ and $V_0 =
\mathbb{K}^q$ (with $q \in 2 \mathbb{N}$), one writes
$\mathfrak{osp}(V) = \mathfrak{osp}_{p|q}(\mathbb{K})\,$.
\end{ex}
Returning to the general case, note that the supertrace of any
bracket $[X,Y]$ in $\mathfrak{gl}(V)$ or $\mathfrak{osp}(V)$
vanishes. Indeed, by $(-1)^{|X| \, |Y|} \mathrm{STr}\, (YX) =
\mathrm{STr}\, (XY)$ one has
\begin{displaymath}
    \mathrm{STr}\, [X,Y] = \mathrm{STr}\, \big( XY - (-1)^{|X|
    \, |Y|} YX \big) = - \mathrm{STr}\, [X,Y] = 0 \;.
\end{displaymath}

Besides the Lie superalgebra $\mathfrak{gl}(V)$, the $\mathbb{Z}
_2$-graded vector space $V = V_0 \oplus V_1$ carries another
canonical structure: this is the graded-commutative algebra of
maps
\begin{displaymath}
    f : V_0 \to \wedge(V_1^\ast) \;,
\end{displaymath}
or functions on $V_0$ with values in the exterior algebra of the dual
vector space of $V_1\,$. We want such maps to be holomorphic for
$\mathbb{K} = \mathbb{C}$ and real-analytic for $\mathbb{K} =
\mathbb{R}\,$. In the latter case, if $x^1, \ldots, x^p$ is a
coordinate system for $V_0\,$, and $\xi^1, \ldots, \xi^q$ is a system
of linear coordinates for $V_1$ (and hence a system of generators of
$\wedge V_1^\ast$), such a mapping is written in the form
\begin{displaymath}
    f = \sum_{k = 0}^q \, \sum_{i_1 < \ldots < i_k}
    f_{i_1 \ldots i_k} (x^1, \ldots, x^p) \,
    \xi^{i_1} \wedge \cdots \wedge \xi^{i_k} \;.
\end{displaymath}
The wedge symbol $\wedge$ means exterior multiplication and will
usually be omitted when its presence is clear from the context. A map
$f : \, V_0 \to \wedge(V_1^\ast)$ is commonly referred to as a
\emph{superfunction on $V$} for short.

Consider now the related situation where the $\mathbb{Z}_2$-graded
vector space is not $V$ but $\mathrm{End}(V)$.  Then there exists
a distinguished superfunction, $\mathrm{SDet}$, called the
\emph{superdeterminant}. To define it, let $B$ be the tautological
mapping (i.e., the identity map)
\begin{displaymath}
    B : \,\, \mathrm{Hom}(V_0\, , V_1)^\ast \to
    \mathrm{Hom}(V_0\, , V_1)^\ast \;.
\end{displaymath}
If $\{ \mathsf{F}_i \}$ is a basis of $\mathrm{Hom} (V_0\, , V_1)$
with dual basis $\{ \varphi^i \}$, this is expressed by
\begin{displaymath}
    B = \sum \varphi^i \otimes \mathsf{F}_i \;.
\end{displaymath}
Of course $B$ does not depend on the choice of basis but is
invariantly defined.

We now re-interpret $B$ in the following way. The basis vectors
$\mathsf{F}_i \in \mathrm {Hom}(V_0\, , V_1)$ are still to be viewed
as linear maps from $V_0$ to $V_1$, and their destiny is to be
composed with other maps; e.g., if $A \in \mathrm{GL}(V_1)$ and $D
\in \mathrm {GL}(V_0)$, then $A^{-1} \mathsf{F}_i \, D^{-1} \in
\mathrm{Hom}(V_0\, , V_1)$ is a meaningful expression. The dual basis
vectors $\varphi^i$, on the other hand, will be regarded as
coordinates for $\mathrm{Hom}(V_0\, , V_1)$; as such they are to be
multiplied in the exterior sense. Thus we view them as a set of
generators of the exterior algebra $\wedge \, \mathrm{Hom} (V_0 \, ,
V_1 )^\ast$.

Similarly, if $\{ \tilde{\mathsf{F}}_j \}$ is a basis of
$\mathrm{Hom}(V_1, V_0)$ with dual basis $\{ \tilde{\varphi}^j \}$,
let the tautological map $C :\, \mathrm{Hom}(V_1, V_0)^\ast \to
\mathrm{Hom}(V_1, V_0)^\ast$ be expressed by
\begin{displaymath}
    C = \sum \tilde{\varphi}^j \otimes \tilde{\mathsf{F}}_j \;,
\end{displaymath}
and let $C$ be given the same re-interpretation as $B$.

In defining the superdeterminant, we assume the following
associative product:
\begin{equation}\label{first envelope}
    (\varphi^i \otimes \mathsf{F}_i) (\tilde{\varphi}^j \otimes
    \tilde{\mathsf{F}}_j) := \varphi^i \tilde{\varphi}^j \otimes
    \mathsf{F}_i \, \tilde{\mathsf{F}}_j \;.
\end{equation}
(Technically speaking, this corresponds to a Grassmann envelope of
the first kind \cite{berezin}.)
\begin{defn}[superdeterminant]\label{SDet}
If $V = V_0 \oplus V_1$ is a $\mathbb{Z}_2$-graded vector space,
$\mathrm{End}(V)$ is understood to carry the canonical
$\mathbb{Z}_2 $-grading given by
\begin{eqnarray*}
    \mathrm{End}(V)_0 &\simeq& \mathrm{End}(V_0) \oplus
    \mathrm{End}(V_1) \;, \\ \mathrm{End}(V)_1 &\simeq&
    \mathrm{Hom}(V_0\, , V_1) \oplus \mathrm{Hom}(V_1, V_0) \;.
\end{eqnarray*}
Let $G = (\mathrm{GL}(V_1) \times \mathrm{GL}(V_0) \hookrightarrow
\mathrm{End}(V)_0)$ be the group of invertibles in $\mathrm{End}
(V)_0 \,$.  Then the superdeterminant on $V$ is the superfunction
\begin{eqnarray*}
    \mathrm{SDet}_V : \quad G &\to& \wedge \, \mathrm{End}(V)_1^\ast
    \;, \\ A,D &\mapsto& \frac{\mathrm{Det}(D)} {\mathrm{Det}(A)} \,
    \mathrm{Det} \left( \mathrm{Id}_{V_1} - \sum\nolimits_{i,j}
    \varphi^i \tilde{\varphi}^j \otimes A^{-1} \mathsf{F}_i\,
    D^{-1} \tilde{\mathsf{F}}_j \right)^{-1} \;,
\end{eqnarray*}
where the last factor is defined by expansion of the determinant
using $\mathrm{Det} = \mathrm{e}^{\mathrm{Tr}\, \circ \ln}$,
\begin{displaymath}
    \mathrm{Det} \left( \mathrm{Id}_{V_1} - \sum\nolimits_{i,j}
    \varphi^i \tilde{\varphi}^j \otimes A^{-1}\mathsf{F}_i\, D^{-1}
    \tilde{\mathsf{F}}_j \right)^{-1} = 1 + \sum\nolimits_{i,j}
    \varphi^i \tilde\varphi^j \, \mathrm{Tr}_{V_1}\, (A^{-1}
    \mathsf{F}_i\, D^{-1} \tilde{\mathsf{F}}_j) + \ldots \;.
\end{displaymath}
\end{defn}
\begin{rem}
The expansion terminates because of the nilpotency of the
$\varphi^i$ and $\tilde{\varphi}^j$.
\end{rem}
It is convenient and standard practice in superanalysis to take the
identifications $\sum \varphi^i \otimes \mathsf{F}_i \equiv B$ and
$\sum \tilde{\varphi}^j \otimes \tilde{\mathsf{F}}_j \equiv C$ for
granted and, with the blocks $A, B, C, D$ assembled to a
\emph{supermatrix} $\Xi$ as in (\ref{M-blocks}), adopt for the
superdeterminant function $\mathrm{SDet}_V : G \to \wedge \,
\mathrm{End} (V)_1^\ast$ the short-hand notation $\mathrm{SDet}
(\Xi)$.
\begin{prop}\label{prop 1.3}
The superdeterminant has the following equivalent expressions:
\begin{displaymath}
    \mathrm{SDet} \begin{pmatrix} A &B \\ C &D \end{pmatrix}
    = \frac{\mathrm{Det}(D)}{\mathrm{Det}(A - B D^{-1} C)}
    = \frac{\mathrm{Det}(D - C A^{-1} B)}{\mathrm{Det}(A)}
    \;.
\end{displaymath}
\end{prop}
\begin{proof}
The first expression results from Def.\ \ref{SDet} on rewriting
$\sum \varphi^i \otimes \mathsf{F}_i = B$ and $\sum \tilde{
\varphi}^j \otimes \tilde{\mathsf{F}}_j = C$ and using the
multiplicativity of the determinant. The second expression follows
from the first one with the help of the relation
\begin{displaymath}
    \mathrm{Det}(\mathrm{Id}_{V_1} - A^{-1} B D^{-1} C)^{-1} =
    \mathrm{Det}(\mathrm{Id}_{V_0} - D^{-1} C A^{-1} B) \;,
\end{displaymath}
which in turn is a straightforward consequence of $\ln \circ
\mathrm{Det} = \mathrm{Tr} \circ \ln$, the Taylor series of the
logarithm $x \mapsto \ln (1 - x)$, and the alternating property of
the exterior product $\wedge\,$.
\end{proof}
\begin{rem}
Note that the reciprocal of a superdeterminant,
\begin{displaymath}
    \mathrm{SDet}^{-1} \begin{pmatrix} A &B \\ C &D \end{pmatrix}
    = \frac{\mathrm{Det}(A - B D^{-1} C)}{\mathrm{Det}(D)}\;,
\end{displaymath}
exists whenever the block $D$ possesses an inverse.
\end{rem}
%

\subsection{The good object to consider}
\label{sect:bestobject}

Consider now the special situation of a $\mathbb{Z}_2$-graded complex
vector space $V = V_1 \oplus V_0$ with tensor-product structure:
\begin{displaymath}
    V = U \otimes \mathbb{C}^N \simeq (U_1 \otimes \mathbb{C}^N)
    \oplus (U_0 \otimes \mathbb{C}^N) \;,
\end{displaymath}
where $U = U_1 \oplus U_0$ is another $\mathbb{Z}_2$-graded vector
space, and $\mathbb{C}^N$ is a fundamental module for $K =
\mathrm{U}_N\,$. If $k \in K$, and $\Xi = \begin{pmatrix} A &B \\ C
&D \end{pmatrix}$ is a supermatrix where $B = \sum \varphi^i \otimes
F_i$ and $C = \sum \tilde{\varphi}^j \otimes \tilde{F}_j$ as above,
the superdeterminant of $\mathrm{Id}_V - \Xi \otimes k$ is defined in
the obvious manner, i.e., by the natural embedding $\mathrm{End}(U)
\otimes \mathrm{End}(\mathbb {C}^N) \hookrightarrow \mathrm{End}(U
\otimes \mathbb{C}^N)$.

For present purposes, let $U_0$ and $U_1$ be Hermitian vector spaces
of equal dimension:
\begin{displaymath}
    U_1 = \mathbb{C}^n \;, \quad U_0 = \mathbb{C}^n \;.
\end{displaymath}
\begin{defn}\label{def 1.4}
If $A = \Xi\vert_{U_1 \to U_1}$ and $D = \Xi\vert_{U_0 \to U_0}$ let
%
%
$\mathcal{D} \subset \mathrm{End}(U)_0$ be the subset where $A$ and
$D$ are invertible and the spectrum of $D$ avoids the unit circle.
The key object of our approach for the case of $K = \mathrm{U}_N$ is
the superfunction $\chi : \mathcal{D} \to \wedge \, \mathrm{End}
(U)_1^\ast$ defined as the integral
\begin{equation}\label{int repn}
    \chi(\Xi) = \int_{K} \mathrm{SDet}
    (\mathrm{Id}_V - \Xi \otimes k)^{-1} \, dk \;.
\end{equation}
\end{defn}
\begin{rem}
By the spectral condition on $D\,$, the linear operator
$\mathrm{Id}_{V_0} - D \otimes k$ has an inverse for all $k \in
\mathrm{U}_N\,$, and this by the previous remark ensures that
(\ref{int repn}) exists.
\end{rem}

To appreciate why the superfunction $\chi$ of (\ref{int repn})
plays such an important role, evaluate its $\mathbb{C}$-number
part on the diagonal matrices $t \in \mathrm{End}(U)$ of (\ref{t
for U(N)}), which gives
\begin{displaymath}
    \chi(t) = \int\limits_{\mathrm{U}_N} \prod_{k = 1}^n
    \frac{\mathrm{Det}(\mathrm{Id}_N -\mathrm{e}^{\mathrm{i}
    \psi_k}\, u)}{\mathrm{Det}(\mathrm{Id}_N -
    \mathrm{e}^{\phi_k}\, u)}\, du \;.
\end{displaymath}
Comparison with the definitions (\ref{concise R}) and (\ref{def chi})
shows that this is exactly the autocorrelation function we want to
compute.

On the other hand, from Howe's theory of dual pairs the superfunction
$\chi$ of Def.\ \ref{def 1.4} can be interpreted as a primitive
character of the Lie supergroup $\mathfrak{G} = \mathrm{GL}_{n
|n}\,$. The task of conveying the precise meaning of this message
will occupy us for much of the present paper. The essential points
are summarized in the following subsection.

Please take note of the following strategic aspect.  We want to prove
Thm.\ \ref{thm0}, which is a statement about the complex-valued
function $t \mapsto \chi(t)$, the autocorrelation function of ratios
of characteristic polynomials defined in (\ref{def chi}). We will
therefore be aiming not at the superfunction $\chi(\Xi)$, but rather
at the ordinary function $\chi(t)$.  Our route to $\chi(t)$, however,
does lead through the extended object $\chi(\Xi)$. Indeed, it is the
very existence of $\chi(\Xi)$ that will put us in a position to
deduce various strong properties of $\chi(t)$ and thereby establish
the expression asserted by Thm.\ \ref{thm0} in the full range $N \in
\mathbb{N}$.

\subsection{Outline of strategy}\label{sect:strategy}

The superfunction $\chi(\Xi)$ defined by (\ref{int repn}) is
analytic, but only piecewise so, because of the singularities that
occur when one or several of the eigenvalues of $D = \Xi\vert_{
\mathrm{End}(U_0)}$ hit the unit circle. Since we want a correlation
of signature $(p , q)$, i.e., of a product of $p$ ratios involving $u
\in \mathrm{U}_N$ and $q$ ratios involving the complex conjugate
$\bar u\,$, we are interested in the domain of analyticity,
$\mathcal{D}_{p,\,q} \,$, where $p$ eigenvalues $\mathrm{e}^{\phi_j}$
of $D \in \mathrm{GL}(\mathbb{C}^{p+q})$ lie inside the unit circle
($\mathfrak{Re}\, \phi_j < 0$), and $q$ eigenvalues
$\mathrm{e}^{\phi_l}$ lie outside ($\mathfrak{Re}\, \phi_l > 0$).

The main step of our approach is to establish a certain system of
differential equations obeyed by $\chi(\Xi)$ and hence by $\chi(t)$.
In this endeavor we have to handle the unbounded operators of a
non-compact group acting on an infinite-dimensional space. For that
reason, most of the discussion of $\chi(\Xi)$ will be carried out not
in the complex space $\mathcal{D}_{p,\,q}$ but in a
\emph{real-analytic domain} $M \subset \mathcal{D}_{p,\,q} \,$, which
is specified as follows.

Let $T = T_1 \times T_0 \subset \mathrm{GL}(U_1) \times
\mathrm{GL}(U_0)$ be the Abelian semigroup with factors
\begin{displaymath}
    T_1 = \mathrm{U}_1^{p+q} \;, \quad
    T_0 = (0,1)^p \times (1,\infty)^q \;,
\end{displaymath}
where $\mathrm{U}_1 \equiv \mathrm{S}^1$ is the unit circle in the
complex number field $\mathbb{C}\,$. We identify this "torus" $T$
with the set of diagonal matrices $t = (t_1,t_0)$ of (\ref{t for
U(N)}),
\begin{displaymath}
    t_1 = \mathrm{diag}\,( \mathrm{e}^{\mathrm{i}\psi_1}, \ldots,
    \mathrm{e}^{\mathrm{i}\psi_n}) \;, \quad
    t_0 = \mathrm{diag}\,( \mathrm{e}^{\phi_1}, \ldots,
    \mathrm{e}^{\phi_n}) \quad (n = p+q) \;,
\end{displaymath}
where all of the variables $\psi_1\, , \ldots, \psi_n\, , \phi_1
\,, \ldots, \phi_n$ are now \emph{real-valued}, and the $\phi_k$
are restricted to the range
\begin{displaymath}
    \phi_j < 0 < \phi_l \quad (1 \le j \le p < l \le n).
\end{displaymath}

Our real-analytic domain $M$ will be a direct product $M = M_1 \times
M_0\,$.  To describe the second space $M_0$ let $\mathrm{U}_{p,\,q}$
be the pseudo-unitary group determined by the signature operator $s =
\mathrm{diag}(\mathrm{Id}_p\, , - \mathrm {Id}_q)$ on the Hermitian
vector space $U_0 \equiv \mathbb{C}^n = \mathbb{C}^p \oplus
\mathbb{C}^q$. Letting $g \cdot t_0 = g t_0 g^{-1}$ denote the
adjoint action of $g \in \mathrm{U}_{p,\,q}$ on $T_0 \subset
\mathrm{GL}(\mathbb{C}^n)$, we define
\begin{displaymath}
    M_0 := \mathrm{U}_{p,\,q} \cdot T_0 \;,
\end{displaymath}
i.e., $M_0$ is the union of all adjoint orbits of
$\mathrm{U}_{p,\,q}$ on the points of $T_0 \,$. Similarly, let
\begin{displaymath}
    M_1 := \mathrm{U}_{p+q} \cdot T_1 = \mathrm{U}_{p+q} \;,
\end{displaymath}
the union of adjoint orbits of the unitary group $\mathrm{U}( U_1) =
\mathrm{U}_{p+q}$ on $T_1\,$, which is the same as
$\mathrm{U}_{p+q}\,$. Note that $M_0 \subset \mathrm{GL}(U_0)$ is a
real-analytic submanifold of dimension $n^2$, and so is $M_1 \subset
\mathrm{GL}(U_1)$.  We shall study the superfunction $\chi(\Xi)$ on
the product manifold
\begin{displaymath}
    M := M_1 \times M_0 \;,
\end{displaymath}
i.e., for $\Xi \vert_{\mathrm{End}(U_1)} \in M_1$ and $\Xi
\vert_{\mathrm{End}(U_0)} \in M_0 \,$.

Now recall that $V = U \otimes \mathbb{C}^N = V_0 \oplus V_1$, and
regarding this as $V = \mathbb{C}^{n|n} \otimes \mathbb{C}^N$, let a
Howe dual pair $(\mathfrak{gl}_{n|n}\, , \mathrm{U}_N)$ \cite{howe}
act on $V$ as its fundamental module.  In Sect.\ \ref{sect:2} it is
shown that the function $Z : \,\, T \times \mathrm{U}_N \to
\mathbb{C}$ defined by
\begin{displaymath}
    Z(t,u) = \mathrm{SDet}(\mathrm{Id}_V - t \otimes u)^{-1}
\end{displaymath}
can be regarded as the character of that pair acting on a
spinor-oscillator module $\mathcal{A}_V \,$ of $V$, i.e., the tensor
product of an exterior algebra (built from $V_1$) with a symmetric
algebra (from $V_0)$. In physics, $\mathcal{A}_V$ is called a Fock
space for bosons and fermions. Since the action of the Howe pair
$(\mathfrak{gl}_{n|n}\,, \mathrm{U}_N)$ on $\mathcal{A}_V$ is known
to be multiplicity-free, the superfunction $\chi(\Xi)$ given by
(\ref{int repn}) is the character of an \emph{irreducible}
$\mathfrak{gl}_{n|n}$-representation.

For better readability, we first establish this statement in Sect.\
\ref{sect:2} for the easy case of a diagonal matrix $t \in T$. In
Sect.\ \ref{sect:4A} we then introduce a certain real-analytic
supermanifold $\mathcal{F}$ inside the complex supergroup $\mathrm
{GL}_{n|n}$ carrying left and right actions of the Lie superalgebra
$\mathfrak{gl}_{n|n}\,$, and we explain how to make sense of the
character $\chi(\Xi)$ as a section of $\mathcal{F}$ over the product
of real-analytic domains $M = M_1 \times M_0\,$.

Although the character $\chi$ is atypical, and is not covered by the
Weyl character formula and its known extensions, it can nonetheless
be computed by a method developed in this paper. In Sect.\
\ref{sect:4A} we accumulate a wealth of information about it:
\begin{itemize}
\item[$\bullet$] The function $\chi: \, T \to \mathbb{C}$ is analytic
and $W$-invariant.
\item[$\bullet$] The character $\chi$ is a linear combination of
exponential functions $\mathrm{e}^{\sum_k (\mathrm{i} m_k \psi_k -
n_k \phi_k)}$ where the weights are in the range $n_j \le 0 \le m_k
\le N \le n_l$ for $1 \le j \le p < l \le n\,$.
\item[$\bullet$] $\chi(\Xi)$ is degenerate with the character of the
trivial representation; thus $\chi(t)$ is annihilated by the radial
parts of all of the Laplace-Casimir operators of $\mathfrak{gl}_{n|n}
\,$.
\end{itemize}
Using the radial parts given by Berezin \cite{berezin}, we show that
the problem posed by properties (A-C) admits one and only one
solution: that stated in Thm.\ \ref{thm0}. Please note that although
the proof is carried out for real variables $\psi_k$ and $\phi_k\,$,
the final statement immediately extends to complex values of these
variables by analytic continuation.

As a historical note, let us mention that our treatment relies on
just two sources: Berezin \cite{berezin}, and Howe \cite{howe} -- and
these have existed in mathematics for 30 years now. Indeed, Berezin's
results are from 1975 and Howe's paper was written in 1976.

\section{Autocorrelation function of ratios as a primitive
character} \label{sect:2} \setcounter{equation}{0}

\subsection{Clifford algebra, spinor module, and character formula}
\label{sect:2.1}

Starting from a complex vector space $V = \mathbb{C}^d$, denote the
dual vector space by $V^\ast$ and form the direct sum $W := V \oplus
V^\ast$. On $W$ define a symmetric bilinear form $S : \, W \times W
\to \mathbb{C}$ by
\begin{equation}\label{symm form}
    S(v + \varphi, v^\prime + \varphi^\prime) = \varphi(v^\prime)
    + \varphi^\prime (v) \;.
\end{equation}
Then let $\mathfrak{c}(W)$ be the Clifford algebra of $W$, i.e.,
the associative algebra generated by $W \oplus \mathbb{C}$ with
anti-commutation relations
\begin{equation}\label{clifford}
    w w^\prime + w^\prime w = S(w\, ,w^\prime) \;.
\end{equation}
$\mathfrak{c}(W)$ is $\mathbb{Z}_2$-graded as a vector space by the
dichotomy of the degree being either even or odd.  Via the relations
(\ref{clifford}) the Clifford algebra $\mathfrak{c}(W)$ carries the
natural structure of a Lie superalgebra, with the fundamental bracket
being $[w , w^\prime] := w w^\prime + w^\prime w$.

\subsubsection{Spinor representation}

Basic to our approach is the spinor representation of $\mathfrak{c}
(W)$. Given the polarization $W = V \oplus V^\ast$ there exist {\it
two} natural realizations of it. Both of them are used in the quantum
field theory of Dirac fermions, where the positive and negative parts
of the Dirac operator call for two different quantization schemes.

Here, too, we need both realizations. To define the first one,
consider the exterior algebra $\wedge (V) = \oplus_{k=0}^d \wedge^k
(V)$. On it one has the operations of exterior multiplication
$\varepsilon\,$,
\begin{displaymath}
    V \times \wedge^k (V) \to \wedge^{k+1} (V) \;, \quad (v\, ,a)
    \mapsto \varepsilon(v)\, a = v\wedge a \;,
\end{displaymath}
and the operation of taking the inner product (or contraction)
$\iota$,
\begin{displaymath}
    V^\ast \times \wedge^k (V) \to \wedge^{k-1} (V) \;,  \quad
    (\varphi,a) \mapsto \iota(\varphi)\, a\;, {\phantom{= v\wedge a}}
\end{displaymath}
where the operator $\iota(\varphi)$ is the anti-derivation of
$\wedge(V)$ defined (for $v, v^\prime \in V$) by
\begin{displaymath}
    \iota(\varphi)\, 1 = 0 \;, \quad
    \iota(\varphi)\, v = \varphi(v) \;, \quad
    \iota(\varphi)\, (v \wedge v^\prime) =
    \varphi(v)\, v^\prime - \varphi (v^\prime)\, v \;,
\end{displaymath}
and so on. The algebraic relations obeyed by these operators of
exterior and interior multiplication are the so-called
\emph{canonical anti-commutation relations} (CAR):
\begin{eqnarray}
    &&\varepsilon(v)\, \varepsilon(v^\prime) + \varepsilon
    (v^\prime) \, \varepsilon(v) = 0 \;, \quad \iota(\varphi)\,
    \iota(\varphi^\prime)+ \iota(\varphi^\prime)\,
    \iota(\varphi) = 0 \;, \nonumber \\ &&\iota(\varphi)\,
    \varepsilon(v) + \varepsilon(v)\, \iota(\varphi) = \varphi(v)
    \, \mathrm{Id}_{\wedge(V)} \;. \label{CAR}
\end{eqnarray}
If one assigns to $w = v + \varphi \in V \oplus V^\ast = W$ a linear
operator on $\wedge(V)$ by
\begin{equation}\label{C1 rep}
    v + \varphi \mapsto \varepsilon(v) + \iota(\varphi) \;,
\end{equation}
then this assignment preserves the Clifford algebra relations
(\ref{clifford}) by virtue of the re\-lations (\ref{CAR}). Therefore,
(\ref{C1 rep}) defines a representation of the Clifford algebra
$\mathfrak{c} (W)$. It is called the spinor representation of
$\mathfrak{c}(W)$ on the spinor module $\wedge(V)$.

Next, consider realizing the same object $v + \varphi$ as an operator
on the exterior algebra $\wedge(V^\ast)$ of the dual vector space
$V^\ast$. This is done by the assignment
\begin{equation}\label{C2_rep}
    W = V \oplus V^\ast \ni
    v + \varphi \mapsto \iota(v) + \varepsilon(\varphi) \;,
\end{equation}
where $\iota(v) : \, \wedge^k(V^\ast) \mapsto \wedge^{k-1} (V^\ast)$
and $\varepsilon(\varphi) : \, \wedge^k(V^\ast) \mapsto \wedge^{k+1}
(V^\ast)$ now mean contraction by the vector $v$ and exterior
multiplication by the linear form $\varphi\,$. These operations still
satisfy the canonical anti-commutation relations and hence give a
second representation of the Clifford algebra $\mathfrak{c}(W)$.

The second representation is isomorphic to the first one. Indeed, if
we fix a generator $\Omega \in \wedge^d (V^\ast)$ then the mapping
$\tau : \, \wedge^k(V) \to \wedge^{d-k} (V^\ast)$ defined for $k = 0,
1, \ldots, d$ by
\begin{displaymath}
    \wedge^0(V) \ni 1 \mapsto \Omega \in \wedge^d(V^\ast) \;,\quad
    \wedge^1 (V) \ni \varepsilon(v) \cdot 1 \mapsto \iota(v) \Omega
    \in \wedge^{d-1}(V^\ast) \;,
\end{displaymath}
and so on, sends $\epsilon(v)$ to $\iota(v)$ and $\iota(\varphi)$ to
$\varepsilon(\varphi)$. It is called a \emph{particle-hole
transformation} in physics. As a small remark, note that $\tau$ is an
isomorphism of $\mathbb{Z}_2$-graded vector spaces only if $d$ is
even. (When $d$ is odd, $\tau$ exchanges the even and odd subspaces.)

\subsubsection{$\mathrm{GL}(V)$ character formula}

Let $\mathrm{GL}(V)$ be the group of invertible complex linear
trans\-formations of $V$. Denoting the action of $\mathrm {GL}(V)$ on
its fundamental module $V$ as $v \mapsto gv\,$, one has an induced
representation $\sigma^k : \mathrm {GL}(V) \to \mathrm{GL}( \wedge^k
V)$ on $\wedge^k (V)$ by
\begin{displaymath}
    \sigma^k(g) (v_1 \wedge v_2 \wedge \cdots \wedge v_k) = g v_1
    \wedge g v_2 \wedge \cdots \wedge g v_k \;.
\end{displaymath}
By taking the direct sum of these representations for $k = 0, 1,
\ldots, d\,$, one gets a canonical $\mathrm{GL}(V)$-repre\-sentation
$\sigma$ on the exterior algebra $\wedge(V)$:
\begin{equation}\label{defn sigma}
    \sigma : \,\, \mathrm{GL}(V) \to \mathrm{GL}(\wedge V) \;.
\end{equation}

It is a standard fact of multi-linear algebra that the alternating
sum of characters of the representations $\sigma^k$ is generated
by a determinant:
\begin{displaymath}
    \sum\nolimits_{k = 0}^d (-t)^k \, \mathrm{Tr}\, \sigma^k (g) =
    \mathrm{Det} (\mathrm{Id}_V - t g ) \qquad
    (t \in \mathbb{C}) \;.
\end{displaymath}
Setting $t = 1$ and denoting by $\mathrm{STr}_{\wedge V}$ the
operation of taking the supertrace over the $\mathbb{Z}_2$-graded
vector space $\wedge (V) = \wedge^\mathrm{even} (V) \oplus
\wedge^\mathrm{odd} (V)$, we write this formula as
\begin{equation}\label{eq:spin-char}
    \mathrm{STr}_{\wedge V}\, \sigma(g)
    = \mathrm{Det} (\mathrm{Id}_V - g) \;.
\end{equation}
If we specialize to the tensor-product situation $V = \mathbb{C}^p
\otimes \mathbb{C}^N$ and $g = t \otimes u\,$, with $u \in \mathrm{U}
(\mathbb{C}^N)$ and $t \in \mathrm{GL}(\mathbb{C}^p)$ a diagonal
transformation $t = \mathrm{diag}\, (t_1, \ldots, t_p)$, then we
obtain
\begin{displaymath}
    \mathrm{STr}_{\wedge V}\, \sigma(t \otimes u) = \prod\nolimits_{
    j = 1}^p \mathrm{Det}(\mathrm{Id}_N - t_j \, u)\;,
\end{displaymath}
which is "half" of the numerator in our basic product of ratios
(\ref{R for U(N)}).

In order to produce the other half, which involves the complex
conjugates $\bar u\,$, we are going to employ a $\mathrm{GL}
(V)$-representation on $\wedge(V^\ast)$. Observing that $g \in
\mathrm{GL}(V)$ acts on the dual vector space $V^\ast$ by $g \varphi
:= \varphi \circ g^{- 1}$, we take $\widetilde{\sigma} : \,
\mathrm{GL}(V) \to \mathrm{GL} (\wedge V^\ast)$ to be the direct sum
$\widetilde{\sigma} = \oplus_{k=0}^d \widetilde{\sigma}^k$ of
representations $\widetilde{ \sigma}^k : \, \mathrm{GL}(V) \to
\mathrm{GL}(\wedge^k V^\ast)$ defined by
\begin{equation}\label{tilde sigma}
    \widetilde{\sigma}^k(g) (\varphi_1 \wedge\cdots\wedge \varphi_k)
    = \mathrm{Det}(g) \, (\varphi_1 \circ g^{-1}) \wedge \cdots
    \wedge (\varphi_k \circ g^{-1}) \;.
\end{equation}
Notice that $\widetilde{\sigma}^d : \, \mathrm{GL}(V) \to \mathrm{GL}
(\wedge^d V^\ast)$ is the trivial representation, $\widetilde{\sigma}
^d (g) \Omega = \Omega\,$, isomorphic to $\sigma^0 : \, \mathrm{GL}
(V) \to \mathrm{GL}(\wedge^0 V)$. As a matter of fact, the insertion
of the determinant multiplier makes the two representations $\sigma$
and $\widetilde{\sigma}$ isomorphic by the particle-hole
transformation $\tau : \, \wedge^k(V) \to \wedge^{d-k} (V^\ast)$. We
will see this at the infinitesimal level below. Since $\tau$ for $d
\in 2\mathbb{N}-1$ interchanges the even and odd subspaces we have
\begin{equation}\label{eq:2.9}
    \mathrm{STr}_{\wedge V^\ast} \widetilde{\sigma}(g) = (-1)^d \,
    \mathrm{Det}(\mathrm{Id}_V - g) = \mathrm{Det}(g - \mathrm{Id}_V)
    \;.
\end{equation}

How are $\sigma$ and $\widetilde{\sigma}$ related to the spinor
representation (\ref{C1 rep}) of $\mathfrak{c}(W)$? To answer that
question, let $\{e_1, \ldots, e_d\}$ be a basis of $V$ with
corresponding dual basis $\{f^1, \ldots, f^d \}$ of $V^\ast$, and
consider the sequence of linear mappings
\begin{displaymath}
    \mathfrak{gl}(V) \to V \otimes V^\ast \to \mathfrak{gl}(\wedge V)
    \;, \quad X \mapsto \sum (X e_i) \otimes f^i \mapsto \sum
    \varepsilon(X e_i) \iota(f^i) \;.
\end{displaymath}
The first map is a Lie algebra isomorphism. The second map turns $v
\otimes \varphi \in V \otimes V^\ast$ into an element $v \varphi$ of
the Clifford algebra $\mathfrak{c}(W)$ by dropping the tensor product
and then sends $v \varphi$ to an operator on $\wedge(V)$ by (\ref{C1
rep}). By the CAR relations (\ref{CAR}), this map $v \otimes \varphi
\mapsto \varepsilon(v) \iota(\varphi)$ is a Lie algebra homomorphism.
Altogether, the composite map $X \mapsto \sum \varepsilon(X e_i)
\iota(f^i)$ is a representation of $\mathfrak {gl}(V)$ on $\wedge
(V)$. By the same token, the linear mapping $X \mapsto \sum \iota(X
e_i) \varepsilon(f^i)$ is a representation of $\mathfrak{gl}(V)$ on
$\wedge(V^\ast)$. These Lie algebra representations are related to
the Lie group representations $\sigma$ and $\widetilde{\sigma}$ as
follows.
\begin{prop}\label{prop:2.1}
Taking the differential of the representations $\sigma : \,
\mathrm{GL}(V) \to \mathrm{GL}(\wedge V)$ and $\widetilde{\sigma} :
\, \mathrm{GL}(V) \to \mathrm{GL}(\wedge V^\ast)$ at the neutral
element one has
\begin{displaymath}
    \sigma_\ast(X) := \frac{d}{dt} \sigma(\mathrm{e}^{tX})
    \Big\vert_{t = 0} = \sum_{i = 1}^d \varepsilon(X e_i)\iota(f^i)
    \;, \qquad \widetilde{\sigma}_\ast(X) = \sum_{i = 1}^d
    \iota(X e_i) \varepsilon(f^i) \;.
\end{displaymath}
\end{prop}
\begin{proof}
Apply the operator $\sigma^\prime(X) := \sum \varepsilon(X e_i)
\iota(f^i)$ to any element $v \in V$:
\begin{displaymath}
    \sigma^\prime(X) v = \sum\nolimits_i \varepsilon(X e_i) f^i(v)
    = \varepsilon(X v) \cdot 1 = X v = \frac{d}{dt} \Big\vert_{t=0}
    \sigma^1(\mathrm{e}^{tX}) v = \sigma_\ast(X) v \;.
\end{displaymath}
Thus $\sigma^\prime(X) = \sigma_\ast(X)$ on $V$. Now the operator
$D \equiv \sigma^\prime(X)$ satisfies the Leibniz rule
\begin{displaymath}
    D(v_1 \wedge v_2) = D v_1 \wedge v_2 + v_1 \wedge D v_2 \;,
\end{displaymath}
as follows easily from the definition of exterior and interior
multiplication. Thus $\sigma^\prime(X)$ is a derivation of the
exterior algebra $\wedge(V)$ just as the differential $\sigma_\ast
(X)$ is. Therefore, since $\sigma^\prime(X)$ and $\sigma_\ast(X)$
agree on the generating vector space $V$, they agree on $\wedge(V)$.

In the case of $\widetilde{\sigma}^\prime(X) := \sum \iota(X e_i)
\varepsilon(f^i)$ one uses CAR to write $\widetilde{ \sigma}^\prime
(X) = \mathrm{Tr}(X) \mathrm{Id}_{\wedge V^\ast} + \sum \varepsilon(-
X^\mathrm{t} f^i) \iota(e_i)$. The first term arises from linearizing
$\mathrm{e}^{tX} \mapsto \mathrm{Det} (\mathrm{e}^{tX})$ at $t =
0\,$. The second term is a derivation of $\wedge(V^\ast)$ and the
rest of the argument goes as above.
\end{proof}
What we have told up to now was the "fermionic" variant (in physics
language) of a story that has to be told twice.

\subsection{Weyl algebra and oscillator module}\label{sect:2.3}

For the second variant of our story -- the "bosonic" one -- consider
again $V = \mathbb{C}^d$ and $W = V \oplus V^\ast$, but now equipped
with the \emph{alternating} bilinear form $A : \, W \times W \to
\mathbb{C}$ given by
\begin{equation}\label{alt form}
    A(v + \varphi , v^\prime + \varphi^\prime) = \varphi(v^\prime)
    - \varphi^\prime(v) \;.
\end{equation}
The associative algebra generated by $W \oplus \mathbb{C}$ with
the relations
\begin{equation}\label{weylalgebra}
    w w^\prime - w^\prime w = A(w\, ,w^\prime)
\end{equation}
is denoted by $\mathfrak{w}(W)$, and is called the Weyl algebra of
$W$.

To construct a representation of $\mathfrak{w}(W)$, let $\mathrm{S}
(V) = \oplus_{k \ge 0} \mathrm{S}^k(V)$ be the symmetric algebra of
$V$, and consider on it the operations of multiplication $\mu$ and
derivation $\delta$:
\begin{eqnarray*}
    &&V \times \mathrm{S}^k(V) \to \mathrm{S}^{k+1}(V) \;, \quad
    (v\, ,a) \mapsto \mu(v)\, a = va = av \;, \\ &&V^\ast \times
    \mathrm{S}^k (V) \to \mathrm{S}^{k-1}(V) \;, \quad (\varphi,a)
    \mapsto \delta(\varphi)\, a \;.
\end{eqnarray*}
The operator $\delta(\varphi)$ obeys the Leibniz rule $\delta(
\varphi)\, v^k = k \, v^{k-1} \varphi(v)$. The algebraic relations
satisfied by $\mu$ and $\delta$ are the \emph{canonical
commutation relations} (CCR):
\begin{eqnarray}
    &&\mu(v)\, \mu(v^\prime) - \mu(v^\prime) \, \mu(v) = 0 \;,
    \quad \delta(\varphi)\, \delta(\varphi^\prime) - \delta(
    \varphi^\prime)\, \delta(\varphi) = 0 \;, \nonumber \\
    &&\delta(\varphi)\, \mu(v) - \mu(v)\,\delta(\varphi) =
    \varphi(v)\, \mathrm{Id}_{\mathrm{S} V} \;. \label{CCR}
\end{eqnarray}
Letting $w = v + \varphi \in V \oplus V^\ast = W$ act on
$\mathrm{S}(V)$ by
\begin{equation}\label{W1 rep}
    v + \varphi \mapsto \mu(v) + \delta(\varphi) \;,
\end{equation}
one gets a representation of the Weyl algebra $\mathfrak{w}(W)$ on
$\mathrm{S}(V)$. The representation space $\mathrm{S}(V)$ is
sometimes referred to as the \emph{oscillator module} of
$\mathfrak{w}(V \oplus V^\ast)$, as it carries the so-called
oscillator representation of the metaplectic group of $W$.

A Hermitian structure $\langle \, , \, \rangle$ on $V$ induces a
canonical Hermitian structure on $\mathrm{S}(V)$ as follows. Let a
complex anti-linear isomorphism $c : \, V \to V^\ast$ be defined by
$c\, v = \langle v\, , \cdot \rangle \,$, and extend this to a
mapping $c :\, \mathrm{S}(V) \to \mathrm{S}(V^\ast)$ in the natural
manner, i.e., by applying the map $c$ to every factor of an element
of $\mathrm{S}(V)$. For example, $c (v v^\prime) = (c v) (c v^\prime)
$. Now, using the canonical pairing of the vector space $\mathrm{S}
(V)$ with its dual vector space $\mathrm{S}(V)^\ast = \mathrm{S}
(V^\ast)$ one defines a Hermitian scalar product $\langle \, , \,
\rangle_{\mathrm{S}V} : \, \mathrm{S}(V) \times \mathrm{S}(V) \to
\mathbb{C}$ by
\begin{displaymath}
    \langle \Phi , \Phi^\prime \rangle_{\mathrm{S} V} =
    (c\, \Phi)(\Phi^\prime) \;.
\end{displaymath}
With respect to this Hermitian structure of $\mathrm{S}(V)$ one has
\begin{equation}\label{eq eps-iota}
    \mu(v)^\dagger = \delta(c\, v) \;,
\end{equation}
and in this sense the operator $\delta$ is adjoint to $\mu$.
(Throughout this paper the symbol $\dagger$ denotes the operation of
taking the adjoint w.r.t.\ a given Hermitian structure.)

There exists a second way of implementing the relations
(\ref{weylalgebra}) of the Weyl algebra $\mathfrak{w}(W)$. Instead of
$\mathrm{S}(V)$ consider the symmetric algebra $\mathrm{S}(V^\ast)$
of the dual vector space $V^\ast$, and let $v + \varphi \in W$
operate on $\mathrm{S}(V^\ast)$ by
\begin{equation}\label{W2 rep}
    v + \varphi \mapsto \delta(v) - \mu(\varphi) \;.
\end{equation}
With the modified sign ($\delta \to \mu$ and $\mu \to -\delta$),
which is forced by the alternating nature of the basic form (\ref{alt
form}), this still defines a representation of $\mathfrak{w}(W)$.

A Hermitian structure of $V$ still induces a canonical Hermitian
structure of $\mathrm{S}(V^\ast)$, and the relations (\ref{eq
eps-iota}) continue to hold in the adapted form $\mu(\varphi)^\dagger
= \delta(c^{-1}\varphi)$, or $\delta(v)^\dagger = \mu(c v)$. Our full
setup below will utilize both representations (\ref{W1 rep}) and
(\ref{W2 rep}).

\subsection{Oscillator character}\label{sect:2.4}

Let the complex Lie group $\mathrm{GL}(V)$ still act on its
fundamental module $V = \mathbb{C}^d$ as $v \mapsto gv\,$, but now
consider the induced $\mathrm{GL}(V)$-representations $\omega^k$ on
the symmetric powers $\mathrm {S}^k(V)$:
\begin{displaymath}
    \omega^k (g) (v_1 \cdots v_k) = (g v_1) \cdots (g v_k) \;.
\end{displaymath}
These combine to a representation $\omega$ on the symmetric algebra
$\mathrm{S}(V)$:
\begin{equation}\label{defn omega}
    \oplus_{k \ge 0}^{\vphantom{\dagger}} \, \omega^k = \omega
    : \,\, \mathrm{GL}(V) \to \mathrm{GL}(\mathrm{S} V) \;.
\end{equation}
Since the sum over symmetric powers is an infinite sum, the
character of this representation does not exist for all $g \in
\mathrm{GL}(V)$. Rather, the following is true.

Equipping $V = \mathbb{C}^d$ with its standard Hermitian structure
$\langle \, , \, \rangle\,$, let $g \mapsto g^\dagger$ denote the
operation of taking the adjoint with respect to $\langle \, , \,
\rangle$. Then consider the set
\begin{equation}\label{defn H<}
    \mathrm{H}^<(V) = \{ h \in \mathrm{GL}(V) \mid
     h^\dagger h < \mathrm{Id}_V \} \;,
\end{equation}
i.e., the set of elements $h \in \mathrm{GL}(V)$ with the property
$\langle h v \, , h v \rangle < \langle v \, , v \rangle$ for all $v
\in V$. Clearly, every eigenvalue $\lambda$ of $h \in \mathrm{H}^<
(V)$ has absolute value $| \lambda | < 1$.  If $g \in \mathrm{GL}(V)$
and $h \in \mathrm{H}^<(V)$, then $h^\dagger h < \mathrm{Id}_V$
implies $(hg)^\dagger hg < g^\dagger g$. Thus $\mathrm{H}^<(V)$ is a
semigroup.

Now, if $g \in \mathrm{GL}(V)$ can be diagonalized and has
eigenvalues $\lambda_1, \ldots, \lambda_d$, the character of the
representation $\omega^k : \, \mathrm{GL}(V) \to \mathrm{GL}
(\mathrm{S}^k V)$ takes on $g$ the value
\begin{displaymath}
    \mathrm{Tr}\, \omega^k(g) = \sum_{1 \le i_1 \le \ldots
    \le i_k \le d} \lambda_{i_1} \cdots \lambda_{i_k} \;.
\end{displaymath}
If the additional requirement $g \in \mathrm{H}^< (V)$ is imposed,
these values can be convergently summed over all symmetric powers
$k \ge 0$ to give
\begin{displaymath}
    \sum_{k = 0}^\infty \mathrm{Tr}\, \omega^k(g) = \sum_{ \{ n_i
    \} \in (\mathbb{N} \cup \{0\})^d} \lambda_1^{n_1} \cdots
    \lambda_d^{n_d} = \prod_{i = 1}^d (1 - \lambda_i)^{-1} \;.
\end{displaymath}
Clearly, the right-hand side is the reciprocal of a determinant,
$\mathrm{Det}^{-1}(\mathrm{Id}_V - g)$.  Thus the character of the
representation (\ref{defn omega}) can be written as
\begin{equation}\label{eq:osc-char}
    \mathrm{Tr}_{\mathrm{S}V}\, \omega(g)
    = \mathrm{Det}^{-1}(\mathrm{Id}_V - g)
\end{equation}
for diagonalizable $g \in \mathrm{H}^<(V)$, in which form the formula
extends to the case of general $g \in \mathrm{H}^<(V)$. Indeed, by
the Jordan decomposition every $g$ is a sum of semisimple (i.e.,
diagonalizable) and nilpotent parts, and the nilpotent part of $g$
contributes neither to the character on the left-hand side nor to the
determinant on the right-hand side.

Besides the $\mathrm{GL}(V)$-representation $\omega$ on
$\mathrm{S}(V)$, we also need the representation $\widetilde{\omega}
: \, \mathrm{GL}(V) \to \mathrm{GL}(\mathrm{S} V^\ast)$ which is
defined as the direct sum $\widetilde{\omega} = \oplus_{k \ge 0}\,
\widetilde{\omega}^k$,
\begin{displaymath}
    \widetilde{\omega}^k(g) (\varphi_1 \cdots \varphi_k) =
    \mathrm{Det}^{-1}(g) (\varphi_1 \circ g^{-1}) \cdots (\varphi_k
    \circ g^{-1}) \;,
\end{displaymath}
of components $\widetilde{ \omega}^k : \,\, \mathrm{GL}(V) \to
\mathrm{GL} (\mathrm{S}^k V^\ast)$. An important observation is that
although we have $\sigma = \tau^{-1} \widetilde{\sigma} \tau$ on the
fermionic side, the representations $\omega$ and $\widetilde{\omega}$
are \emph{not} isomorphic. (There is no such thing as a particle-hole
transformation $\tau$ for bosons.) Indeed, while the character of
$\omega$ exists on $\mathrm{H}^<(V)$, that of $\widetilde{\omega}$
exists on the opposite semigroup $\mathrm{H}^>(V) = \{h\in\mathrm{GL}
(V) \mid h^\dagger h > \mathrm{Id}_V \}$, where it has the value
\begin{equation}\label{eq:2.19}
    \mathrm{Tr}_{\mathrm{S} V^\ast}\, \widetilde{\omega}(h) =
    \mathrm{Det}^{-1}(h) \mathrm{Det}^{-1} (\mathrm{Id}_V - h^{-1})
    = \mathrm{Det}^{-1}(h - \mathrm{Id}_V) \;.
\end{equation}

The following statement is an analog of Prop.\ \ref{prop:2.1} and is
proved in the same way.
\begin{prop}\label{prop:2.4}
Taking the differential of the representations $\omega : \,
\mathrm{GL}(V) \to \mathrm{GL}(\mathrm{S} V)$ and $\widetilde{\omega}
: \, \mathrm{GL}(V) \to \mathrm{GL}(\mathrm{S} V^\ast)$ at the
neutral element one has
\begin{displaymath}
    \omega_\ast(X) := \sum_{i = 1}^d \mu(X e_i) \delta(f^i)
    \;, \quad \widetilde{\omega}_\ast(X) = - \sum_{i = 1}^d
    \delta(X e_i) \mu(f^i) \;.
\end{displaymath}
\end{prop}

\subsection{Enter supersymmetry}\label{sect:2.5}

Drawing on the material of Sects.\ \ref{sect:2.1}--\ref{sect:2.4}, we
can now express the product of ratios of characteristic polynomials
(\ref{R for U(N)}) as a product of characters. Let $V$ be a
$\mathbb{Z}_2$-graded vector space which decomposes as
\begin{equation}\label{susy V}
    V = V_1 \oplus V_0 = (V_1^+ \oplus V_1^-) \oplus (V_0^+ \oplus
    V_0^-) \;,
\end{equation}
and introduce the following tensor product of spinor and oscillator
modules:
\begin{equation}\label{spin-osc module}
    \mathcal{V} := \wedge (V_1^+) \otimes \wedge({V_1^-}^\ast)
    \otimes \mathrm{S}(V_0^+) \otimes \mathrm{S}({V_0^-}^\ast)\;.
\end{equation}
Let a product of representations
\begin{displaymath}
    R : \,\, \mathbf{G} \equiv \mathrm{GL}(V_1^+) \times
    \mathrm{GL}(V_1^-)\times \mathrm{GL}(V_0^+) \times
    \mathrm{GL}(V_0^-)\to \mathrm{GL}(\mathcal{V})
\end{displaymath}
be defined by
\begin{equation}\label{eq:2.22}
    R(g_1^+, g_1^-, g_0^+, g_0^-) = \sigma(g_1^+)\widetilde{\sigma}
    (g_1^-) \omega(g_0^+) \widetilde{\omega}(g_0^-) \;,
\end{equation}
where each of the factors $\sigma$, $\widetilde{\sigma}$, $\omega$,
and $\widetilde{\omega}$ operates on the corresponding factor in the
tensor product $\mathcal{V}$ (and is trivial on all other factors).
The character formulas (\ref{eq:spin-char}), (\ref{eq:2.9}),
(\ref{eq:osc-char}), and (\ref{eq:2.19}) then combine to the
following statement for the character $\mathrm{STr}_\mathcal{V} R =
\mathrm{Tr}_{ \mathcal{V}_0} R - \mathrm{Tr}_{\mathcal{V}_1} R$ of
the $\mathbb{Z}_2$-graded representation space $\mathcal{V} =
\mathcal{V}_0 \oplus \mathcal{V}_1\,$.
\begin{prop}\label{prop:6707}
The character of the $\mathbb{Z}_2$-graded representation $R$ of Eq.\
(\ref{eq:2.22}) exists for $(x,y,w,z) \in \mathrm{GL}(V_1^+) \times
\mathrm{GL}(V_1^-) \times \mathrm{H}^<(V_0^+) \times \mathrm{H}^>
(V_0^-)$ where it has the value
\begin{displaymath}
    \mathrm{STr}_\mathcal{V} R(x,y,w,z) =
    \frac{\mathrm{Det}(\mathrm{Id}_{V_1^+} - x)
    \, \mathrm{Det}(y - \mathrm{Id}_{V_1^-})}
    {\mathrm{Det}(\mathrm{Id}_{V_0^+} - w)
    \, \mathrm{Det}(z - \mathrm{Id}_{V_0^-})} \;.
\end{displaymath}
\end{prop}
We now take the components of the $\mathbb{Z}_2$-graded vector space
$V$ of (\ref{susy V}) to be
\begin{displaymath}
    V_\tau^\pm = U_\tau^\pm \otimes \mathbb{C}^N \quad (\tau = 0,1)
\end{displaymath}
with Hermitian vector spaces $U_0^+ = U_1^+ = \mathbb{C}^p$ and
$U_0^- = U_1^- = \mathbb{C}^q$. We also introduce $t := (t_1^+,
t_1^-, t_0^+, t_0^-)$ where the components $t_\tau^\pm$ are diagonal
operators in $\mathrm {End}(U_\tau^\pm)$,
\begin{displaymath}
\begin{array}{ll}
    t_1^+ = \mathrm{diag}( \mathrm{e}^{\mathrm{i} \psi_1},
    \ldots, \mathrm{e}^{\mathrm{i} \psi_p}) \;, \quad &t_1^- =
    \mathrm{diag}(\mathrm{e}^{\mathrm{i}\psi_{p+1}}, \ldots,
    \mathrm{e}^{\mathrm{i}\psi_{p+q}}) \;, \\ t_0^+ = \mathrm{diag}
    (\mathrm{e}^{\phi_1}, \ldots, \mathrm{e}^{\phi_p}) \;,
    \quad &t_0^- = \mathrm{diag}(\mathrm{e}^{\phi_{p+1}},\ldots,
    \mathrm{e}^{\phi_{p+q}})\;,
\end{array}
\end{displaymath}
with complex parameters $\psi_1, \ldots, \phi_{p+q}\,$. Let $u\in
\mathrm{U}_N\,$, the unitary group of the Hermitian vector space
$\mathbb{C}^N$. With every pair $(t,u)$ we associate an element $t
\otimes u \in \mathbf{G}$ by
\begin{displaymath}
    t \otimes u \equiv (t_1^+ \otimes u \, , \, t_1^- \otimes
    u\, , \, t_0^+ \otimes u\, , \, t_0^- \otimes u) \;.
\end{displaymath}
\begin{cor}\label{cor 2.5}
If the parameters are restricted to the range $\mathfrak {Re}\,
\phi_j < 0 < \mathfrak{Re}\, \phi_l$ for $j = 1, \ldots, p$ and $l =
p+1, \ldots, p+q \,$, the operator $R(t \otimes u) : \, \mathcal{V}
\to \mathcal{V}$ is trace class and
\begin{displaymath}
    \mathrm{STr}_\mathcal{V} R(t \otimes u) = \prod_{j=1}^p
    \frac{\mathrm{Det}(\mathrm{Id}_N - \mathrm{e}^{\mathrm{i}\psi_j}
    \, u)}{\mathrm{Det}(\mathrm{Id}_N - \mathrm{e}^{\phi_j}\, u)}\,
    \prod_{l = p+1}^{p+q} \frac{\mathrm{Det}(\mathrm{e}^{\mathrm{i}
    \psi_l}\,\mathrm{Id}_N -\bar{u})}{\mathrm{Det}(\mathrm{e}^{\phi_l}
    \, \mathrm{Id}_N - \bar{u})} \;.
\end{displaymath}
\end{cor}
\begin{proof}
By the restriction on the range of the parameters $\phi_j$ and
$\phi_l\,$, the operator $t_0^+ \otimes u$ lies in $\mathrm{H}^<
(U_0^+ \otimes \mathbb{C}^N)$ and $t_0^- \otimes u$ lies in
$\mathrm{H}^> (U_0^- \otimes \mathbb{C}^N)$. Therefore the formula of
Prop.\ \ref{prop:6707} applies with $(x\,,y\,,w\,,z) = (t_1^+ \otimes
u\, ,\, t_1^- \otimes u\, ,\, t_0^+ \otimes u\, ,\, t_0^- \otimes
u)$. In particular, it follows that the supertrace of $R(t \otimes
u)$ converges absolutely. By the elementary manipulation
\begin{displaymath}
    \frac{\mathrm{Det}(\mathrm{e}^{\mathrm{i}\psi_l}\, u -
    \mathrm{Id}_N)} {\mathrm{Det}(\mathrm{e}^{\phi_l}\, u -
    \mathrm{Id}_N)} = \frac{\mathrm{Det}(\mathrm{e}^{\mathrm{i}
    \psi_l}\,\mathrm{Id}_N - \bar{u})}{\mathrm{Det}(
    \mathrm{e}^{\phi_l} \,\mathrm{Id}_N - \bar{u})} \;,
\end{displaymath}
the expression for $\mathrm{STr}_\mathcal{V} R(t \otimes u)$ is
brought into the stated form.
\end{proof}
\begin{rem}
Note that the expression for $\mathrm{STr}_\mathcal{V} R(t \otimes
u)$ can be written as
\begin{displaymath}
    \mathrm{STr}_\mathcal{V} R(t \otimes u) = \prod_{k=1}^{p+q}
    \frac{\mathrm{Det}(\mathrm{Id}_N-\mathrm{e}^{\mathrm{i}\psi_k}\,u)}
    {\mathrm{Det}(\mathrm{Id}_N - \mathrm{e}^{\phi_k} \, u)} \;.
\end{displaymath}
This simpler expression hides the fact that complex conjugates
$\bar{u}$ are lurking here by the implicit condition on the range of
the parameters $\phi_k \,$.  Nonetheless, it is this last expression
for $\mathrm{STr}_\mathcal{V} R(t \otimes u)$ which will be quoted in
Prop.\ \ref{howe-char} below.
\end{rem}
Cor.\ \ref{cor 2.5} says that (\ref{R for U(N)}) is a character.
While of no immediate use by itself, this observation becomes a
powerful fact when combined with the following message.

Since $(t,u) = (t,\mathrm{Id}) \cdot (\mathrm{Id}\, ,u) = (\mathrm
{Id}\, ,u) \cdot (t,\mathrm{Id})$ and the sequence of maps
\begin{displaymath}
    (t,u) \mapsto t \otimes u \mapsto R(t \otimes u)
\end{displaymath}
is a homomorphism, we have a decomposition
\begin{equation}\label{factorization}
    R(t \otimes u) = \rho(t) r(u) = r(u) \rho(t) \;,
\end{equation}
where the factors are
\begin{equation}\label{howe factors}
    \rho(t) := R(t \otimes \mathrm{Id}) \;, \quad
    r(u) := R(\mathrm{Id} \otimes u) \;.
\end{equation}
It will turn out that these factors correspond to a so-called
\emph{Howe pair} $(\mathfrak{gl}_{n|n}\, , \mathrm{U}_N)$ where $n =
p+q\,$. To appreciate this message and reap full benefit from its
representation-theoretic impact, we are now going to broaden our
framework.

\subsection{Supersymmetric framework}\label{sect:susyframe}

Recall from Sect.\ \ref{sect:basics} that a $\mathbb{Z}_2$-graded
complex vector space $V = V_1 \oplus V_0$ determines a complex Lie
superalgebra $\mathfrak{gl}(V)$, whose Lie superbracket is given by
the supercommutator $[X,Y] = XY - (-1)^{|X| |Y|} YX$ for homogeneous
elements $X\, , Y \in \mathfrak {gl}(V)$.

Another algebraic structure associated with $V = V_1\oplus V_0\,$,
or rather with
\begin{displaymath}
    W = V \oplus V^\ast = (V_1^{\vphantom{\ast}} \oplus V_1^\ast)
    \oplus (V_0^{\vphantom{\ast}} \oplus V_0^\ast) =: W_1 \oplus
    W_0 \;,
\end{displaymath}
is the \emph{Clifford-Weyl algebra} of $W$, which unifies the
Clifford algebra $\mathfrak{c}(W_1)$ of the odd component $W_1$
with the Weyl algebra $\mathfrak{w}(W_0)$ of the even component
$W_0\,$.
\begin{defn}
Let $Q :\, W \times W \to \mathbb{C}$ be the non-degenerate complex
bilinear form for which $W = W_1 \oplus W_0$ is an orthogonal
decomposition and which restricts to the canonical symmetric form $S$
on $W_1^{\vphantom{\ast}} = V_1^{\vphantom{\ast}} \oplus V_1^\ast$
and the canonical alternating form $A$ on $W_0^{\vphantom{ \ast}} =
V_0^{\vphantom{\ast}} \oplus V_0^\ast$. Then the Clifford-Weyl
algebra of the $\mathbb{Z}_2$-graded vector space $W$ is defined to
be the associative algebra generated by $W \oplus \mathbb{C}$ with
relations
\begin{equation}\label{clifford-weyl}
    w w^\prime - (-1)^{|w| \, |w^\prime|} w^\prime w =
    Q(w,w^\prime)
\end{equation}
for homogeneous $w, w^\prime \in W$.  We denote it by $\mathfrak
{q}(W)$ (with $\mathfrak{q}$ as in "quantum").
\end{defn}
\begin{rem}
Although $\mathfrak{q}(W)$ is primarily defined as an associative
algebra, it carries a natural Lie superalgebra structure. Indeed,
$\mathfrak{q}(W)$ as a vector space inherits from the Clifford
algebra $\mathfrak{c}(W_1)$ a $\mathbb{Z}_2$-grading by the
isomorphism
\begin{eqnarray*}
    \mathfrak{q}(W) &\simeq& \mathfrak{c}(W_1) \otimes
    \mathfrak{w}(W_0) \\ &\simeq& \big(\mathfrak{c}^\mathrm{even}
    (W_1) \otimes \mathfrak{w}(W_0) \big) \oplus \big(
    \mathfrak{c}^\mathrm{odd}(W_1) \otimes \mathfrak{w}(W_0) \big)
     \;,
\end{eqnarray*}
and the supercommutator (\ref{sbracket}) determined by this
$\mathbb{Z}_2$-grading is compatible with the defining relations
(\ref{clifford-weyl}).
\end{rem}
There is a canonical way in which the Lie superalgebra $\mathfrak
{gl}(V)$ is realized inside the Clifford-Weyl algebra $\mathfrak
{q}(V \oplus V^\ast)$. To describe this realization, fix any
homogeneous basis $\{ e_i \}_{i = 1, \ldots, \mathrm{dim} \, V}$
of $V$, and denote the dual basis of $V^\ast$ by $\{ f^i \}$. Then
consider the isomorphism of $\mathbb{Z}_2$-graded vector spaces
\begin{equation}\label{isomorph}
    \mathfrak{gl}(V) \to V \otimes V^\ast \;, \quad X \mapsto
    \sum (X e_i) \otimes f^i \;.
\end{equation}
Of course the tensor product $V \otimes V^\ast$ is equipped with
the induced $\mathbb{Z}_2$-grading
\begin{eqnarray*}
    (V \otimes V^\ast)_1 &=& (V_0^{\vphantom{\ast}} \otimes
    V_1^\ast) \oplus (V_1^{\vphantom{\ast}} \otimes V_0^\ast)
    \;, \\ (V \otimes V^\ast)_0 &=& (V_0^{\vphantom{\ast}}
    \otimes V_0^\ast) \oplus (V_1^{\vphantom{\ast}} \otimes
    V_1^\ast) \;,
\end{eqnarray*}
and with the bracket (for homogeneous elements $v\, , v^\prime \in V$
and $\varphi, \varphi^\prime \in V^\ast$)
\begin{equation}\label{sbracket tensor}
    [ v \otimes \varphi , v^\prime \otimes \varphi^\prime ] :=
    \varphi(v^\prime)\, v \otimes \varphi^\prime - (-1)^{|v \otimes
    \varphi| \, |v^\prime \otimes \varphi^\prime|}
    \varphi^\prime(v)\, v^\prime \otimes \varphi \;.
\end{equation}
With these conventions one immediately verifies that
\begin{displaymath}
    \left[ \sum (X e_i) \otimes f^i \, , \, \sum (Y e_j)
    \otimes f^j \right] = \sum ([X,Y] e_i) \otimes f^i \;.
\end{displaymath}
Thus (\ref{isomorph}) is an isomorphism of Lie superalgebras.
\begin{lem}\label{lem 2.7}
If $W = V \oplus V^\ast$, the linear mapping
\begin{displaymath}
    \mathfrak{gl}(V) \to \mathfrak{q}(W)\;, \quad
    X \mapsto \sum (X e_i) f^i \equiv \hat{X}\;,
\end{displaymath}
is a homomorphism of Lie superalgebras.
\end{lem}
\begin{proof}
We regard our mapping $\mathfrak{gl}(V) \to \mathfrak{q}(V \oplus
V^\ast)$ as a sequence of two mappings: first is the map
$\mathfrak{gl}(V) \to V \otimes V^\ast$ of (\ref{isomorph}), and
this is followed by
\begin{displaymath}
    V \otimes V^\ast \to \mathfrak{q}(V \oplus V^\ast)
    \;, \quad v \otimes \varphi \mapsto v \varphi \;.
\end{displaymath}
Since the first map is an isomorphism, $X \mapsto \hat{X}$ is a
homomorphism if $v \otimes \varphi \mapsto v \varphi$ is.  To show
that the latter is true, we use the defining Clifford-Weyl
relations (\ref{clifford-weyl}) to do the following computation
(for $\mathbb{Z}_2$-homogeneous factors):
\begin{eqnarray*}
    v \varphi v^\prime \varphi^\prime &=& \varphi(v^\prime) \, v
    \varphi^\prime + (-1)^{|\varphi| \, |v^\prime|} v v^\prime \varphi
    \varphi^\prime \\ &=& \varphi(v^\prime) \, v \varphi^\prime +
    (-1)^{|\varphi|\, |v^\prime| + |v|\, |v^\prime| + |\varphi| \,
    |\varphi^\prime|} v^\prime v \varphi^\prime \varphi \;.
\end{eqnarray*}
Subtraction of $(-1)^{(|v|+|\varphi|)(|v^\prime| +
|\varphi^\prime|)} v^\prime \varphi^\prime v \varphi$ on both
sides gives
\begin{displaymath}
    [ v \varphi \, , \, v^\prime \varphi^\prime ] =
    \varphi(v^\prime) \, v \varphi^\prime -
    (-1)^{(|v| + |\varphi|)(|v^\prime| +
    |\varphi^\prime|)} \varphi^\prime(v)\, v^\prime \varphi \;,
\end{displaymath}
which in fact agrees with the bracket (\ref{sbracket tensor})
since $|v \otimes \varphi| = |v| + |\varphi|$ (mod 2).
\end{proof}
As an important corollary to Lem.\ \ref{lem 2.7} one concludes
that every representation of the Clifford-Weyl algebra
$\mathfrak{q}(W)$ for $W = V \oplus V^\ast$ induces a
representation of the Lie superalgebra $\mathfrak{gl}(V)$. For
this, one just sends $X \in \mathfrak{gl}(V)$ to its image in
$\mathfrak{q}(W)$ under the mapping of Lem.\ \ref{lem 2.7}, and
then applies the given representation of $\mathfrak{q}(W)$.

We now adopt the following supersymmetric framework, which unifies
and extends the algebraic structures underlying Sects.\
\ref{sect:2.1}--\ref{sect:2.4}.
\begin{defn}\label{def 2.8}
Given a $\mathbb{Z}_2$-graded vector space
\begin{displaymath}
    V = V_1 \oplus V_0 = (V_1^+ \oplus V_1^-) \oplus (V_0^+
    \oplus V_0^-)
\end{displaymath}
and its Clifford-Weyl algebra $\mathfrak{q}(W)$ for $W = V \oplus
V^\ast$, define $\mathcal{A}_V \subset \mathfrak{q}(W)$ to be the
graded-commutative subalgebra generated by the $Q$-isotropic
vector subspace
\begin{displaymath}
    (V_1^+ \oplus ({V_1^-})^\ast) \oplus
    (V_0^+ \oplus ({V_0^-})^\ast) \subset W \;.
\end{displaymath}
On $\mathcal{A}_V$ let the operators of alternating multiplication
$\varepsilon$, alternating contraction $\iota$, symmetric
multiplication $\mu$, and symmetric contraction $\delta$, be defined
as usual.  By the spinor-oscillator representation $\mathbf{q}:\,\,
\mathfrak{q}(W) \to \mathrm{End}(\mathcal{A}_V)$ we mean the
representation which is given by letting the elements of $W_\tau =
V_\tau^+ \oplus (V_\tau^+)^\ast \oplus V_\tau^- \oplus
(V_\tau^-)^\ast$ operate as
\begin{eqnarray*}
    \mathbf{c}(v_1^+ + \varphi_1^+ + v_1^- + \varphi_1^-) &=&
    \varepsilon(v_1^+) \, + \, \iota(\varphi_1^+) \, + \,
    \iota(v_1^-)\, + \,\varepsilon(\varphi_1^-)\quad (\tau=1)
    \;, \\ \mathbf{w}(v_0^+ + \varphi_0^+ + v_0^- + \varphi_0^-)
    &=& \mu(v_0^+) + \delta(\varphi_0^+) + \delta(v_0^-) -
    \mu(\varphi_0^-) \quad (\tau = 0) \;.
\end{eqnarray*}
\end{defn}
\begin{rem}
The operators $\varepsilon (v_1^+)$, $\iota(\varphi_1^+)$, and
$\iota(v_1^-)$, $\varepsilon (\varphi_1^-)$ obey CAR (\ref{CAR})
transcribed to $\mathcal{A}_V$; the operators $\mu(v_0^+)$,
$\delta(\varphi_0^+) $, and $\delta(v_0^-)$, $-\mu(\varphi_0^-)$ obey
CCR (\ref{CCR}) transcribed to $\mathcal{A}_V$; and the two sets of
operators commute. Note that the algebra $\mathcal{A}_V$ is
isomorphic to the spinor-oscillator module $\mathcal{V}$ of
(\ref{spin-osc module}):
\begin{displaymath}
    \mathcal{A}_V \simeq \mathcal{V} \simeq \sum \wedge^k (V_1^+ \oplus
    {V_1^-}^\ast)\otimes \sum \mathrm{S}^l (V_0^+ \oplus {V_0^-}^\ast)\;,
\end{displaymath}
and carries a natural $\mathbb{Z}$-grading by the total degree $n
= k + l$.  By this isomorphism, we will regard the linear operator
$R(t \otimes u)$ of Cor.\ \ref{cor 2.5} from now on as an operator
on $\mathcal{A}_V$.
\end{rem}
Let $R_\ast : \, \mathfrak{gl}(V) \to \mathfrak{gl}(\mathcal{A}_V)$
denote the Lie superalgebra representation induced by $\mathbf{q} :\,
\mathfrak{q}(W) \to \mathfrak{gl}(\mathcal{A}_V)$. It follows
immediately from Def.\ \ref{def 2.8} that
\begin{displaymath}
    R_\ast \vert_{V_1^+ \to V_1^+} = \sigma \;, \quad
    R_\ast \vert_{V_1^- \to V_1^-} = \widetilde{\sigma} \;, \quad
    R_\ast \vert_{V_0^+ \to V_0^+} = \omega \;, \quad
    R_\ast \vert_{V_1^- \to V_1^-} = \widetilde{\omega} \;.
\end{displaymath}

\subsection{Howe duality and a consequence}\label{sect:howe}

We now adapt the tensor-product situation of Cor.\ \ref{cor 2.5}
to our expanded framework. Introducing a $\mathbb{Z}_2$-graded
vector space $U$,
\begin{equation}\label{def U}
    U = U_1 \oplus U_0 = (U_1^+ \oplus U_1^-) \oplus (U_0^+
    \oplus U_0^-) \;,
\end{equation}
with $U_0^+ = U_1^+ = \mathbb{C}^p$ and $U_0^- = U_1^- =
\mathbb{C}^q$, we make in Def.\ \ref{def 2.8} the identifications
\begin{equation}
    V = U \otimes \mathbb{C}^N \;, \quad V_\tau^\pm =
    U_\tau^\pm \otimes \mathbb{C}^N \quad (\tau = 0, 1) \;.
\end{equation}
This tensor-product decomposition of $V$ gives rise to two
distinguished subalgebras of $\mathfrak{gl}(V)$: there is a Lie
superalgebra $\mathfrak{gl}(U)$ which acts on the $\mathbb
{Z}_2$-graded vector space $U$ as its fundamental module and operates
trivially on the second factor, $\mathbb{C}^N$; and there is a Lie
algebra $\mathfrak{gl}_N$ which acts on $\mathbb{C}^N$ and is trivial
on the first factor, $U$.

R.\ Howe, in 1976, wrote an insightful article "Remarks on Classical
Invariant Theory" \cite{howe} which has much bearing on our present
situation. The main concept is this.
\begin{defn}
Given an orthosymplectic $\mathbb{Z}_2$-graded vector space $W$, let
$(\Gamma, \Gamma^\prime)$ be a pair of Lie (sub-)superalgebras in
$\mathfrak{osp}(W)$. Such a pair is called a reductive supersymmetric
Howe dual pair if $\Gamma^\prime$ acts reductively on $W$ and
$\Gamma$ is the centralizer of $\Gamma^\prime$ in
$\mathfrak{osp}(W)$.  The pair is called classical if $\Gamma^\prime$
is a classical Lie algebra.
\end{defn}
\begin{prop}
If $V = U \otimes \mathbb{C}^N$ as above, the pair $(\mathfrak{gl}(U)
, \mathfrak{gl}_N)$ is a classical reductive supersymmetric Howe dual
pair in $\mathfrak{osp}(V \oplus V^\ast)$.
\end{prop}
\begin{proof}
The action of the classical Lie algebra $\mathfrak{gl}_N$ on $W := V
\oplus V^\ast$ is reductive by the assumed tensor product structure
$V = U \otimes \mathbb{C}^N$.

$\mathfrak{gl}_N$ and $\mathfrak{gl}(U)$ act on different factors of
a tensor product $V = U \otimes \mathbb{C}^N$; therefore the two
actions commute. Conversely, if $X \in \mathfrak{osp}(W)$ commutes
with every $Y \in \mathfrak{gl}_N\,$, then $X$ must be in
$\mathfrak{gl}(V)$, since the two $\mathfrak{gl}_N$-representation
spaces $V$ and $V^\ast$ belong to different isomorphism classes.
Because $\mathbb{C}^N$ is $\mathfrak{gl}_N$-irreducible, it follows
that $X \in \mathfrak{gl}(U)$.  Thus $\mathfrak{gl}(U)$ indeed is the
centralizer of $\mathfrak{gl}_N$ in $\mathfrak{osp}(W)$.
\end{proof}
Now set $n = p + q$ and note the identifications $U = \mathbb{C}^{n|
n}$ and $\mathfrak{gl}(U) \simeq \mathfrak{gl}_{n|n}\,$. As before,
let $R_\ast :\, \mathfrak {gl}(V) \to \mathfrak{gl}(\mathcal{A}_V)$
be the representation which is induced by the spinor-oscillator
representation of the Clifford-Weyl algebra $\mathfrak{q}(V \oplus
V^\ast)$ on $\mathcal{A}_V$. Given $R_\ast$ and the identification $V
\simeq \mathbb{C}^{n|n} \otimes \mathbb {C}^N$, one immediately has a
representation $\rho_\ast$ of the Lie superalgebra $\mathfrak
{gl}_{n|n} \hookrightarrow \mathfrak{gl} (V)$ on $\mathcal{A}_V$:
\begin{equation}
    \rho_\ast: \,\,\mathfrak{gl}_{n|n} \to
    \mathfrak{gl}(\mathcal{A}_V) \;, \quad
    x \mapsto R_\ast (x \otimes \mathrm{Id}_N) \;.
\end{equation}
Similarly, the $\mathfrak{gl}(V)$-representation $R_\ast$ induces
a representation $r_\ast$ of $\mathfrak{gl}_N \hookrightarrow
\mathfrak{gl}(V)$ by
\begin{equation}
    r_\ast : \,\,\mathfrak{gl}_N \to \mathfrak{gl}(\mathcal{A}_V)
    \;, \quad y \mapsto R_\ast(\mathrm{Id}_{n|n} \otimes y) \;.
\end{equation}
The latter arises also from the $\mathrm{U}_N$-representation $r$ of
(\ref{howe factors}), by linearization at $u = \mathrm{Id}_N$ and
complexification. This representation is unitary if $\mathcal{A}_V$
is given its canonical Hermitian structure, in which
\begin{eqnarray*}
    &&\mu(v_0^+)^\dagger = \delta(c\, v_0^+) \;, \quad
    \varepsilon(v_1^+)^\dagger = \iota(c\, v_1^+) \;, \\
    &&\delta(v_0^-)^\dagger = \mu(c\, v_0^-) \;, \quad
    \iota(v_1^-)^\dagger = \varepsilon(c\, v_1^-) \;.
\end{eqnarray*}
(See Sect.\ \ref{sect:2.3} for explanation in the bosonic case; the
fermionic case is similar).  Indeed, using these relations it is
readily seen that $r_\ast(y)^\dagger = - r_\ast(y)$ for $y \in
\mathfrak{u}_N = \mathrm{Lie}\, \mathrm{U}_N\,$.  Note also that $y
\in \mathfrak{u}_N$ acts on the spinor-oscillator module $\mathcal
{A}_V$ by operators of type $\varepsilon\iota$, $\iota\varepsilon$,
$\mu\delta$, and $\delta\mu$, none of which changes the degree of the
$\mathbb{Z}$-graded algebra $\mathcal{A}_V$.

What we have at hand, then, is a Howe pair $(\mathfrak{gl}_{n|n}\, ,
\mathfrak{gl}_N)$ where the second member $\mathfrak{gl}_N =
\mathfrak{u}_N + \mathrm{i} \mathfrak{u}_N$ acts on the
spinor-oscillator module $\mathcal{A}_V$ by complex linear extension
of the representation of its compact real form $\mathfrak{u}_N\,$. In
such a situation one prefers the notation $(\mathfrak{gl}_{n|n}\, ,
\mathrm{U}_N) = (\Gamma, K)$ and speaks of a Howe dual pair with
compact group $K$.

For present purposes, a major result of Howe's theory of dual pairs
$(\Gamma,K)$ with compact group $K$ is Thm.\ 8 of \cite{howe}, which
is concerned with the question of how the spinor-oscillator
representation decomposes with respect to the action of $(\Gamma,K)$.
In our case of $(\mathfrak{gl}_{n|n}\, , \mathrm {U}_N)$ represented
by
\begin{displaymath}
    \rho_\ast :\,\, \mathfrak{gl}_{n|n}\to\mathfrak{gl}(\mathcal{A}_V)
    \;, \quad r : \,\, \mathrm{U}_N \to \mathrm{U} (\mathcal{A}_V) \;,
\end{displaymath}
the theorem implies that the $\mathrm{U}_N$-trivial isotypic
component of $\mathcal{A}_V$, i.e., the space of $\mathrm{U}_N
$-invariant vectors in $\mathcal{A}_V$, is an \emph{irreducible}
representation space for $\mathfrak{gl}_{n|n}\,$.

To make use of this fact, recall from (\ref{howe factors}) the
definition $\rho(t) = R(t \otimes \mathrm{Id}_N)$ for the diagonal
operators $t = (t_1^+ , t_1^- , t_0^+ , t_0^-)$ specified for Cor.\
\ref{cor 2.5}. If $t$ is reinterpreted via
\begin{displaymath}
    \mathrm{GL}(U_1^+) \times \mathrm{GL}(U_1^-) \times
    \mathrm{GL}(U_0^+) \times \mathrm{GL}(U_0^-) \hookrightarrow
    \mathrm{End}(U)
\end{displaymath}
as a linear operator on $U$ with logarithm $\ln t \in \mathfrak{gl}
(U)$, the infinitesimal representation $\rho_\ast : \, \mathfrak{gl}
_{n|n} \to \mathfrak{gl}(\mathcal{A}_V)$ is related to $\rho$ by
$\rho(t) = \mathrm{e}^{\, \rho_\ast (\ln t)}$.
\begin{prop}\label{howe-char}
Denoting by $du$ the Haar measure of $K \equiv \mathrm{U}_N$ with
total mass equal to 1, define a function $t \to \chi(t)$ by
\begin{displaymath}
    t \mapsto \chi(t) := \int_{\mathrm{U}_N} \prod_{k = 1}^n\frac{
    \mathrm{Det}(\mathrm{Id}_N - \mathrm{e}^{\mathrm{i}\psi_k}\,u)}
    {\mathrm{Det}(\mathrm{Id}_N -\mathrm{e}^{\phi_k}\, u)}\,du\;.
\end{displaymath}
In the range of parameters $t = t(\psi_k,\phi_k)$ specified in Cor.\
\ref{cor 2.5}, this function has an alternative expression as a
character:
\begin{displaymath}
    \chi(t) = \mathrm{STr}_{\mathcal{A}_V^K} \, \rho(t) \equiv
    \mathrm{STr}_{\mathcal{A}_V^K} \, R(t \otimes \mathrm{Id}_N)\;,
\end{displaymath}
which is an absolutely convergent sum over $\mathcal{A}_V^K$, the
irreducible $\mathfrak{gl}_{n|n} $-representation space spanned by
the $K$-invariant vectors in the spinor-oscillator module
$\mathcal{A}_V$.
\end{prop}
\begin{proof}
Take the $\mathcal{A}_V$-supertrace of the factorization
(\ref{factorization}) and integrate over $K$:
\begin{displaymath}
    \int_{K} \mathrm{STr}_{\mathcal{A}_V} \, R(t \otimes u) \, du
    = \int_{K} \mathrm{STr}_{\mathcal{A}_V}\, \rho(t) r(u) \, du \;.
\end{displaymath}
On inserting the explicit form of $\mathrm{STr}_{\mathcal{A}_V} \, R(
t \otimes u)$ stated in the remark after Cor.\ \ref{cor 2.5}, the
left-hand side of this equality turns into the defining expression
for $\chi(t)$.

In the prescribed range for $t$, where $\mathrm{STr}_{\mathcal{A}_V}
\, \rho(t) r(u)$ converges absolutely and uniformly in $u \,$, one
may interchange the order of taking the trace over $\mathcal {A}_V$
and integrating over the compact group $K$. The result of doing the
latter integral is another absolutely convergent sum, which coincides
with $\mathrm{STr}_{\mathcal{A}_V^K}\, \rho(t)$. Indeed, since $r$ is
a unitary representation of $K$ on $\mathcal{A} _V$, Haar-averaging
the operator $r(u)$ over $u \in K$ yields the projector onto
$\mathcal{A}_V^K$, the subspace of $K$-invariants in $\mathcal{A}_V$.

The irreducibility of $\mathcal{A}_V^K$ w.r.t.\ the $\mathfrak
{gl}_{n|n}$-action is implied by Thm.\ 8 of \cite{howe}.
\end{proof}
\begin{rem}
Since the $K$-action on $\mathcal{A}_V$ preserves the $\mathbb
{Z}$-grading of this graded-commu\-tative algebra, the subalgebra of
$K$-invariants $\mathcal{A}_V^K$ is still $\mathbb{Z}$-graded. For
the same reason, the $\mathbb{Z}_2$-grading of $\mathcal{A}_V$
defines a $\mathbb{Z}_2$-grading of $\mathcal{A}_V^K$.
\end{rem}
Our function $t \mapsto \chi(t)$ is identical to the function on the
left-hand side of the statement of Thm.\ \ref{thm0}. We have thus
established that left-hand side to be a character associated with the
irreducible representation $(\mathcal{A}_V^K , \rho_\ast)$ of
$\mathfrak {gl}_{n|n}\,$. This completes the first step of the
programme outlined in Sect.\ \ref{sect:strategy}.

We shall see that the character $\chi(t)$ is expressed by a
generalization of the Weyl character formula for irreducible
representations of compact Lie groups.  If that generalization
were available in the published literature, we would already be
done and this paper could end right here. However, the required
generalization apparently has not been worked out before; it will
keep us busy for another three main sections.

\subsection{Weight expansion of $\chi\,$}\label{sect:weights}

We begin our investigation of the character $\chi(t)$ by gathering
some standard facts about the Lie superalgebra $\mathfrak{g}\equiv
\mathfrak{gl}(U) = \mathfrak{gl}_{n|n}\,$.

Let $\mathfrak{h} \subset \mathfrak{g}$ denote the maximal Abelian
subalgebra which is generated over $\mathbb{C}$ by the projection
operators $E_{\tau,i}^\pm \in \mathrm{End}(U_\tau^ \pm)$ of Cor.\
\ref{cor 2.5}.  Although the variables $\psi_1 , \ldots, \phi_n$ were
originally introduced as \emph{parameters}, it is now natural to
reinterpret them as \emph{linear coordinate functions} on
$\mathfrak{h}\,$. Adopting this new perspective we expand $H \in
\mathfrak{h}$ as
\begin{displaymath}
    H = \mathrm{diag}\big(\mathrm{i}\psi_1(H), \ldots, \mathrm{i}
    \psi_{p+q}(H), \phi_1(H), \ldots, \phi_{p+q}(H)\big) \;.
\end{displaymath}
In Lie theory, the non-zero eigenvalues, $\alpha$, of the adjoint
action of $\mathfrak{h}$ on $\mathfrak{g}$ are called the
\emph{roots} of the pair $(\mathfrak{h} \, , \mathfrak{g})$. In the
present context, a root is called \emph{even} or \emph{odd} depending
on whether the eigenvector $X$ in the eigenvalue equation $[H , X] =
\alpha(H) X$ is an even or odd element of $\mathfrak{g}\,$.  The even
roots of our Lie superalgebra $\mathfrak{g} = \mathfrak{gl}(U)$ are
$\mathrm{i}\psi_k - \mathrm{i}\psi_{k^\prime}$ and $\phi_k -
\phi_{k^\prime}$ (for $k \not= k^\prime$), while the odd roots are
$\mathrm{i}\psi_k - \phi_{k^\prime}$ and $\phi_k -
\mathrm{i}\psi_{k^\prime}$ (for any $k,k^\prime$).

By general principles, if $\alpha$ is a root, then so is $-\alpha$.
Let us fix a system of \emph{positive roots}, $\Delta^+$.  For that,
we arrange the coordinate functions as an ordered set:
\begin{equation}\label{order coords}
    \phi_1 \, , \ldots, \phi_p \, , \mathrm{i}\psi_1 \, , \ldots,
    \mathrm{i}\psi_p \, , \mathrm{i}\psi_{p+1} \, , \ldots,
    \mathrm{i}\psi_{p+q} \, , \phi_{p+1} \, , \ldots, \phi_{p+q} \;,
\end{equation}
and take $\Delta^+$ to be the set of differences $x - y$ where $x$
and $y$ are any two entries from this sequence subject to the
requirement that $x$ occurs later than $y$. To illustrate: $\phi_2
- \phi_1$ and $\mathrm{i}\psi_1 - \phi_1$ are two examples of
positive roots.  If $\mathfrak{g}^\alpha \subset \mathfrak{g}$
denotes the root space of the root $\alpha$, i.e., the $\mathrm
{ad}(\mathfrak{h})$-eigenspace with eigenvalue $\alpha$, then
\begin{displaymath}
    \mathfrak{g} = \mathfrak{n}^- \oplus \mathfrak{h} \oplus
    \mathfrak{n}^+ \;, \qquad \mathfrak{n}^+ = \sum_{\alpha \in
    \Delta^+} \mathfrak{g}^\alpha \;, \quad \mathfrak{n}^- =
    \sum_{\alpha \in \Delta^+} \mathfrak{g}^{-\alpha} \;.
\end{displaymath}

While all this is standard and applies to semisimple Lie algebras
and superalgebras in general, the situation at hand is actually
ruled by a coarser structure. Rearranging the components of the
vector space $U$ of (\ref{def U}) as
\begin{equation}
    U = U^+ \oplus U^- \;, \quad U^\pm = U_1^\pm \oplus U_0^\pm \;,
\end{equation}
our Lie superalgebra $\mathfrak{g} = \mathfrak{gl}(U)$ is
$\mathbb{Z}$-graded by a direct-sum decomposition
\begin{equation}\label{Z grading}
    \mathfrak{g} = \mathfrak{g}^{(-2)} \oplus
    \mathfrak{g}^{(0)} \oplus \mathfrak{g}^{(2)} \;,
\end{equation}
where the middle summand is a Lie (sub-)superalgebra in
$\mathfrak{g}$:
\begin{equation}
    \mathfrak{g}^{(0)} = \mathfrak{gl}(U^+) \oplus
    \mathfrak{gl}(U^-) = \mathfrak{gl}_{p|p} \oplus
    \mathfrak{gl}_{q|q} \;,
\end{equation}
while the first and last summands are $\mathbb{Z}_2$-graded vector
spaces
\begin{equation}
    \mathfrak{g}^{(-2)} = \mathrm{Hom}(U^+,U^-) \;, \quad
    \mathfrak{g}^{(2)} = \mathrm{Hom}(U^-,U^+) \;,
\end{equation}
which are $\mathfrak{g}^{(0)}$-modules by the adjoint action. Note
the inclusions
\begin{equation}
    \mathfrak{h} \subset \mathfrak{g}^{(0)} \;, \quad
    \mathfrak{g}^{(-2)} \subset \mathfrak{n}^+ \;, \quad
    \mathfrak{g}^{(2)} \subset \mathfrak{n}^- \;.
\end{equation}
The first one is obvious.  The last two follow from our choice of
positive root system by noticing that the functions $\psi_1, \ldots,
\phi_{p+q}$ have been arranged in (\ref{order coords}) so that the
first $2p$ of them vanish on $\mathfrak{h} \cap \mathfrak{gl}(U^-)$
and the last $2q$ vanish on $\mathfrak{h} \cap \mathfrak{gl}(U^+)$.
(The apparent sign inconsistency is forced by our desire to follow
mathematical conventions and treat the \emph{lowest}-degree subspace
of $\mathcal{A}_V^K$ as a \emph{highest}-weight space.)

The decomposition (\ref{Z grading}) reflects the way in which
$\mathfrak{gl}(U) \hookrightarrow \mathfrak{gl}(V) = \mathfrak{gl} (U
\otimes \mathbb{C}^N)$ is represented on the spinor-oscillator module
$\mathcal{A}_V$ of Def.\ \ref{def 2.8}: the elements in
$\mathfrak{g}^{(2)}$ act as operators of type $\varepsilon
\varepsilon$, $\varepsilon\mu$, or $\mu\mu$, which raise the degree
in $\mathcal{A}_V$ and $\mathcal{A}_V^K$ by two; the elements of
$\mathfrak{g}^{ (-2)}$ act as operators of type $\iota\iota$,
$\iota\delta$, or $\delta\delta$, which lower the degree by two; and
$\mathfrak{g}^{(0)}$ is represented by degree-preserving operators of
type $\varepsilon\iota$, $\varepsilon\delta$, $\mu\iota$, or
$\mu\delta$.

The following proposition means that $(\mathcal{A}_V^K , \rho_\ast)$
is a highest-weight (actually, lowest-weight) representation of
$\mathfrak{gl}_{n|n}\,$. As usual, the eigenvalues and eigenspaces of
the $\mathfrak{h}$-action by $\rho_\ast$ on $\mathcal{A}_V^K$ are
called the \emph{weights} and \emph{weight spaces} of the
representation.

Notice that the subspace of constants $\mathbb{C} \subset \mathcal
{A}_V$ (the "vacuum" in physics language) is invariant under the
action of $K = \mathrm{U}_N$ and hence lies in $\mathcal{A}_V^K$.
\begin{prop}
The one-dimensional subspace of constants $\mathbb{C} \subset
\mathcal{A}_V^K$ is a cyclic subspace for $\mathcal{A}_V^K$ under the
action of the algebra $\mathfrak{g}^{(2)}$ of raising operators. It
is stabilized by the subalgebra $\mathfrak{g}^{(0)} \oplus
\mathfrak{g}^{(-2)}$. The weight of this subspace is
\begin{displaymath}
    \lambda_N = N \sum\nolimits_{l=p+1}^{p+q}
    (\mathrm{i} \psi_l - \phi_l) \;.
\end{displaymath}
\end{prop}
\begin{proof}
The degree-zero subspace of $\mathcal{A}_V$ is one-dimensional,
being given just by the constants, and this remains of course true
on passing to $\mathcal{A}_V^K$. Because zero is the smallest
possible degree and the action of $\mathfrak{g}^{(k)}$ changes the
degree by $k$ units, $\mathfrak{g}^{(-2)}$ annihilates the
degree-zero subspace $\mathbb{C} \subset \mathcal{A}_V^K$, and
$\mathfrak{g} ^{(0)}$ stabilizes it.

The cyclic property, i.e., the fact that every vector of $\mathcal
{A}_V^K$ is generated by the successive application of raising
operators in $\mathfrak{g}^{(2)}$, is asserted by Thm.\ 9 (iii) of
\cite{howe}.

Let now $z \in \mathbb{C} \subset \mathcal{A}_V^K$ be any non-zero
constant. Applying to it the operator $\rho_\ast(H)$,
\begin{eqnarray*}
    \rho_\ast(H) \, z &=& \sum_{a=1}^N \sum_{l=1}^q \big(
    \mathrm{i}\psi_{l+p}(H)\, \iota(e_{1,l}^- \otimes e_a)\,
    \varepsilon(f_-^{1,l} \otimes f^a) \\
    &&\hspace{1cm} - \phi_{l+p}(H)\, \delta(e_{0,l}^-
    \otimes e_a)\, \mu(f_-^{0,l} \otimes f^a) \big) \, z \;,
\end{eqnarray*}
and using the relations $\iota(v) \varepsilon(\varphi) \cdot 1 =
\varphi(v)$ and $\delta(v) \mu(\varphi) \cdot 1 = \varphi(v)$, one
sees that $z \in \mathbb{C}$ is an eigenvector with the stated
eigenvalue $\lambda_N(H)$ for any $H \in \mathfrak{h}\,$.
\end{proof}
Now define a real form $\mathfrak{h}_\mathbb{R}^\prime$ of
$\mathfrak{h}$ by demanding that all of the $\mathbb{C}$-linear
coordinate functions $\mathrm{i}\psi_k : \, \mathfrak{h} \to
\mathbb{C}$ and $\phi_k :\, \mathfrak{h} \to \mathbb{C}$ take
\emph{imaginary} values on $\mathfrak{h}_\mathbb{R}^\prime\,$. The
diagonal transformations $\exp H$ for $H \in \mathfrak{h}_
\mathbb{R}^\prime$ then form a compact Abelian group, and the
$\mathfrak{gl}_{n|n} $-module $\mathcal{A}_V^K$ decomposes as an
orthogonal direct sum of weight spaces $V_\gamma$ with respect to the
unitary action $\exp H \mapsto \exp \rho_\ast(H)$:
\begin{equation}\label{sum over mu}
    \mathcal{A}_V^K = \bigoplus\limits_{\mathrm{weights}\,\gamma}
    V_\gamma \;.
\end{equation}
\begin{prop}\label{prop 2.13}
Expressing the weights $\gamma$ of the decomposition (\ref{sum over
mu}) as
\begin{displaymath}
    \gamma = \sum\nolimits_{k=1}^{p+q} (\mathrm{i} m_k \psi_k
    - n_k \phi_k) \;,
\end{displaymath}
their coefficients are integers in the range
\begin{displaymath}
    n_j \le 0 \le m_k \le N \le n_l \quad
    (1 \le j \le p < l \le p+q) \;.
\end{displaymath}
\end{prop}
\begin{proof}
From general theory \cite{knapp} one knows that the weights of an
irreducible representation with highest weight $\lambda$ are of
the form $\lambda + \sum c_\alpha \alpha$ where the coefficients
$c_\alpha$ are non-negative integers, and the sum over roots
excludes those $\alpha$ whose root vectors annihilate or stabilize
the highest weight space.  In the present case, where the highest
weight space is annihilated and stabilized by $\mathfrak{g}^{
(-2)}$ resp.\ $\mathfrak{g}^{(0)}$, we introduce the set
\begin{equation}\label{eq 2.40}
    \Delta_\lambda^+ := \{ \alpha \in \Delta^+ \mid
    \mathfrak{g}^{-\alpha} \subset \mathfrak{g}^{(2)} \} \;,
\end{equation}
and the weights of the representation $(\mathcal{A}_V^K ,
\rho_\ast)$ are then of the form
\begin{displaymath}
    \gamma = \lambda_N - \sum\nolimits_{\alpha \in \Delta_\lambda^+}
    \, c_\alpha \, \alpha
\end{displaymath}
with integers $c_\alpha \ge 0\,$. Since $\mathfrak{g}^{(2)} =
\mathrm{Hom} (U^- , U^+)$, the choice of ordering in the list
(\ref{order coords}) means that the set $\Delta_\lambda^+$ consists
of the roots obtained by subtracting an entry in the second half of
the list from an entry in the first half.  These are the roots
\begin{equation}\label{eq 2.41}
    \phi_l - \phi_j \;, \quad \mathrm{i}\psi_l - \mathrm{i}\psi_j \;,
    \quad \phi_l - \mathrm{i}\psi_j \;, \quad \mathrm{i}\psi_l - \phi_j
    \quad (1 \le j \le p < l \le p+q) \;.
\end{equation}
Subtracting from the highest weight $\lambda_N$ a linear combination
of these roots with positive integers as coefficients, one gets
weights $\gamma$ with coefficients in the range $n_j \le 0$, $n_l \ge
N$, $m_j \ge 0$, and $m_l \le N$ (for $j,l$ still subject to $1 \le j
\le p < l \le p+q$).

The stronger condition $0 \le m_k \le N$ (for all $k = 1, \ldots,
p+q$) is best seen by going back to the integral formula for the
character $\chi$ of $(\mathcal{A}_V^K , \rho_\ast)$ in Prop.\
\ref{howe-char}. Indeed, since $\mathrm{Det}(\mathrm {Id}_N -
\mathrm{e}^{\mathrm{i}\psi_k}\, u)$ is a polynomial in $\mathrm
{e}^{\mathrm{i}\psi_k}$ of degree $N$, the weight-space decomposition
of $\mathcal{A}_V^K$ must be such that only the powers $\mathrm
{e}^{\mathrm{i}m_k \psi_k}$ with $0 \le m_k \le N$ appear.
\end{proof}
\begin{rem}
The weights $\gamma$ of the weight-space decomposition (\ref{sum
over mu}) are \emph{analytically integral}, i.e., whenever $H \in
\mathfrak{h}$ is such that $\exp H \in \mathrm{End}(U)$ is unity,
then $\mathrm{e}^{\gamma(H)} = 1\,$.
\end{rem}
Let now $\Gamma_\lambda$ denote the set of weights of our
representation $(\mathcal{A}_V^K , \rho_\ast)$ with highest weight
$\lambda_N$ and $\mathbb{Z}_2$-grading $\mathcal{A}_V^K =
(\mathcal{A}_V^K )_0 \oplus (\mathcal{A}_V^K)_1$. For $\gamma \in
\Gamma_\lambda$ put $|\gamma| := 0$ if $V_\gamma \subset
(\mathcal{A}_V^K)_0$ and $|\gamma| := 1$ if $V_\gamma \in
(\mathcal{A}_V^K)_1\,$. (One of these must be true, as roots are
either even or odd and weights are generated additively and
integrally from roots.)
\begin{cor}\label{cor 2.14}
The $\mathfrak{gl}_{n|n}$-character $\chi$ has an expansion as a
sum over weights,
\begin{displaymath}
    \chi(t) = \sum\nolimits_{\gamma\in \Gamma_\lambda} (-1)^{|\gamma|}
    \mathrm{dim}(V_\gamma) \, \mathrm{e}^{\,\gamma(\ln t)} \;,
\end{displaymath}
which converges absolutely in the domain for $t$ defined by
\begin{displaymath}
    \mathfrak{Re}\, \phi_j(\ln t) < 0 < \mathfrak{Re}\,
    \phi_l (\ln t) \quad (1 \le j \le p < l \le p+q) \;.
\end{displaymath}
The coefficients $m_k\,$, $n_k$ of the weights $\gamma \in
\Gamma_\lambda$ lie in the range specified by Prop.\ \ref{prop 2.13}.
\end{cor}
\begin{proof}
Based on Prop.\ \ref{howe-char}, compute $\chi(t) = \mathrm{STr}_{
\mathcal{A}_V^K} \, \rho(t)$ by using the weight-space decomposition
of $\mathcal{A}_V^K$ and the fact that $\rho(t) v = \mathrm{e}^{\,
\rho_\ast(\ln t)} v = \mathrm {e}^{\gamma(\ln t)} v$ for $v \in
V_\gamma \,$. The sign factor arises because one is computing a
supertrace and the vector space $V_\gamma$ has parity $(-1)^{
|\gamma|}$. Since $| \mathrm{e}^{\gamma(\ln t)} | = \mathrm{e}^{
\mathfrak{Re} \, \gamma(\ln t)}$, absolute convergence in the stated
domain for $t$ follows from the limited range of the coefficients
$m_k\,$, $n_k$ as given in Prop.\ \ref{prop 2.13}.
\end{proof}
\begin{rem}
Finding the multiplicities $\mathrm{dim}(V_\gamma)$ is a
combinatorial task which we shall side-step in this paper by
digging deeper into the structure of the problem.
\end{rem}

\section{Background material from superanalysis}
\label{sect:3}\setcounter{equation}{0}

Our task is to compute the character $t \mapsto \chi(t)$ and show
that it is given by the right-hand side in the statement of Thm.\
\ref{thm0}. For that purpose it is not prudent to cling to the
limited view of $\chi$ as a function of the diagonal operators
$t\,$. Rather, to get enough analytical control we are going to
exploit the fact that $\chi$ extends to a function on (a certain
subspace of) the Lie supergroup $\mathrm {GL}_{n|n}\,$. The
desired statement will then follow rather easily from some
standard results in superanalysis.

Since superanalysis is not a widely known subject, we inject the
present section to collect most of the necessary material for the
convenience of the reader.  (Experts in superanalysis should of
course skip over this section.)

\subsection{The vector bundle underlying a classical Lie supergroup}
\label{sect:superG}

Lie supergroups have been defined and studied by Berezin
\cite{berezin}. Here we will review a simplified definition, which
uses nothing but canonical constructions in geometry.

Recall from Sect.\ \ref{sect:basics} the notion of a Lie superalgebra
$\mathfrak{g}\,$. In the sequel we will be concerned only with
$\mathfrak{g} = \mathfrak{gl}$ and $\mathfrak{g} = \mathfrak{osp}\,$.
Our number field $\mathbb{K}$ is either $\mathbb{R}$ or
$\mathbb{C}\,$.

Let now $G$ be a Lie group with Lie algebra $\mathrm {Lie}(G) =
\mathfrak{g}_0 \subset \mathfrak{g}\,$.  In the case of $\mathfrak{g}
= \mathfrak{gl}_{p|q}$ we take $G \simeq \mathrm{GL} (\mathbb{K}^p)
\times \mathrm{GL}(\mathbb{K}^q)$, i.e., the set of linear
transformations
\begin{displaymath}
    g = \begin{pmatrix} g_1 &0\\ 0 &g_0 \end{pmatrix} \;,
\end{displaymath}
with $g_1 \in \mathrm{GL}(\mathbb{K}^p)$ and $g_0 \in \mathrm{GL}
(\mathbb{K}^q)$. If $\mathfrak{g} = \mathfrak{osp}_{p|q}$ we take $G
\simeq \mathrm{SO}(\mathbb{K}^p) \times \mathrm{Sp}(\mathbb{K}^q)$.

To construct the Lie supergroups $\mathfrak{G} = \mathrm{GL}_{p|q}$
and $\mathfrak{G} = \mathrm{OSp}_{p|q}$ one should in principle
invoke a Grassmann algebra of anti-commuting parameters to
exponentiate the odd part $\mathfrak{g}_1$ of the Lie superalgebra
$\mathfrak{g}$ \cite{berezin}. The difficult step in this approach is
to explain exactly what is meant by the group multiplication law.  We
will therefore adopt the viewpoint offered by Deligne and Morgan in
their Notes on Supersymmetry (following J.\ Bernstein) \cite{QFT
math}; see also \cite{varadarajan}. In that approach, supergroup
multiplication is described only at the infinitesimal level, as
follows.

With the aim of constructing a certain important vector bundle
(see below), consider the mapping
\begin{displaymath}
    P := G \times G \to G \;, \quad (h^\prime , h) \mapsto
    h^\prime h^{-1} \;,
\end{displaymath}
and view $P$ as a $G$-principal bundle by the right $G$-action
$(h^\prime , h) \mapsto (h^\prime g , h g)$. The structure group $G$
of $P$ also acts on the odd part $\mathfrak{g}_1$ of the Lie
superalgebra $\mathfrak{g}$ by the adjoint representation,
$\mathrm{Ad}\,$; in the cases at hand the adjoint action of $G$ on
$\mathfrak{g}$ is
\begin{displaymath}
    \mathrm{Ad} \begin{pmatrix} g_1 &0\\ 0 &g_0 \end{pmatrix}:\quad
    \begin{pmatrix} \mathsf{A} &\mathsf{B}\\ \mathsf{C} &\mathsf{D}
    \end{pmatrix} \mapsto \begin{pmatrix} g_1^{\vphantom{-1}}
    \mathsf{A}\, g_1^{-1} &g_1^{\vphantom{-1}} \mathsf{B}\, g_0^{-1}\\
    g_0^{\vphantom{-1}} \mathsf{C}\, g_1^{-1} &g_0^{\vphantom{-1}}
    \mathsf{D}\, g_0^{-1} \end{pmatrix} \;,
\end{displaymath}
and the $G$-action on $\mathfrak{g}_1$ is obtained by restricting to
$\mathsf{A} = \mathsf{D} = 0\,$.

It is then a standard construction of differential geometry to
associate to the principal bundle $P \to G$ a vector bundle $F \to G$
with total space
\begin{displaymath}
    F = P \times_{G}^{\vphantom{\dagger}} \mathfrak{g}_1 \;,
\end{displaymath}
whose elements are the orbits of the diagonal right $G$-action on $P
\times \mathfrak{g}_1\,$. In other words, the fibre of $F$ over a
point $x = h^\prime h^{-1} \in G$ is the vector space $F_x \simeq
\mathfrak{g}_1$ of $G$-orbits
\begin{displaymath}
    [(h^\prime,h)\,;\,Y] \equiv \{h^\prime g\,, h g\,,\mathrm{Ad}
    (g^{-1})Y \mid g \in G \} \quad (Y \in \mathfrak{g}_1) \;.
\end{displaymath}

The adjoint representation $\mathrm{Ad}$ of $G$ on $\mathfrak {g}_1$
induces a representation $(\mathrm{Ad}^\ast)^{-1}$ of $G$ on $\wedge
(\mathfrak{g}_1^\ast)$, the exterior algebra of the dual of
$\mathfrak{g}_1\,$. Thus, along with the vector bundle $F \to G$ we
have a vector bundle
\begin{displaymath}
    \wedge (F^\ast) \to G \;,
\end{displaymath}
whose fibre over a point $h^\prime h^{-1} = x \in G$ is $\wedge
F_x^\ast \simeq \wedge \mathfrak{g}_1^\ast$, i.e., the vector space
of $G$-orbits
\begin{displaymath}
    [(h^\prime,h)\, ; \, a] \equiv \{h^\prime g \, , h g\,
    ,\mathrm{Ad}^\ast(g) a \mid g \in G \} \quad (a \in
    \wedge \mathfrak{g}_1^\ast) \;.
\end{displaymath}
Of course the exterior multiplication in the fibre $\wedge F_x^\ast$
over $x = h^\prime h^{-1}$ is defined by
\begin{displaymath}
    [(h^\prime,h)\,;\,a] \wedge [(h^\prime,h)\,;\,b] =
    [(h^\prime,h)\,;\,a \wedge b] \qquad
    (a, b \in \wedge\mathfrak{g}_1^\ast)\;.
\end{displaymath}

Given the vector bundle $\wedge F^\ast \to G$, we will now
consider the graded-commutative algebra, $\mathcal{F}$, of smooth
sections of $\wedge F^\ast$:
\begin{displaymath}
    \mathcal{F} := \Gamma(G , \wedge F^\ast) \;.
\end{displaymath}
Here the basic idea is that a Lie group (just like any manifold) can
be described either by its points, or by the algebra of its functions
(more precisely: the sheaf of algebras). In the case of a Lie
supergroup, the latter viewpoint is the good one, i.e., one uses a
description by the algebra of superfunctions, and this algebra is
none other than $\mathcal{F}$.

\subsection{$\mathfrak{g}$-action on a Lie supergroup}
\label{sect:g-action}

A very useful fact about $\mathcal{F}$ is that its elements are in
one-to-one correspondence with maps $\Phi \in \mathrm{C}^\infty (P,
\wedge \mathfrak{g}_1^\ast)^{G}$, i.e., with smooth functions $\Phi$
on $P = G\times G$ that take values in $\wedge (\mathfrak{g}_1^\ast)$
and are equivariant w.r.t.\ $G:$
\begin{displaymath}
    \Phi(h^\prime,h) = \mathrm{Ad}^\ast (g^{-1}) \,
    \Phi(h^\prime g,hg) \quad (h^\prime,h,g \in G) \;.
\end{displaymath}
Indeed, such a function $\Phi$ determines a unique section $s \in
\mathcal{F}$ by
\begin{displaymath}
    s \big(h^\prime h^{-1} ) = [(h^\prime,h)\, ; \,
    \Phi(h^\prime,h)] = [ (h^\prime g , h g) \, ;
    \, \Phi(h^\prime g, h g) ] \;.
\end{displaymath}
In the sequel we assume the identification $\mathrm{C}^ \infty (P,
\wedge \mathfrak{g}_1^\ast)^G \simeq \mathcal{F}$ by this bijection
to be understood, and we will often abuse language to call $\Phi \in
\mathrm{C} ^\infty(P, \wedge \mathfrak{g}_1^\ast)^G$ a section of
$\mathcal{F}$.

The algebra $\mathcal{F}$ has a natural $\mathbb{Z}_2$-grading
$\mathcal{F} = \mathcal{F}_0 \oplus \mathcal{F}_1$ by
\begin{displaymath}
    \mathcal{F}_0 = \bigoplus\nolimits_{\ell \, \, \text{even}}
    \Gamma(G , \wedge^\ell F^\ast) \;, \quad \mathcal{F}_1 =
    \bigoplus\nolimits_{\ell \, \, \text{odd}}
    \Gamma(G , \wedge^\ell F^\ast) \;.
\end{displaymath}
The numerical (or $\mathbb{C}$-number) part of a section $s \in
\mathcal{F}$ will be denoted by $\mathrm{num} (s)$.

An \emph{even derivation} of $\mathcal{F}$ is a first-order
differential operator $D : \mathcal{F}_0 \to \mathcal{F}_0$ and $D
: \mathcal{F}_1 \to \mathcal{F}_1$. An \emph{odd derivation} is a
first-order differential operator $D : \mathcal{F}_0 \to
\mathcal{F}_1$ and $D : \mathcal{F}_1 \to \mathcal{F}_0$
satisfying the anti-Leibniz rule $D\, (st) = (Ds)\, t + (-1)^{|s|}
s\, Dt$.

\subsubsection{The action of $\mathfrak{g}_0\,$}
\label{sect:g0-action}

We are now ready to specify half of the Lie supergroup structure. The
Lie group $G$ acts on $\mathcal{F} \simeq \mathrm{C}^\infty (P,
\wedge \mathfrak{g}_1^\ast)^{G}$ in the obvious manner: there is a
$G$-action on the left by
\begin{displaymath}
    (L_g \Phi)(h,h^\prime) = \Phi(g^{-1}h, h^\prime) \;,
\end{displaymath}
and a $G$-action on the right by
\begin{displaymath}
    (R_g \Phi)(h,h^\prime) = \Phi(h, g^{-1} h^\prime) \;.
\end{displaymath}
Specializing these to the infinitesimal level we get the
corresponding canonical actions of the Lie algebra
$\mathfrak{g}_0$ by even derivations of $\mathcal{F}$:
\begin{displaymath}
    \widehat{X}^L := \frac{d}{dt} L_{\exp (tX)} \Big\vert_{t=0}
    \;, \quad \widehat{X}^R := \frac{d}{dt} R_{\exp (tX)}
    \Big\vert_{t = 0} \;.
\end{displaymath}
Note the representation property $[\widehat{X}_1^L , \widehat
{X}_2^L ] = \widehat{[X_1,X_2]}^L$ and $[ \widehat{X}_1^R ,
\widehat{X}_2^R ] = \widehat{[X_1,X_2]}^R$.

\subsubsection{Grassmann envelope}\label{sect:envelope}

The odd part $\mathfrak{g}_1$ of the Lie superalgebra $\mathfrak{g}$
acts on the graded-commutative algebra $\mathcal{F}$ by odd
derivations. To describe this action one needs the notion of
Grassmann envelope of a Lie superalgebra.

First, consider $\mathrm{End}(V)$ for some basic
$\mathbb{Z}_2$-graded vector space $V$. Picking any parameter space
$\mathbb{K}^d$ and denoting its exterior algebra by
\begin{displaymath}
    \Omega = \Omega_0 \oplus \Omega_1 = \wedge^\mathrm{even}
    (\mathbb {K}^d) \oplus \wedge^\mathrm{odd} (\mathbb{K}^d) \;,
\end{displaymath}
the \emph{Grassmann envelope} of $\mathrm{End}(V)$ by $\Omega$ is
the vector space
\begin{displaymath}
    \mathrm{End}_\Omega(V) = \bigoplus\nolimits_{\tau =
    0, 1} \Omega_\tau \otimes \mathrm{End}(V)_\tau \;.
\end{displaymath}
Thus an element $\xi \in \mathrm{End}_\Omega(V)$ consists of an
even part with commuting parameters as coefficients and an odd
part with anti-commuting parameters as coefficients.  When
$\mathrm{End}(V)$ is realized by matrices, $\xi$ is called a
\emph{supermatrix}.

There exists more than one associative multiplication law to turn
the vector space $\mathrm{End}_\Omega(V)$ into an associative
algebra. To achieve consistency, we have to adopt here the
convention of a Grassmann envelope of the \emph{second kind}
\cite{berezin}. This means that the associative product for
homogeneous $\alpha,\beta \in \Omega$ and $X,Y \in \mathrm{End}
(V)$ is defined by
\begin{displaymath}
    (\alpha \otimes X) \, (\beta \otimes Y) := (-1)^{|X| \,
    |\beta|} \alpha\beta \otimes X Y \;,
\end{displaymath}
where we follow the general sign rule \cite{QFT math} which tells
us to insert a minus sign (due to reversing the order) when both
$X$ and $\beta$ are odd.

Now transcribe this construction to the case of a Lie superalgebra
$\mathfrak{g} = \mathfrak{gl}(V)$ or $\mathfrak{g} = \mathfrak
{osp}(V)$.  The Grassmann envelope of $\mathfrak{g}$ by $\Omega$
then is the vector space
\begin{displaymath}
    \tilde{\mathfrak{g}}(\Omega) := \bigoplus\nolimits_{\tau =
    0, 1} \Omega_\tau \otimes \mathfrak{g}_\tau \;,
\end{displaymath}
and this carries the natural structure of a Lie algebra with
commutator bracket defined (still for homogeneous $\alpha,\beta
\in \Omega$ and $X,Y \in \mathfrak{g}$) by
\begin{eqnarray*}
    [\alpha \otimes X , \beta \otimes Y ] &:=& (-1)^{|X|
    |\beta|} \alpha\beta \otimes X Y - (-1)^{|Y| |\alpha|}
    \beta\alpha \otimes Y X \\ &=& (-1)^{|\alpha| |\beta|}
    \alpha\beta \otimes ( X Y - (-1)^{|X| |Y|} Y X) =
    \beta\alpha \otimes [X,Y] \;,
\end{eqnarray*}
where $[X,Y]$ is the bracket in $\mathfrak{g}\,$.  Note that our Lie
group $G$ naturally acts on the Grassmann envelope
$\tilde{\mathfrak{g}}(\Omega)$ by the adjoint action $\mathrm{Ad}(g)
(\alpha \otimes X) := \alpha \otimes \mathrm{Ad}(g)X\,$.

\subsubsection{The action of $\mathfrak{g}_1\,$}
\label{sect:g1-action}

Now fix some basis $\{ \mathsf{F}_i \}$ of $\mathfrak{g}_1\,$, and
denote the dual basis of $\mathfrak{g}_1^\ast$ by $\{ \varphi^i
\}$. (The construction to be made does not depend on which basis
is chosen.) Then let $\xi \in \mathfrak{g}_1^\ast \otimes
\mathfrak{g}_1^{\vphantom{\ast}}$ be the tautological object
\begin{displaymath}
    \xi = \sum \varphi^i \otimes \mathsf{F}_i \;,
\end{displaymath}
and repeat the change of interpretation that was made for the
operator $B$ in Sect.\ \ref{sect:basics}: the $\mathsf{F}_i \in
\mathfrak{g}_1\,$ are viewed as linear trans\-formations and
multiplied as linear transformations; whereas the $\varphi^i$, being
linear functions on the odd space $\mathfrak{g}_1$, are multiplied
via the wedge product and viewed as a set of generators of the
exterior algebra $\wedge(\mathfrak{g}_1^\ast)$.

Notice that, when $\mathfrak{g} = \mathfrak{gl}(V)$ or $\mathfrak
{g} = \mathfrak {osp}(V)$ acts on its fundamental module $V$, the
exponential $\mathrm{e}^\xi$ makes sense as an element of the
Grassmann envelope $\mathrm{End}_{\wedge \mathfrak{g}_1^\ast}(V)$:
\begin{displaymath}
    \mathrm{e}^\xi = \mathrm{e}^{\sum \varphi^i \otimes \mathsf{F}_i}
    = 1 \otimes \mathrm{Id}_V + \sum \varphi^i \otimes \mathsf{F}_i
    + {\textstyle{\frac{1}{2}}} \sum \varphi^j \varphi^i \otimes
    \mathsf{F}_i \, \mathsf{F}_j + \ldots \;.
\end{displaymath}
Further, the mapping $P = G \times G \to \mathrm{End}_{\wedge
\mathfrak{g}_1^\ast}(V)$ by $(h^\prime,h) \mapsto (1 \otimes
h^\prime)\, \mathrm{e}^{\sum \varphi^i \otimes \mathsf{F}_i} \, (1
\otimes h)^{-1}$ is equivariant with respect to $G$. Indeed, dropping
the trivial factors of unity, one has
\begin{displaymath}
    h^\prime \, \mathrm{e}^{\sum \varphi^i \otimes \mathsf{F}_i}
    \, h^{-1} = (h^\prime g)\, \mathrm{e}^{\,\sum \varphi^i
    \otimes \mathrm{Ad}(g^{-1}) \mathsf{F}_i} (h g)^{-1} \;,
\end{displaymath}
and since $\xi = \sum \varphi^i \otimes \mathsf{F}_i$ is
$G$-invariant by definition, one may trade the right $G$-action $g
\mapsto \mathrm{Ad}(g^{-1})$ on $\mathfrak {g}_1$ for the induced
action $g \mapsto \mathrm{Ad}^\ast(g^{-1})$ on $g_1^\ast \,$.

Using the notion of Grassmann envelope for $\mathfrak{g}\,$, we are
now going to describe the action $Y \mapsto \widehat{Y}^R$ of an odd
generator $Y \in \mathfrak{g}_1$ on the sections $\Phi \in
\mathcal{F}$. For the purpose of differentiation one extra
anti-commuting parameter is needed, and we therefore enlarge the
parameter Grassmann algebra to be
\begin{displaymath}
    \Omega = \wedge ( \mathfrak{g}_1^\ast \oplus \mathbb{K} )
    = \oplus_k \, \Omega^k \;.
\end{displaymath}
Let $\sigma \in \mathbb{K}$ denote a generator of the second summand,
$\mathbb{K}$, and fix some basis $\{ \mathsf{E}_j \}$ of the Lie
algebra $\mathfrak{g}_0 \subset \mathfrak{g}\,$.  The main step
toward defining $\widehat{Y}^R$ is the following lemma.
\begin{lem}[Berezin]\label{odd rearrange}
There exist even- and odd-type functions
\begin{eqnarray*}
    \alpha^i :\,\, \mathfrak{g}_1 \times G \to \wedge^\mathrm{even}
    (\mathfrak{g}_1^\ast) \;, &&(Y,g) \mapsto \alpha_Y^i(g) \quad
    (i = 1,\, \ldots, \,\mathrm{dim}\, \mathfrak{g}_1) \;,\\
    \beta^j :\,\, \mathfrak{g}_1 \times G \to \wedge^{\,\mathrm{odd}}
    (\mathfrak{g}_1^\ast) \;, &&(Y,g) \mapsto \beta_Y^j(g) \quad
    (j = 1,\, \ldots, \,\mathrm{dim}\, \mathfrak{g}_0)
\end{eqnarray*}
with linear (resp.\ differentiable) dependence on their first (resp.\
second) argument, such that for all $g^\prime, g \in G$ one has
\begin{displaymath}
    (g^\prime \mathrm{e}^\xi g^{-1}) \mathrm{e}^{\sigma \otimes
    Y} \equiv g^\prime \mathrm{e}^{\sum \varphi^i \otimes
    \mathsf{F}_i} g^{-1} \mathrm{e}^{\sigma \otimes Y} =
    g^\prime \mathrm{e}^{ \sum (\varphi^i + \alpha_Y^i(g)
    \sigma) \otimes \mathsf{F}_i} g^{-1} \mathrm{e}^{\sum
    \beta_Y^j(g) \sigma \otimes \mathsf{E}_j} \;.
\end{displaymath}
\end{lem}
\begin{proof}
The Lie algebra $\tilde{\mathfrak{g}} \equiv \tilde{\mathfrak{g}}
(\Omega)$ inherits from $\mathfrak{g}$ a $\mathbb{Z}_2$-grading
$\tilde{\mathfrak{g}} = \tilde{\mathfrak{g}}^\mathrm{even} \oplus
\tilde{\mathfrak{g}}^\mathrm{odd}$ by
\begin{displaymath}
    \tilde{\mathfrak{g}}^\mathrm{even} = \bigoplus_{k \ge 0}
    \tilde{\mathfrak{g}}^{(2k)}\;,\quad \tilde{\mathfrak{g}}^
    \mathrm{odd} = \bigoplus_{k \ge 0} \tilde{\mathfrak{g}}^{(2k+1)}
    \;, \quad \tilde{\mathfrak{g}}^{(2k+\tau)} = \Omega^{2k+\tau}
    \otimes \mathfrak{g}_\tau \quad (\tau = 0 , 1)\;.
\end{displaymath}
Notice that $[\tilde{\mathfrak{g}}^{(k)},\tilde{\mathfrak{g}}^{(l)} ]
\subset \tilde{\mathfrak{g}}^{(k+l)}$, so that $\oplus_{k \ge 1}
\tilde{\mathfrak{g}}^{(k)} =: \mathfrak{n}$ is a nilpotent Lie
subalgebra.

Proving the lemma essentially amounts to showing that, for any
$g^\prime, g \in G$, there exist unique elements $\Xi_1 \in \tilde{
\mathfrak{g}}^\mathrm{odd}$ and $\Xi_0 \in \tilde{\mathfrak{g}
}^\mathrm{even} \cap \mathfrak{n}$ with the property that
\begin{displaymath}
    g^\prime \mathrm{e}^{\xi} g^{-1} \mathrm{e}^{\sigma \otimes Y}
    = g^\prime \mathrm{e}^{\xi + \Xi_1} g^{-1} \mathrm{e}^{\Xi_0}
    \;.
\end{displaymath}
For the purpose of establishing such an identity, we translate by
$(g^\prime \mathrm{e}^\xi)^{-1}$ on the left and by $g$ on the right,
and then take the logarithm:
\begin{displaymath}
    \sigma \otimes \mathrm{Ad}(g)^{-1} Y = \ln \big(
    \mathrm{e}^{-\xi} \mathrm{e}^{\xi + \Xi_1}
    \mathrm{e}^{\mathrm{Ad}(g)^{-1} \Xi_0} \big) \;.
\end{displaymath}
This equation is equivalent to the previous one, as the exponential
map from the nilpotent Lie algebra $\mathfrak{n} = \oplus_{k \ge 1}
\tilde{\mathfrak{g}}^{(k)}$ to the nilpotent Lie group
$\exp(\mathfrak{n})$ is a bijection.

It is clear that both unknowns $\Xi_1$ and $\Xi_0$ must contain at
least one factor of the generator $\sigma$ (and hence exactly one
such factor, since $\sigma^2 = 0)$. Using a standard formula from Lie
theory we now combine the first two factors under the logarithm:
\begin{displaymath}
    \mathrm{e}^{-\xi} \mathrm{e}^{\xi + \Xi_1} = 1 + T_\xi(\Xi_1)
    = \mathrm{e}^{T_\xi(\Xi_1)} \;, \quad T_\xi = \sum_{k \ge 0}
    \frac{\mathrm{ad}^k(-\xi)}{(k+1)!} \;.
\end{displaymath}
Since $\mathrm{e}^{T_\xi(\Xi_1)} \mathrm{e}^{\mathrm{Ad}(g)^{-1}
\Xi_0} = \mathrm{e}^{T_\xi(\Xi_1) + \mathrm{Ad}(g)^{-1} \Xi_0}$ by
$\sigma^2 = 0\,$, our equation becomes
\begin{displaymath}
    \sigma \otimes \mathrm{Ad}(g)^{-1} Y =
    T_\xi(\Xi_1) + \mathrm{Ad}(g)^{-1} \Xi_0 \;,
\end{displaymath}
in which form it can be solved for $\Xi_1 = \sum_{k \ge 1}
\Xi_1^{(2k-1)}$ and $\Xi_0 = \sum_{k \ge 1} \Xi_0^{(2k)}$ by
re\-cursion in $k = 0, 1, 2, \ldots$.  The first few terms of the
solution are
\begin{displaymath}
    \Xi_1^{(1)} = \sigma \otimes \mathrm{Ad}(g)^{-1} Y \;, \quad
    \Xi_0^{(2)} = {\textstyle{\frac{1}{2!}}} \mathrm{Ad}(g) [ \xi ,
    \Xi_1^{(1)} ] \;, \quad \Xi_1^{(3)} = - {\textstyle{\frac{1}{3!}}}
    [ \xi , [ \xi , \Xi_1^{(1)} ] ] \;,
\end{displaymath}
and so on.  The recursion terminates by nilpotency of the Lie algebra
$\mathfrak{n}\,$.

All components $\Xi_{\tau}^{(2k-\tau)}$ of the solution are linear in
$Y$ and depend differentiably (in fact, analytically) on $g \in G$.
By going back to the original equation $g^\prime \mathrm{e}^{\xi}
g^{-1} \mathrm{e}^{\sigma \otimes Y} = g^\prime \mathrm{e}^{\xi +
\Xi_1} g^{-1} \mathrm{e}^{\Xi_0}$, and expanding the solution for
$\Xi_1$ and $\Xi_0$ in the chosen bases,
\begin{displaymath}
    \Xi_1 = \sum \alpha_Y^i(g) \sigma \otimes \mathsf{F}_i \;,
    \quad \Xi_0 = \sum \beta_Y^j(g)\sigma \otimes \mathsf{E}_j \;,
\end{displaymath}
we arrive at the statement of the lemma.
\end{proof}
\begin{prop}\label{prop 3.2p}
The odd component $\mathfrak{g}_1 \subset \mathfrak{g}$ acts on
sections $\Phi \in \mathcal{F}$ by $Y \mapsto \widehat{Y}^R$,
\begin{displaymath}
    (\widehat{Y}^R \Phi)(g^\prime,g) := \sum \alpha_Y^i(g)
    \frac{\partial} {\partial\varphi^i} \Phi(g^\prime,g) - \sum
    \beta_Y^j(g) \,(\widehat{\mathsf{E}}_j^R \Phi)(g^\prime,g) \;,
\end{displaymath}
where $\alpha^i$ and $\beta^j$ are the functions of Lem.\ \ref{odd
rearrange}, the operator $\partial/\partial\varphi^i := \iota(\mathsf
{F}_i)$ is the odd derivation of $\wedge (\mathfrak{g}_1^\ast )$ by
contraction, and $\widehat{\mathsf{E}}_j^R$ is the left-invariant
differential operator determined by $\mathsf{E}_j \in \mathfrak{g}_0
\,$.
\end{prop}
\begin{proof}
We interpret the identity of Lem.\ \ref{odd rearrange} as saying that
a right multiplication of the supermatrix $g^\prime \mathrm{e}^\xi
g^{-1}$ by $\mathrm{e}^{\sigma \otimes Y}$ is the same as shifting
the generators $\varphi^i \to \varphi^i + \sigma \alpha_Y^i$ and
making a right translation by $\mathrm{e}^{\sum \beta_Y^j \sigma
\otimes \mathsf{E}_j}$.  This observation motivates the proposed
formula for $\widehat{Y}^R$, by linearization in the parameter
$\sigma$ (the sign in the second term of the expression for
$\widehat{Y}^R$ stems from $\beta^j \sigma = - \sigma \beta^j$). It
is easy to check that $\widehat{Y}^R$ really does map $\mathcal{F}$
into $\mathcal{F}$, i.e., if $\Phi$ is a $G$-equivariant map $\Phi :
P \to \wedge (\mathfrak{g}_1^\ast)$, then so is $\widehat{Y}^R \Phi$.

Let now $\sigma_1, \sigma_2$ be \emph{two} anti-commuting
parameters, and let $Y_1 , Y_2 \in \mathfrak{g}_1\,$. Then, since
\begin{displaymath}
    \mathrm{e}^{\sigma_1 \otimes Y_1} \mathrm{e}^{\sigma_2
    \otimes Y_2} \mathrm{e}^{- \sigma_1 \otimes Y_1}
    \mathrm{e}^{- \sigma_2 \otimes Y_2} = \mathrm{e}^{\sigma_2
    \sigma_1 \otimes [Y_1 , Y_2 ]} \;,
\end{displaymath}
it immediately follows that the correspondence $Y \mapsto
\widehat{Y}^R$ preserves the Lie superbracket of $\mathfrak{g}
\,$, i.e., one has the anti-commutation relations
\begin{displaymath}
    [ \widehat{Y}_1^R , \widehat{Y}_2^R ] \equiv
    \widehat{Y}_1^R \widehat{Y}_2^R + \widehat{Y}_2^R
    \widehat{Y}_1^R = \widehat{[Y_1 , Y_2]}^R \;.
\end{displaymath}
(Please be advised that if we had used a Grassmann envelope of the
first kind, as in (\ref{first envelope}), then there would have been
a sign inconsistency in this equation.) Similarly, for $X \in
\mathfrak{g}_0$ and $Y \in \mathfrak{g}_1$ one has $[ \widehat{X}^R ,
\widehat{Y}^R ] = \widehat{[X,Y]}^R$.
\end{proof}
What we have described is the action on the right. But the same
construction can be done on the left, and of course the resulting
correspondence $Y \mapsto \widehat{Y}^L$ again is an odd
derivation which preserves the bracket relations of $\mathfrak{g}
\,$.

\subsubsection{Summary}

Thus for every $Y \in \mathfrak {g}_1$ we are given two odd
first-order differential operators $\widehat{Y}^L$ and
$\widehat{Y}^R$ on sections $\Phi \in \mathcal{F}$ such that the
superbracket relations
\begin{displaymath}
    [ \widehat{Y}_1^i , \widehat{Y}_2^i ] = \widehat{[Y_1,Y_2]}^i
    \;, \quad [\widehat{X}^i , \widehat{Y}^i ] = \widehat{[X,Y]}^i
\end{displaymath}
hold for $i = L, R$ and all $X \in \mathfrak{g}_0$ and $Y, Y_1,
Y_2 \in \mathfrak{g}_1\,$. The actions on the left and right
commute in the graded-commutative sense.

Unlike the action of the Lie algebra $\mathfrak{g}_0\,$, that of
the odd part $\mathfrak {g}_1$ does not readily exponentiate. We
therefore leave this at the infinitesimal level.

Let us now summarize the educational material of this subsection.
\begin{defn}\label{LSG}
Given a Lie superalgebra $\mathfrak{g} = \mathfrak{g}_0 \oplus
\mathfrak{g}_1\,$, and a Lie group $G$ (with Lie algebra
$\mathfrak{g}_0$) acting on the vector space $\mathfrak {g}_1$ by the
adjoint representation $\mathrm{Ad}\,$, there is a principal bundle
$P = G \times G \to G$ and an associated vector bundle $F \to G$ with
total space $F = P \times_{G}^{ \vphantom{\dagger}}
\mathfrak{g}_1\,$. If $\mathfrak{g} = \mathfrak {gl}$ or
$\mathfrak{g} = \mathfrak{osp}\,$, then by the Lie supergroup
$\mathfrak{G} = (\mathfrak{g} , G)$ we mean the graded-commutative
algebra of sections $\mathcal{F} = \Gamma(G, \wedge F^\ast)$ carrying
the canonical actions of $\mathfrak{g}$ by left- or right-invariant
derivations. A section of $\mathcal{F}$ is also referred to as a
(super)function on $\mathfrak{G}\,$. The component in
$\mathrm{C}^\infty(G)$ of a section $s \in \mathcal{F}$ is called the
numerical part of $s$ and is denoted by $\mathrm{num}(s)$.
\end{defn}
\begin{rem}
Supermanifolds, of which Lie supergroups are special examples, are in
principle constructed by using the language of sheaves, i.e., by
joining together locally defined functions (here with values in an
exterior algebra) by means of transition functions on overlapping
domains. The transition functions for a general supermanifold respect
only the $\mathbb{Z}_2$-grading (not the $\mathbb{Z}$-grading) of
exterior algebras.

What is special about the supermanifolds defined above is that
their structure sheaf \emph{does} respect a $\mathbb{Z}$-grading,
as it reduces to that of a vector bundle. Therefore it was not
compulsory to use sheaf-theoretic language in formulating Def.\
\ref{LSG}. Nevertheless, we will sometimes refer to $\mathcal{F} =
\Gamma(G,\wedge F^\ast)$ as a sheaf of graded-commutative
algebras.
%
%
\end{rem}
In the case of $\mathfrak{g} = \mathfrak{gl}_{p|q}$ and $G =
\mathrm{GL}_p \times \mathrm{GL}_q$ one writes $\mathfrak{G} =
\mathrm{GL}_{p|q}$ for short.

\subsection{What's a representation in the supergroup setting?}

If $G$ is a Lie group (or, for that matter, any group), a
representation of $G$ is given by a $\mathbb{K}$-vector space $V$ and
a homomorphism from $G$ into the group of $\mathbb{K}$-linear
transformations of $V$, i.e., a mapping $\rho : G \to \mathrm{GL}(V)$
which respects the group multiplication law:
\begin{displaymath}
    \rho(g_1 g_2) = \rho(g_1) \rho(g_2) \;.
\end{displaymath}
In the Lie supergroup setting, since we have purposely avoided
defining what is meant by the group multiplication law, we have to
give meaning to the word "representation" in an alternative (but
equivalent) way.

Recall that the basic data of a Lie supergroup $\mathfrak {G}$ are a
Lie superalgebra $\mathfrak{g} = \mathfrak{g}_0 \oplus
\mathfrak{g}_1$ and a Lie group $G$ with $\mathrm {Lie}(G) =
\mathfrak{g}_0 \,$.  To specify a representation of $\mathfrak{G}$,
one first of all needs a representation space, which in the current
context has to be a $\mathbb{Z}_2$-graded $\mathbb{K}$-vector space
$V = V_0 \oplus V_1$. For present use recall from Sect.\
\ref{sect:basics} the definition of the Lie superalgebra
$\mathfrak{gl}(V) = \mathfrak{gl}(V)_0 \oplus \mathfrak{gl}(V)_1\,$,
and let $e \in G$ be the neutral element.
\begin{defn}
A representation of a Lie supergroup $\mathfrak{G} = (\mathfrak{g},
G)$ on a $\mathbb{Z}_2$-graded vector space $V = V_0 \oplus V_1$ is
given by a homomorphism $\rho_\ast$ of Lie superalgebras
\begin{displaymath}
    \rho_\ast :  \,\, \mathfrak{g}_0 \to \mathfrak{gl}(V)_0 \;,
    \quad \rho_\ast :  \,\, \mathfrak{g}_1 \to \mathfrak{gl}(V)_1
    \;,
\end{displaymath}
and a homomorphism of Lie groups
\begin{displaymath}
    \rho : \,\, G \to \big( \mathrm{GL}(V_0) \times
    \mathrm{GL}(V_1) \hookrightarrow \mathrm{GL}(V) \big) \;,
\end{displaymath}
which are compatible in the sense that $(\mathrm{d} \rho)_e =
\rho_\ast \vert_{\mathfrak{g}_0}\,$.
\end{defn}
\begin{rem}
Note that for the case of a finite-dimensional representation
space $V$, compatibility immediately implies that
\begin{displaymath}
    \rho(g) \, \rho_\ast (Y) \, \rho(g^{-1}) =
    \rho_\ast \big( \mathrm{Ad}(g) Y \big) \quad
    (g \in G \, , \,\, Y \in \mathfrak{g}_1) \;,
\end{displaymath}
by the formulas $\rho(\mathrm{e}^{tX}) = \mathrm{e}^{ t\, (\mathrm{d}
\rho)_e (X)}$ (for $t \in \mathbb{K}$) and $\mathrm{Ad} \circ \exp =
\exp \circ \mathrm{ad}\,$.
\end{rem}

\subsection{What's a character of $G$?}\label{sect:3.4}

If we are given a representation $(V,\rho_\ast,\rho)$ of the Lie
supergroup $\mathfrak{G} = (\mathfrak{g}, G)$, we can say precisely
-- in the language of Sect.\ \ref{sect:superG} -- what is meant by
its character $\chi$. The precise meaning of $\chi$ comes from its
definition as an element of $\mathcal{F} = \Gamma(G, \wedge F^\ast)$,
i.e., as a section of the vector bundle
\begin{displaymath}
    \wedge (F^\ast) = P \times_{G}^{\vphantom{\dagger}}
    \wedge (\mathfrak{g}_1^\ast) \,\, \to \,\, G \;.
\end{displaymath}
This section $\chi$ is constructed from the representation $(V,
\rho_\ast , \rho)$ as follows.

As before, fix some basis $\{ \mathsf{F}_i \}$ of $\mathfrak {g}_1$,
denote the dual basis of $\mathfrak{g}_1^\ast$ by $\{ \varphi^i \}$,
and recall the tautological object $\xi = \sum \varphi^i \otimes
\mathsf{F}_i\,$. Then, fixing some pair $(h^\prime,h) \in G \times G
= P$, consider the task of representing the formal object $\Xi =
h^\prime \, \mathrm{e}^{\xi} h^{-1}$ as an operator on $V$, with
coefficients in $\wedge (\mathfrak{g}_1^\ast)$. The obvious choice of
operator is
\begin{displaymath}
    \rho(h^\prime) \, \mathrm{e}^{\sum \varphi^i
    \rho_\ast(\mathsf{F}_i)} \rho(h^{-1}) \;,
\end{displaymath}
which is to be viewed as an element of the Grassmann envelope
$\mathrm{End}_{\wedge \mathfrak{g}_1^\ast}(V)$.

Since $\mathrm{End}_{\wedge \mathfrak{g}_1^\ast}(V)$ is of the
second kind, the corresponding supertrace is defined by
\begin{displaymath}
    \mathrm{STr}\, (\alpha \otimes X) (\beta \otimes Y) :=
    \beta\alpha \, \mathrm{STr}_V (X Y) \;.
\end{displaymath}
For later use, note that the supertrace on a Grassmann envelope of
first or second kind always has the cyclic property: $\mathrm{STr}\,
(\Xi \Theta) = \mathrm{STr}\, (\Theta \Xi)\,$.

Now take the supertrace of the operator above to define
\begin{eqnarray*}
    \chi(h^\prime,h) &:=& \mathrm{STr}_V \, \rho(h^\prime)\,
    \mathrm{e}^{\sum \varphi^i \rho_\ast(\mathsf{F}_i)}
    \rho(h^{-1})\\ &=& \mathrm{STr}_V \, \rho(h^\prime h^{-1})
    + \sum \varphi^i \,\mathrm{STr}_V \,\rho(h^\prime)
    \rho_\ast(\mathsf{F}_i) \rho(h^{-1})\\
    &+&{\textstyle{\frac{1}{2}}} \sum \varphi^j \varphi^i \,
    \mathrm{STr}_V \, \rho(h^\prime) \rho_\ast(\mathsf{F}_i)
    \rho_\ast(\mathsf{F}_j) \rho(h^{-1}) + \ldots\;,
\end{eqnarray*}
where the Taylor expansion of the exponential function terminates at
finite order since the $\varphi^i$ are nilpotent.  If the dimension
of $V$ is finite, this definition makes perfect sense. In the
infinite-dimensional case, the convergence of $\mathrm{STr}_V \,
\rho(h^\prime h^{-1})$ etc.\ is an issue and $\chi(h^\prime, h)$ will
typically exist only for $(h^\prime,h)$ in a suitable subset of $G
\times G\,$; for simplicity of the exposition, we here assume this
complication to be absent.

When $h$ and $h^\prime$ are allowed to vary, $\chi$ becomes a
function on $G \times G = P$ which takes values in $\wedge
(\mathfrak{g}_1^\ast)$ and is equivariant with respect to $G$.
Indeed,
\begin{displaymath}
    \chi(h^\prime g, h g) = \mathrm{STr}_V \, \rho(h^\prime)
    \, \mathrm{e}^{\sum \varphi^i \, \rho(g) \rho_\ast(
    \mathsf{F}_i) \rho(g^{-1})} \rho(h^{-1}) \;.
\end{displaymath}
By the compatibility of $\rho$ and $\rho_\ast$,
\begin{displaymath}
    \sum \varphi^i \, \rho(g) \rho_\ast(\mathsf{F}_i)\rho(g^{-1}) =
    \sum \varphi^i \, \rho_\ast \big(\mathsf{F}_j \, \mathrm{Ad}
    (g)_{\phantom{j}i}^{j\phantom{i}}\big) = \sum \varphi^i \,
    \mathrm{Ad}^\ast(g)_{i\phantom{j}}^{\phantom{i}j}\,\rho_\ast
    (\mathsf{F}_j) \;,
\end{displaymath}
and therefore
\begin{displaymath}
    \chi(h^\prime,h) = \mathrm{Ad}^\ast(g^{-1}) \,
    \chi(h^\prime g, h g) \;,
\end{displaymath}
where $g \mapsto \mathrm{Ad}^\ast(g^{-1})$ is the induced
$G$-representation on $\wedge (\mathfrak{g}_1^\ast)$.  Hence $\chi$
is an element of $\mathrm{C}^\infty(P, \wedge\mathfrak {g}_1^\ast)
^{G}$ or, equivalently, a section of $\Gamma(G, \wedge F^\ast)$.
\begin{defn}
The character determined by the representation $(V, \rho_\ast, \rho)$
of a Lie supergroup $\mathfrak{G} = (\mathfrak{g}, G)$ is defined to
be the section $\chi \in \Gamma(G, \wedge F^\ast)$ given by
\begin{displaymath}
    \chi(g h^{-1}) = \big[ (g,h)\,;\,\mathrm{STr}_V \, \rho(g)\,
    \mathrm{e}^{\sum \varphi^i \, \rho_\ast(\mathsf{F}_i)}
    \rho(h^{-1}) \big] \;,
\end{displaymath}
whenever this exists. We also write $\Xi = g \, \mathrm {e}^\xi
h^{-1}$ and $\chi(\Xi) = \mathrm{STr}_V \, \rho (\Xi)$ for short.
\end{defn}
\begin{rem}
The numerical part $\mathrm{num}(\chi) \in \Gamma(G, \wedge^0
F^\ast)$ coincides with the character of the $\mathbb{Z}_2$-graded
$G$-representation $(V_0 \oplus V_1, \rho)$ in the $\mathbb{Z}
_2$-graded sense:
\begin{displaymath}
    \mathrm{num}(\chi)(x) = \mathrm{Tr}_{V_0}\, \rho(x) -
    \mathrm{Tr}_{V_1} \, \rho(x) \;.
\end{displaymath}
\end{rem}

\subsection{$\chi$ is an eigenfunction of all Laplace-Casimir
operators}\label{sect:eigenfctn}

Primitive characters, i.e., characters of irreducible
representations, are special functions with special properties.
Foremost among these is their being joint eigenfunctions of the ring
of invariant differential operators. Let us review this general
property in the superalgebra setting, which is where it will be
exploited below.  In this subsection we put $\mathbb{K} =
\mathbb{C}\,$.

If $\mathfrak{g}$ is a Lie superalgebra, the universal enveloping
algebra $\mathsf{U}(\mathfrak{g})$ is the associative algebra
generated by $\mathfrak{g}$ with the bracket relations of
$\mathfrak{g}$ being understood. A \emph{Casimir invariant} of
$\mathfrak{g}$ then is an element $I$ in the center of
$\mathsf{U}(\mathfrak{g})$, i.e, a polynomial $I$ in the
generators of $\mathfrak{g}$ with the property $[I , X] = 0$ for
all $X \in \mathfrak{g}\,$. For example, if $E_1, \ldots, E_d$
(with $d = \mathrm{dim}\, \mathfrak{g}$) is a basis of
$\mathfrak{g}$ and $Q : \mathfrak{g} \times \mathfrak {g} \to
\mathbb{C}$ is a non-degenerate $\mathfrak {g}$-invariant
quadratic form, the quadratic Casimir invariant is
\begin{displaymath}
    I_2 = \sum Q^{ij} E_i E_j \;,
\end{displaymath}
where the coefficients $Q^{ij}$ are determined by $\sum_j Q^{ij}
Q_{jk} = \delta_k^i$ and $Q_{ij} = Q(E_i \, , E_j)$.

To describe the Casimir invariants of the Lie superalgebra
$\mathfrak{g} = \mathfrak{gl}_{p|q}\,$, let $E_{ij}^{\sigma\tau}$
with $\sigma, \tau = 0, 1$ be a standard basis of $\mathrm{End}
(\mathbb{C}^{p|q})$. This means that the parity of $E_{ij}^{\sigma
\tau}$ is $\sigma + \tau$ (mod 2), the range of a lower index is
understood to be $i \in \{ 1, 2, \ldots, p \}$ if the corresponding
upper index is $\sigma = 1$ and $i \in \{ 1, 2, \ldots, q \}$ if
$\sigma = 0$, and one has
\begin{displaymath}
    E_{ij}^{\sigma\tau} E_{i^\prime j^\prime}^{\sigma^\prime
    \tau^\prime} = \delta^{\tau \sigma^\prime} \delta_{j i^\prime}
    E_{i j^\prime}^{\sigma \tau^\prime} \;.
\end{displaymath}
\begin{lem}[Berezin]\label{Ber0}
There exists a degree-$\ell$ Casimir invariant $I_\ell$ of
$\mathfrak {gl}_{p|q}$ for every $\ell \in \mathbb{N}\,$.  Its
expression in terms of a standard basis of generators
$E_{ij}^{\sigma\tau}$ is
\begin{displaymath}
    I_\ell = \sum E_{j_1 j_2}^{\sigma_1 \sigma_2} (-1)^{\sigma_2}
    E_{j_2 j_3}^{\sigma_2 \sigma_3} (-1)^{\sigma_3} \cdots
    E_{j_{\ell-1} j_\ell}^{\sigma_{\ell-1} \sigma_\ell}
    (-1)^{\sigma_\ell} E_{j_\ell j_1}^{\sigma_\ell \sigma_1} \;.
\end{displaymath}
\end{lem}
\begin{rem}
This formula for $I_\ell$ is Eqs.\ (4.35-37) of \cite{berezin}.
From the basic brackets
\begin{equation}\label{basic brackets}
   [ E_{ij}^{\sigma\tau} ,
     E_{i^\prime j^\prime}^{\sigma^\prime \tau^\prime}]
   = E_{i j^\prime}^{\sigma \tau^\prime} \delta_{j {i^\prime}}
   \delta^{\tau \sigma^\prime} - (-1)^{(\sigma + \tau)
   (\sigma^\prime + \tau^\prime)} E_{i^\prime j}^{\sigma^\prime
   \tau} \delta_{j^\prime i} \, \delta^{\tau^\prime \sigma} \;,
\end{equation}
it is in fact easy to check that $[X , I_\ell] = 0$ for all $X \in
\mathfrak{gl}_{p|q}\,$.
\end{rem}
Now turn to the functions on a Lie supergroup $\mathfrak{G} =
(\mathfrak{g}, G)$, i.e., to the sections $\Phi$ of the algebra
$\mathcal{F} = \Gamma(G, \wedge F^\ast)$. Every Casimir invariant $I
\in \mathsf{U}( \mathfrak{g} )$ determines an invariant differential
operator $D(I)$ -- also called a Laplace-Casimir operator -- on such
functions.  In the case of the quadratic Casimir invariant $I_2$ this
is
\begin{displaymath}
    D(I_2) \Phi  = \sum Q^{ij} \widehat{E}_i^L \widehat{E}_j^L \,
    \Phi = \sum Q^{ij} \widehat{E}_i^R \widehat{E}_j^R\, \Phi \;,
\end{displaymath}
where $E_i \mapsto \widehat{E}_i^{L,R}$ are the canonical right-
and left-invariant actions of the Lie superalgebra $\mathfrak{g}$
on the Lie supergroup $\mathfrak{G}$ (see Def.\ \ref{LSG}). More
generally, $D(I)$ is obtained by replacing in the polynomial
expression for $I$ each generator of $\mathfrak{g}$ by the
corresponding right-invariant or left-invariant differential
operator.

Assume now an irreducible representation $(V,\rho_\ast,\rho)$ of
the Lie supergroup $\mathfrak{G}$ to be given. Applying any
Laplace-Casimir operator $D(I)$ to the character $\chi(\Xi)$ one
gets
\begin{displaymath}
    \big( D(I) \chi \big)(\Xi) = \mathrm{STr}_V \, \rho_\ast(I)
    \rho(\Xi) \;,
\end{displaymath}
where $\rho_\ast$ has been extended from $\mathfrak{g}$ to
$\mathsf{U}(\mathfrak{g})\,$. Recall that we now are working over
$\mathbb{K} = \mathbb{C}\,$. Since the representation $\rho_\ast$ is
irreducible and the Casimir invariant $I$ commutes with all
generators of $\mathfrak{g}\,$, it follows that the operator
$\rho_\ast(I)$ is a multiple of unity by Schur's lemma: $\rho_\ast(I)
= \lambda(I) \, \mathrm {Id}_V$ with $\lambda(I) \in \mathbb{C}\,$.
Thus $\chi$ is an eigenfunction:
\begin{displaymath}
    D(I) \chi = \lambda(I) \, \chi \;.
\end{displaymath}
This fact will be exploited in Sect.\ \ref{sect:4.2}.

\subsection{Radial functions}\label{sect:3.5}

An important fact about the characters of a group $G$ is that they
are constant on conjugacy classes:  $\chi(g)= \chi(h g h^{-1} )$
for $g \, , h \in G\,$. Functions $f: \, G \to \mathbb {C}$ with
this property are called \emph{radial}. In the case of a Lie group
$G$ the infinitesimal version of the radial property is
\begin{displaymath}
    0 = \frac{d}{ds} f(\mathrm{e}^{-sX} g \, \mathrm{e}^{sX})
    \Big|_{s = 0} = \big( \widehat{X}^L + \widehat{X}^R \big)
    f(g) \;.
\end{displaymath}
Notice that the notation used here is consistent with our earlier
definition of the actions on the left and right, $\widehat{X}^{L,
R}$, by the identification $f_2(g,h) \equiv f_1(g h^{-1})$.
\begin{defn}
A section $\Phi \in \mathcal{F}$ of a Lie supergroup $\mathfrak{G}
= (\mathfrak{g},G)$ is called radial if $\Phi(g x g^{-1}) =
\Phi(x)$ for all $g, x \in G\,$, and if
\begin{displaymath}
    \big( \widehat{Y}^L + \widehat{Y}^R \big) \Phi = 0
\end{displaymath}
holds for all $Y \in \mathfrak{g}_1 \,$, the odd part of the Lie
superalgebra $\mathfrak{g}\,$.
\end{defn}
\begin{rem}
An important example of a radial section for the case of
$\mathfrak{g} = \mathfrak{gl}(V_1 \oplus V_0)$ and $G =
\mathrm{GL}(V_1) \times \mathrm{GL}(V_0)$ is the superdeterminant
(see Def.\ \ref{SDet}).
\end{rem}
The material reviewed in Sects.\ \ref{sect:superG}-\ref{sect:3.4} did
not depend on the choice of number field $\mathbb{K} = \mathbb{C}$ or
$\mathbb{K} = \mathbb{R}$. However, to do analysis with radial
functions we now wish to abandon the basic case of a complex Lie
superalgebra $\mathfrak{g}$ with complex Lie group $G$ and restrict
the discussion to a \emph{real} subgroup $G_\mathbb{R} \subset G\,$.

Hence, let $G_\mathbb{R} \equiv K$ be a connected compact Lie
group. The set of conjugacy classes then is parameterized by the
elements $t$ in a Cartan subgroup, or maximal torus, $T \subset
K$. It follows that a radial function $\Phi$ on $K$ determines a
function $t \mapsto f(t)$ by restriction, $f := \Phi \big\vert_T
\,$. Conversely, a function $f$ on $T$ is known to extend to a
radial function $\Phi$ on $K$ provided that $f$ is invariant under
the action of the so-called Weyl group $W$ of $K$.

Turning to the case of a complex Lie supergroup $\mathfrak{G} =
(\mathfrak{g}\, ,G)$, let $K \subset G$ be a connected compact Lie
group whose Lie algebra is a compact real form of $\mathfrak{g}_0 =
\mathrm {Lie}(G)$.  In the case of $\mathfrak{G} = \mathrm{GL}_{m|n}$
for example, take $K = \mathrm{U}_m \times \mathrm{U}_n \,$. Then let
$\mathcal{F}^\prime$ denote the sheaf of algebras $\mathcal{F}$
restricted to $K\,$. Note that the restriction $\mathcal{F}^\prime$
still carries the action of the complex Lie superalgebra
$\mathfrak{g}$ by left-invariant and right-invariant differential
operators.

Now fix some maximal torus $T$ of $K$. A radial section $\Phi \in
\mathcal{F}^\prime$ determines a function $f : T \to \mathbb{C}$
by restriction and truncation to the numerical part,
\begin{displaymath}
    f := \mathrm{num}(\Phi) \big\vert_T \;,
\end{displaymath}
and this function $f$ again is invariant under the Weyl group $W$ of
$K$. (Please be advised that the converse is no longer true! In order
for a $W$-invariant function $f$ on $T$ to extend to a radial section
$\Phi \in \mathcal{F}^\prime$ some extra regularity conditions must
be fulfilled; see Part II, Chapter 3, Thm.\ 3.1 of \cite{berezin}.
However, in the application we are aiming at, namely the character
$\chi$ of Sect.\ \ref{sect:2}, extendability will be automatic by
construction. Hence there is no need to discuss the full set of
conditions for extendability here.)

For every Casimir invariant $I \in \mathsf{U}(\mathfrak {g})$ we
have a Laplace-Casimir operator $D(I)$ on $\mathcal{F}$. If $\Phi
\in \mathcal{F}$ is radial, i.e., $\Phi = L_g R_g \Phi$ for all $g
\in G$ and $(\widehat{Y}^L + \widehat {Y}^R) \Phi = 0$ for all $Y
\in \mathfrak{g}_1$, then so is $D(I) \Phi$, since $D(I)$ commutes
with all of the $L_g$ and $R_g$ and $\widehat{Y}^{L,R}$. Thus the
subalgebra of radial sections in $\mathcal{F}$ is invariant under
the application of $D(I)$. The restriction $\dot {D}(I)$ of $D(I)$
to this invariant subalgebra is called the \emph{radial part}.

Given a compact real form $K = G_\mathbb{R} \subset G$, one can
transfer this discussion to the subalgebra of radial sections of the
restriction $\mathcal{F}^\prime = \mathcal{F} \vert_K$, which is
still invariant under the application of the Laplace-Casimir
operators $D(I)$.  The radial parts $\dot{D}(I)$ for $\mathfrak{G} =
\mathrm{GL}$ or $\mathfrak {G} = \mathrm{OSp}$ can then be described
as differential operators on the algebra of extendable $W$-invariant
functions on a maximal torus $T \subset K\,$.  We will write them
down for the case of $\mathfrak{G} = \mathrm{GL}_{m|n}$ with $K =
\mathrm{U}_m \times \mathrm{U}_n$ in Sect. \ref{sect:radialpart}.

\subsection{Block diagonalization of supermatrices}

Berezin has given a description \cite{berezin} of the radial parts of
the Laplace-Casimir operators for the cases of $\mathfrak{g} =
\mathfrak {gl}$ and $\mathfrak{g} = \mathfrak{osp}\,$.  Since our
paper will make heavy use of it, we now give a flavor of the first
step of that theory.  (For this first step, the choice of $K \subset
G$ is not yet relevant.)

Recalling the material of Sect.\ \ref{sect:envelope}, let $\tilde{
\mathfrak{g}} (\Omega)$ be the Grassmann envelope (of first or second
kind) of the Lie superalgebra $\mathfrak{g} = \mathfrak{g}_0 \oplus
\mathfrak{g}_1$ by the Grassmann algebra $\Omega = \wedge(\mathfrak
{g}_1^\ast)$. The Lie algebra $\tilde{\mathfrak{g}} \equiv \tilde{
\mathfrak{g}}(\Omega)$ carries a $\mathbb{Z}_2$-grading $\tilde{
\mathfrak{g}} = \tilde{\mathfrak{g}}^\mathrm{even} \oplus \tilde{
\mathfrak{g}}^\mathrm{odd}$ and also a compatible $\mathbb
{Z}$-grading $\tilde{\mathfrak{g}} = \oplus_{k \ge 0}\, \tilde{
\mathfrak{g}}^{(k)}$. From $\mathrm{Ad}(g) :\, \mathfrak{g} \to
\mathfrak{g}$ one has an induced adjoint action of $G$ on $\tilde{
\mathfrak{g}}$ and each of its subspaces $\tilde{\mathfrak {g}}^{
(k)}$ for $k \ge 0$ by $\alpha \otimes X \mapsto \alpha \otimes
\mathrm {Ad}(g)\,X$.
\begin{prop}\label{block diag}
Let $g \in G$ be regular in the sense that $\mathrm{Ad}(g) -
\mathrm{Id}$ has zero kernel as a linear operator on $\mathfrak
{g}_1\,$. Then for any $\Theta_1 \in \tilde{\mathfrak{g}}^
\mathrm{odd}$ the supermatrix $g\, \mathrm{e}^{\Theta_1}$ can be
block diagonalized by nilpotents, i.e., there exist unique elements
$\Xi_1^{(2k-1)} \in \tilde{\mathfrak{g}}^{(2k-1)}$ and $\Xi_0^{(2k)}
\in \tilde{\mathfrak{g}}^{(2k)}$ for $k \ge 1$ such that
\begin{displaymath}
    g\, \mathrm{e}^{\Theta_1} = \mathrm{e}^{\Xi_1} g \,
    \mathrm{e}^{\Xi_0} \mathrm{e}^{- \Xi_1}
\end{displaymath}
holds with $\Xi_1 = \sum_{k \ge 1} \Xi_1^{(2k-1)} \in \tilde{
\mathfrak{g}}^\mathrm {odd}$ and $\Xi_0 = \sum_{k \ge 1} \Xi_0^{(2k)}
\in \tilde{\mathfrak{g}}^\mathrm{even}$.
\end{prop}
\begin{proof}
Translate by $g^{-1}$ on the left and take the logarithm:
\begin{displaymath}
    \Theta_1 = \ln \big( \mathrm{e}^{ \mathrm{Ad}(g^{-1})
    \Xi_1} \, \mathrm{e}^{\Xi_0} \mathrm{e}^{- \Xi_1} \big) \;.
\end{displaymath}
This equation is equivalent to the previous one, as the exponential
map from the nilpotent Lie algebra $\mathfrak{n} = \oplus_{k \ge 1}
\tilde{\mathfrak{g}}^{(k)}$ to the nilpotent Lie group
$\exp(\mathfrak{n})$ is a bijection. Using the
Baker-Campbell-Hausdorff formula to compute the logarithm, one gets
an expansion
\begin{displaymath}
    \Theta_1 = \mathrm{Ad}(g^{-1}) \Xi_1 - \Xi_1 + \Xi_0
    - {\textstyle{\frac{1}{2}}} [ \Xi_0 , \Xi_1 ] +
    {\textstyle{\frac{1}{2}}} [ \mathrm{Ad}(g^{-1}) \Xi_1 ,
    \Xi_0 - \Xi_1 ] + ... \;,
\end{displaymath}
which terminates at finite order by nilpotency. The equation in this
form can be solved for the unknowns $\Xi_1$ and $\Xi_0$ by recursion
in the $\mathbb{Z}$-degree. Indeed, expanding $\Theta_1 = \sum_{k \ge
1} \Theta_1^{(2k-1)}$ the degree-1 equation is
\begin{displaymath}
    \Theta_1^{(1)} = \left( \mathrm{Ad}(g^{-1}) - \mathrm{Id}
    \right) \Xi_1^{(1)} \;,
\end{displaymath}
which has a unique solution for $\Xi_1^{(1)}$ because $\mathrm{Ad}
(g^{-1})-\mathrm{Id} = \mathrm{Ad}(g^{-1}) (\mathrm{Id} - \mathrm{Ad}
(g))$ is invertible by assumption. The degree-2 equation determines
$\Xi_0^{(2)}$ from $\Xi_1^{(1)}$:
\begin{displaymath}
    0 = \Xi_0^{(2)} - {\textstyle{\frac{1}{2}}}
    [ \mathrm{Ad}(g^{-1}) \Xi_1^{(1)} , \Xi_1^{(1)} ] \;.
\end{displaymath}
The degree-3 equation determines $\Xi_1^{(3)}$ from $\Theta_1^{ (3)}$
and known quantities of lower order, the degree-4 equation determines
$\Xi_0^{(4)}$, and so on.
\end{proof}

\subsection{Radial parts of the Laplacians for $\mathrm{U}_{m|n}\,$}
\label{sect:radialpart}

We finish our exposition of background material by writing down
the radial parts of the Laplace-Casimir operators for the case of
$\mathfrak{G} = \mathrm{GL}_{m|n}$ with $K = \mathrm{U}_m \times
\mathrm{U}_n \,$. This case is referred to as the \emph{unitary
Lie supergroup} $\mathrm{U}_{m|n}\,$.  The Casimir invariants
$I_\ell$ of $\mathsf{U}( \mathfrak {gl}_{m|n})$ were given in
Lem.\ \ref{Ber0}.

Fixing some choice of standard basis $E_{i \, i^\prime}^{\sigma
\sigma^\prime}$ of $\mathrm{End} (\mathbb{C}^{m|n})$, let $T \simeq
\mathrm{U}_1^{m+n}$ be the corresponding maximal torus of diagonal
transformations $t\,$,
\begin{displaymath}
    t =  \sum\nolimits_{j=1}^m \mathrm{e}^{\mathrm{i}\psi_j}\,
    E_{jj}^{11} + \sum\nolimits_{l=1}^n \mathrm{e}^{\mathrm{i}
    \phi_l} \, E_{ll}^{00}
\end{displaymath}
where the angular variables $\psi_j$ and $\phi_l$ are real-valued
local coordinates for $T\,$, and denote by $D_k$ the degree-$k$
radial differential operator
\begin{displaymath}
    D_k = \sum_{j=1}^m \frac{\partial^k}{\partial\psi_j^k} -(-1)^k
    \sum_{l = 1}^n \frac{\partial^k}{\partial \phi_l^k} \;.
\end{displaymath}
\begin{thm}[Berezin]\label{Ber1}
If $J$ is the function defined (on a dense open set in $T$) by
\begin{displaymath}
    J = \frac{ \prod_{1 \le j < j^\prime \le m} \sin^2 \big(
    \frac{1}{2} (\psi_j - \psi_{j^\prime}) \big) \prod_{1 \le
    l < l^\prime \le n} \sin^2 \big( \frac{1}{2} (\phi_l -
    \phi_{l^\prime}) \big)} {\prod_{j = 1}^m \prod_{l = 1}^n
    \sin^2 \big( \frac{1}{2} (\psi_j - \phi_l) \big)} \;,
\end{displaymath}
the degree-$\ell$ Laplace-Casimir operator $D(I_\ell)$ for
$\mathrm{U}_{m|n}$ has the radial part
\begin{displaymath}
    \dot{D}(I_\ell) = J^{-1/2} P_\ell \circ J^{1/2} \;,
\end{displaymath}
where $P_\ell = D_\ell + Q_{\ell-1}$ and $Q_{\ell-1}$ is some
polynomial in the homogeneous operators $D_k$ of total degree less
than $\ell\,$.
\end{thm}
\begin{rem}
The expression for the highest-order term, $\dot{D}(I_\ell) =
J^{-1/2} D_\ell \circ J^{1/2} + \ldots \,$, is from Thm.\ 3.2,
Eq.\ (3.47), of \cite{berezin}, specialized to the case of
$\mathrm{U}_{m|n}\,$. The lower-order terms $Q_{\ell-1}$ are known
to be polynomial in the $D_k$ by Thm.\ 4.4 of \cite{berezin}.

The square root $J^{1/2}$ does not exist as a single-valued
function on $\mathrm{U}_1^{m+n}$ for every dimension $m|n\,$, but
the square root ambiguity cancels in the differential operator
$\dot{D}(I_\ell)$.
\end{rem}
In the special case where dimensions match, one has the following
simplification.
\begin{lem}[Berezin]\label{simpleJ}
For $m = n$ the function $J$ can be put in the form $J =
\mathrm{Det}^2 (A)$ where $A$ is the $n \times n$ matrix with
entries
\begin{displaymath}
    A_{jl} = \frac{1}{\sin \big( \frac{1}{2}
    (\psi_j - \phi_l) \big)} \;.
\end{displaymath}
\end{lem}
\begin{proof}
In the Cauchy determinant formula
\begin{displaymath}
    \frac{\prod_{j < l} (x_j - x_l)(y_l - y_j)}
    {\prod_{j,l} (x_j - y_l)} = \mathrm{Det} \left(
    \frac{1}{x_j - y_l} \right)_{j,l = 1, \ldots, n} \;,
\end{displaymath}
make the substitution $x_j = \mathrm{e}^{\mathrm{i}\psi_j}$ and
$y_l = \mathrm{e}^{\mathrm{i}\phi_l}$.  The statement then
immediately follows from Euler's formula $2\mathrm{i} \sin z =
\mathrm{e}^{ \mathrm{i}z} - \mathrm{e}^{-\mathrm{i} z}$ on
dividing both sides by a suitable factor.
\end{proof}
\begin{cor}\label{no consts}
For $m = n$ the constant term of the lower-order differential
operator $Q_{\ell - 1}(\partial/\partial\psi_1 \, , \ldots ,
\partial /\partial\psi_n \, , \partial/\partial\phi_1 \, ,
\ldots,\partial/\partial \phi_n)$ in the radial part $\dot{D}
(I_\ell)$ vanishes for all $\ell \in \mathbb{N}:$ $Q_{\ell -
1}(\mathbf{0}, \mathbf{0}) = 0 $.
\end{cor}
\begin{proof}
The first step is to show that for $m = n$ the square root
$J^{1/2}$ is annihilated by each of the homogeneous radial
differential operators $D_k\,$. For that purpose, write
\begin{displaymath}
    J^{1/2} := \mathrm{Det}(A) = \mathrm{Det} \left(
    \frac{2\mathrm{i} \, \mathrm{e}^{-\frac{\mathrm{i}}{2}
    (\psi_j - \phi_l)}}{1 - \mathrm{e}^{-\mathrm{i}(\psi_j
    - \phi_l)} } \right)_{j,l = 1, \ldots, n} \;.
\end{displaymath}
Now express the determinant as a sum over permutations, move the
$\phi_l$ into the upper half of the complex plane, and expand the
factors $(1 - \mathrm{e}^{-\mathrm{i}(\psi_j - \phi_l)})^{-1}$ as
geometric series. The result is an absolutely convergent series, each
term of which is readily seen to be annihilated by every one of the
$D_k\,$. Thus $D_k J^{1/2} = 0$ for all $k \in \mathbb{N}\,$.

Now, since $Q_{\ell-1}$ is a polynomial in the operators $D_k \,$,
it follows that when $Q_{\ell-1}$ is applied to $J^{1/2}$, only
the zero-order term in $Q_{\ell-1}$ can give a nonzero answer:
$Q_{\ell-1} J^{1/2} = J^{1/2} Q_{\ell-1}(\mathbf{0}, \mathbf{0})$.
But the differential operator $\dot{D}(I_\ell)$ annihilates the
constants; thus one has $\dot{D}(I_\ell) \cdot 1 = 0$ and Thm.\
\ref{Ber1} implies $0 = J^{-1/2} ( D_\ell + Q_{\ell - 1} ) J^{1/2}
= Q_{\ell-1}(\mathbf{0},\mathbf{0})$.
\end{proof}
Finding the radial part of an invariant differential operator is
an algebraic problem (as opposed to an analytical one).  It should
therefore be clear that the formulas given in Thm.\ \ref{Ber1}
hold not only for the case of $K = \mathrm{U}_m \times
\mathrm{U}_n$ with maximal torus $\mathrm{U}_1^{m+n}$, but admit
analytic continuation to other real-analytic domains. It is in
such a domain, $M$ (with another maximal torus $T$), that we will
use them to prove Thm.\ \ref{thm0}.

\section{Determination of the character $\chi$}
\label{sect:4A}\setcounter{equation}{0}

In Sect.\ \ref{sect:howe} we considered a $\mathbb{Z}_2$-graded
complex vector space $V = U \otimes \mathbb{C}^N$ where
\begin{eqnarray*}
    V &=& V_1 \oplus V_0 = (V_1^+ \oplus V_1^-) \oplus
    (V_0^+ \oplus V_0^-) \;,\\ U &=& U_1 \oplus U_0 =
    (U_1^+ \oplus U_1^-) \oplus (U_0^+ \oplus U_0^-) \;,\\
    V_\tau^\pm &=& U_\tau^\pm \otimes \mathbb{C}^N \;, \quad U_1^+
    = U_0^+ = \mathbb{C}^p\;,\quad U_1^- = U_0^- = \mathbb{C}^q \;.
\end{eqnarray*}
In this tensor-product situation there is a Howe dual pair of a Lie
group $\mathrm{U}_N$ and a Lie superalgebra $\mathfrak {gl}(U) =
\mathfrak{gl}_{n|n}$ (with $n = p+q$) acting on the spinor-oscillator
module
\begin{displaymath}
    \mathcal{A}_V \simeq \wedge (V_1^+ \oplus {V_1^-}^\ast)
    \otimes \mathrm{S} (V_0^+ \oplus {V_0^-}^\ast) \;.
\end{displaymath}
Prop.\ \ref{howe-char} states that the subspace $\mathcal{A}_V^{
\mathrm{U}_N}$ of $\mathrm{U}_N$-invariants in $\mathcal{A}_V$ is an
irreducible representation space for $\mathfrak{gl}(U)$ and that,
moreover, the autocorrelation function of ratios (\ref{def chi})
coincides with the character associated with that representation.

In the current section we use the simplified notation
\begin{equation}
    \mathcal{V}_\lambda \equiv \mathcal{A}_V^{\mathrm{U}_N} \;.
\end{equation}
This notation expresses the two properties of $\mathcal
{V}_\lambda$ being (i) a graded-commuta\-tive algebra built from
$V$ and (ii) an irreducible $\mathfrak{gl}(U)$-module with highest
weight $\lambda \equiv \lambda_N\,$.

Recalling the definition of the superdeterminant, and specializing it
to the case of $V = U \otimes \mathbb{C}^N$, we can cast the equality
of expressions of Prop.\ \ref{howe-char} in the following form, where
$m \otimes u$ is viewed as an even element of $\mathrm {End}(V)\simeq
\mathrm{End}(U) \otimes \mathrm{End}(\mathbb{C}^N)$.
\begin{cor}\label{cor 2.12}
\begin{displaymath}
    \chi(m) = \mathrm{STr}_{\mathcal{V}_\lambda} \, \rho(m) =
    \int_{\mathrm{U}_N} \mathrm{SDet}(\mathrm{Id}_V - m \otimes
    u)^{-1} \, du \;.
\end{displaymath}
\end{cor}
All of our considerations of $\chi$ in Sect.\ \ref{sect:2} were
restricted to the diagonal transformations $m = t\,$; it is, in fact,
only the values on the diagonal that we ultimately want to know.
Nevertheless, in order to establish good analytical control and
actually compute these values, our next goal is to extend the
function $t \mapsto \chi(t)$ to a section of the Lie supergroup
$\mathfrak{G} = \mathrm {GL}_{n|n}\,$.  We wish to do this by
following Sect.\ \ref{sect:3.4}, where we explained how to construct
a $\mathfrak{G}$-representation from a representation $\rho_\ast$ of
the Lie superalgebra and a compatible representation $\rho$ of the
Lie group $G$ underlying $\mathfrak{G}$.

In this quest, however, we are facing a difficulty: since the
representation space $\mathcal{V}_\lambda$ is infinite-dimensional
and some of the Lie algebra elements are represented by unbounded
operators, the representation $\rho_\ast$ does not exponentiate to
all of the complex Lie group of linear transformations of the
$\mathbb{Z}_2$-graded complex vector space $U\,$. We therefore
require a more elaborate variant of the construction given in Sect.\
\ref{sect:3.4}.

\subsection{The good real structure}
\label{sect:csm}

Our first step toward extending the character function $t \mapsto
\chi(t)$ is to identify a suitable real structure inside the complex
Lie supergroup $\mathrm {GL}_{n|n}$ of the $\mathbb{Z} _2$-graded
vector space $U \simeq \mathbb{C}^{n|n}$. Let $\mathrm{GL}(U_1)
\times \mathrm{GL}(U_0) =: G$ be the group underlying the Lie
supergroup $\mathrm{GL}$ of $U$. As should be clear from Prop.\
\ref{prop:6707} (making a statement about semigroup elements), this
complex Lie group is not the appropriate space on which to consider
our character $\chi\,$. Rather, the fact that $\mathrm{dim}\,
\mathcal{V}_\lambda = \infty \,$ forces us to get a certain real
subgroup $G_\mathbb{R} \subset G$ into play.

To introduce $G_\mathbb{R}\,$, let the decomposition $U_0^{ \vphantom
{+}} = U_0^+ \oplus U_0^-$ be encoded in an involution $s : \, U_0
\to U_0\,$ which has $U_0^+ = \mathbb{C}^p$ and $U_0^-= \mathbb{C}^q$
for its two eigenspaces:
\begin{equation}\label{def involn}
    s\, (u_+ + u_-) = u_+ - u_- \quad (u_\pm \in U_0^\pm) \;,
\end{equation}
and let $U_0^s \equiv (U_0 \, , s)$ mean $U_0$ equipped with the
pseudo-unitary structure
\begin{displaymath}
    U_0 \times U_0 \to \mathbb{C} \;, \quad
    (u,v) \mapsto \langle u , s v \rangle \;,
\end{displaymath}
where $\langle \, , \, \rangle$ is the Hermitian scalar product of
the Hermitian vector space $U_0 = \mathbb{C}^{p+q}$. The group of
isometries of $U_0^s$ is the non-compact Lie group
\begin{equation}\label{def U(p,q)}
    \mathrm{U}_{p,\,q} := \{ g \in \mathrm{GL}(U_0) \mid
    g^\dagger s g = s \} \;.
\end{equation}

Also, let $\mathrm{U}_n$ denote the isometry group of the Hermitian
vector space $U_1 = \mathbb{C}^n$. For our purposes, the ``good''
real subgroup of $G$ will turn out to be
\begin{equation}
    G_\mathbb{R} = \mathrm{U}_n \times \mathrm{U}_{p,\,q} \;.
\end{equation}

\subsubsection{$G_\mathbb{R}$-principal bundle}
\label{sect:P-to-M}

We now describe the $G_\mathbb{R}$-principal bundle which is to
assume the role formerly played by $G \times G \to G$ in Sect.\
\ref{sect:superG}. Beginning with the bosonic sector, let
$\mathrm{H}_{p,\,q}$ be the set
\begin{equation}\label{def semigroup}
    \mathrm{H}_{p,\,q} := \{ g \in \mathrm{GL}(U_0) \mid
    g^\dagger s g < s \} \;.
\end{equation}
If two group elements $g, h \in \mathrm{GL}(U_0)$ satisfy
$g^\dagger s g < s$ and $h^\dagger s h < s\,$, then
\begin{displaymath}
    (gh)^\dagger s (gh) = h^\dagger (g^\dagger s g) h <
    h^\dagger s h < s \;.
\end{displaymath}
Thus $\mathrm{H}_{p,\,q}$ is a semigroup w.r.t.\ multiplication by
composition, $(g,h) \mapsto gh$. Note also that $\mathrm{H}_{p,\,q}$
contains the pseudo-unitary group $\mathrm{U}_{p,\,q}$ in its closure
$\overline{\mathrm{H}}_{p,\,q}\,$.

Now define an involutory automorphism $\sigma :\, \mathrm{GL}(U_0)
\to \mathrm{GL}(U_0)$ by
\begin{equation}\label{def sigma}
    \sigma(g) = s {g^{-1}}^\dagger s \;.
\end{equation}
The set of $\sigma$-fixed points in $\mathrm{GL}(U_0)$ is the
pseudo-unitary group $\mathrm{U}_{p,\,q}\,$.

Since the inequality $x^\dagger s x < s$ is invariant under left and
right translations of $x$ by the elements of $\mathrm{U}_{p,\,q}\,$,
the semigroup $\mathrm {H}_{p,\,q}$ is a left and right
$\mathrm{U}_{p,\,q}$-space. In Cor.\ \ref{cor 5.4} we will prove that
$\mathrm{H}_{p,\,q}$ has the following closure property: if $x \in
\mathrm{H}_{p,\,q}$ then $\sigma(x)^{-1} \in \mathrm{H}_{p,\,q}\,$.

Next, define a $\mathrm{U}_{p,\,q}$-principal bundle $P_0 \to M_0$ as
follows.  Let $P_0$ be the graph of the mapping $\sigma :\,
\mathrm{H}_{p,\,q} \to \mathrm{GL}(U_0)$:
\begin{equation}
    P_0 = \{ (x,y) \in \mathrm{H}_{p,\,q} \times \mathrm{GL}(U_0)
    \mid y = \sigma(x) \} \;,
\end{equation}
and take $M_0 \subset \mathrm{GL}(U_0)$ to be the set
\begin{equation}
    M_0 = \{ x \, \sigma(x)^{-1} \mid x \in
    \mathrm{H}_{p,\,q} \} \;.
\end{equation}
Note $M_0 \subset \mathrm{H}_{p,\,q}$ from $\sigma(x)^{-1} \in
\mathrm{H}_{p,\,q}$ and the fact that $\mathrm{H}_{p,\,q}$ is a
semigroup. Since $\mathrm{U}_{p,\,q} \subset \mathrm{GL}(U_0)$ is the
subgroup of fixed points of the automorphism $\sigma$, the map
\begin{displaymath}
    P_0 \to M_0 \simeq P_0 / \mathrm{U}_{p,\,q} \;,
    \quad (x,\sigma(x)) \mapsto x \, \sigma(x)^{-1} \;,
\end{displaymath}
defines a $\mathrm{U}_{p,\,q}$-principal bundle by the right $\mathrm
{U}_{p,\,q}$-action $(x,\sigma(x)) \mapsto (xg,\sigma(xg))$.

The corresponding construction for the fermionic sector built on $U_1
= \mathbb{C}^n$ is simpler: let a $\mathrm{U}_n$-principal bundle
$P_1 \to M_1$ be defined by the projection map
\begin{equation}
    P_1 := \mathrm{U}_n \times \mathrm{U}_n \to \mathrm{U}_n
    =: M_1 \;, \quad (x,y) \mapsto x \, y^{-1} \;.
\end{equation}
The base manifold $M_1$ is a real-analytic submanifold of
$\mathrm{GL}(U_1) = \mathrm{GL}(\mathbb{C}^n)$ of dimension $n^2 =
(p+q)^2$. The same is true of $M_0 \subset \mathrm{GL}(U_0) =
\mathrm{GL}(\mathbb{C}^{p+q})$.

We now introduce the direct products $P := P_1 \times P_0\,$ and $M
:= M_1 \times M_0\,$, and view these as defining a principal bundle
with structure group $G_\mathbb{R} = \mathrm{U}_n \times
\mathrm{U}_{p,\,q}:$
\begin{equation}
    P \to M \simeq P / G_\mathbb{R} \;.
\end{equation}
Since $G_\mathbb{R}$ acts on the odd part $\mathfrak{g}_1 \equiv
\mathfrak{gl}(U)_1$ of the Lie superalgebra $\mathfrak{gl}(U)$ by
$\mathrm{Ad}\,$, one could associate with $\mathfrak{g}_1$ a complex
vector bundle $P \times_{G_\mathbb{R}} \mathfrak{g}_1 \to M$, which
would give rise to a cs-manifold \cite{QFT math} (meaning a
supermanifold which is complex in the odd direction). A more concise
framework is achieved, however, by passing to a \emph{real} subspace
of $\mathfrak{g}_1\,$.

\subsubsection{Real form of $\mathfrak{gl}(U)$}

Write $X \in \mathfrak{gl}(U) = \mathfrak{gl}(U_1 \oplus U_0)$ in
block decomposition with respect to the Hermitian vector spaces $U_1$
and $U_0$ as
\begin{displaymath}
    X = \begin{pmatrix} \mathsf{A} &\mathsf{B} \\ \mathsf{C}
    &\mathsf{D} \end{pmatrix} \;,
\end{displaymath}
where $\mathsf{A} \in \mathrm{End}(U_1)$, $\mathsf{B} \in
\mathrm{Hom}(U_0\, , U_1)$, $\mathsf{C} \in \mathrm{Hom}(U_1 , U_0)$,
and $\mathsf{D} \in \mathrm{End}(U_0)$. Then define a
$\mathbb{Z}_2$-graded real vector space $\mathfrak{u}(U^s)$ as the
subset of elements $X \in \mathfrak{gl}(U)$ whose blocks satisfy the
set of conditions
\begin{eqnarray}\label{def u_n|p,q}
    &&\langle \mathsf{A} v_1^{\vphantom{\prime}} , v_1^\prime
    \rangle + \langle v_1^{\vphantom{\prime}} , \mathsf{A}
    v_1^\prime \rangle = 0 \;, \quad \langle \mathsf{D}
    v_0^{\vphantom{\prime}} \, , s v_0^\prime \rangle +
    \langle v_0^{\vphantom{\prime}}\, , s \mathsf{D} v_0^\prime
    \rangle = 0 \;, \nonumber \\ &&\langle \mathsf{C} v_1 , s v_0
    \rangle + \mathrm{i} \langle v_1 , \mathsf{B} v_0 \rangle = 0
    \quad (v_1^{\vphantom{\prime}} , v_1^\prime \in U_1 ;\,
    v_0^{\vphantom{\prime}}\, , v_0^\prime \in U_0) \;.
\end{eqnarray}
This subspace $\mathfrak{u}(U^s) \subset \mathfrak{gl}(U)$ is readily
seen to close under the supercommutator, and thus is a Lie
sub-superalgebra of $\mathfrak{gl}(U)$. The alternative notation
\begin{equation}
    \mathfrak{u}(U^s) \equiv \mathfrak{u}_{n|p,\,q}
\end{equation}
makes it evident that $\mathfrak{u}(U^s)$ is a pseudo-unitary real
form of $\mathfrak{gl}(U) = \mathfrak{gl}_{n|n}$ with mixed signature
$(p,q)$ in the non-compact sector $\mathrm{End}(U_0)$.

The even part of the Lie superalgebra $\mathfrak{u}(U^s) =
\mathfrak{u}_{n|p,\,q}$ is the real Lie algebra $\mathfrak{u}_n
\oplus \mathfrak{u}_{p,\,q}$ where $\mathfrak{u}_n = \mathrm{Lie}(
\mathrm{U}_n)$ and $\mathfrak{u}_{p,\,q} = \mathrm{Lie}(
\mathrm{U}_{p,\,q})$. We now focus on the odd component,
$\mathfrak{u}(U^s)_1\,$.  From the third condition in (\ref{def
u_n|p,q}) an element $X$ of the odd space $\mathfrak{u}(U^s)_1$ is
already determined by its block $\mathsf{B} \in \mathrm{Hom}(U_0\, ,
U_1)$. We therefore have a canonical isomorphism
\begin{equation}
    \mathfrak{u}(U^s)_1 \simeq \mathrm{Hom}(U_0\, , U_1) \simeq
    U_1^{\vphantom{\ast}} \otimes U_0^\ast \simeq
    U_1 \otimes U_0 \;,
\end{equation}
where the last identification is by the pseudo-unitary structure of
$U_0\,$.  This isomorphism $\mathfrak{u}(U^s)_1 \simeq U_1 \otimes
U_0$ makes it particularly clear that $\mathfrak{u}(U^s)_1$ is a
module for our real Lie group $G_\mathbb{R} = \mathrm{U}(U_1) \times
\mathrm{U}(U_0^s) = \mathrm{U}_n \times \mathrm{U}_{p,\,q}\,$.  In
the following we use the simplified notation $\mathfrak{g} :=
\mathfrak {gl}(U)$ and $\mathfrak{g}_\mathbb{R} := \mathfrak{u}(U^s)$
and
\begin{equation}
    \mathfrak{g}_{\mathbb{R},\tau} := \mathfrak{u}(U^s)_\tau
    \quad (\tau = 0, 1) \;.
\end{equation}
Note that $\mathfrak{g}_{\mathbb{R},0} = \mathrm{Lie}(G_\mathbb{R}) =
\mathfrak{u}_n \oplus \mathfrak{u}_{p,\,q}\,$.

\subsubsection{Lie supergroup structure}

Given the $G_\mathbb{R}$-principal bundle $P \to M$ and the
$G_\mathbb{R}$-module $\mathfrak{g}_{\mathbb{R},1}\,$, we form the
associated vector bundle
\begin{equation}
    F = P \times_{G_\mathbb{R}}^{\vphantom{\dagger}}
    \mathfrak{g}_{\mathbb{R},1} \to M \;.
\end{equation}
Then, following the principles laid down in Sect.\ \ref{sect:3} we
are led to consider $\Gamma(M,\wedge F^\ast)$, the graded-commutative
algebra of sections of the bundle $\wedge(F^\ast) \to M$.

The key idea now is to extend our function $t \mapsto \chi(t)$ to a
radial section of $\Gamma(M, \wedge F^\ast)$ and exploit the argument
of Sect.\ \ref{sect:eigenfctn}, by which the primitive character
$\chi$ is an eigen\-function of the ring of $\mathfrak{g}$-invariant
differential operators.  This argument, however, relies on Schur's
lemma (which holds over the complex numbers) while the vector bundle
$F \to M$ is a real vector bundle over the real manifold $M$. Thus
the good object to consider isn't $\Gamma(M,\wedge F^\ast)$ but the
bundle of \emph{complex} exterior algebras,
\begin{equation}
    \mathcal{F} := \Gamma(M,\mathbb{C} \otimes \wedge F^\ast) \;.
\end{equation}
There, we can implement the basic setting of Sect.\
\ref{sect:eigenfctn}, namely independent left and right actions by
the complex Lie superalgebra $\mathfrak{g} = \mathfrak{gl}(U) =
\mathfrak{gl}_{n|n}\,$, as follows.

Recall $M = M_1 \times M_0$ with $M_1 = \mathrm{U}_n\,$, and let
$\Phi$ be a section of $\mathcal{F}$. If $X \in \mathfrak{u}_n$ and
$\mathrm{e}^{tX} = g \in \mathrm{U}_n$ then from
\begin{displaymath}
    (L_g \Phi)(m_1 , m_0) = \Phi(g^{-1} m_1 , m_0) \;,
    \quad (R_g \Phi)(m_1 , m_0) = \Phi(m_1 g , m_0) \;,
\end{displaymath}
one gets $\mathfrak{u}_n$-actions on the left and right by taking the
differential as usual:
\begin{displaymath}
    \widehat{X}^L := \frac{d}{dt} L_{\exp(tX)} \Big\vert_{t = 0}
    \;, \quad \widehat{X}^R := \frac{d}{dt} R_{\exp(tX)}
    \Big\vert_{t = 0} \;.
\end{displaymath}
These extend complex linearly to left and right actions of
$\mathfrak{gl}(U_1) = \mathfrak{gl}_n = \mathfrak{u}_n + \mathrm{i}
\mathfrak{u}_n :$
\begin{equation}
    \widehat{X + \mathrm{i}Y}^j := \widehat{X}^j + \mathrm{i}
    \widehat{Y}^j \quad (j = L, R) \;.
\end{equation}

Next let $h \in \mathrm{H}_{p,\,q}\,$.  The twisted $\mathrm{H}_{p ,
q}$-action on $M_0$ by $m_0 \mapsto h m_0 \sigma(h)^{-1}$ induces an
action $h \mapsto T_h$ on sections $\Phi \in \mathcal{F}$ by
\begin{displaymath}
    (T_h \Phi)(m_1 , m_0) = \Phi(m_1 , h^{-1} m_0 \sigma(h)) \;.
\end{displaymath}
To pass to the infinitesimal action, notice that the semigroup
$\mathrm{H}_{p,\,q}$ is open. Therefore, having fixed any point $h
\in \mathrm{H}_{p,\,q}$ one can find $\varepsilon > 0$ so that $g h$
is in $\mathrm{H}_{p,\,q}$ for every $g$ in the $\varepsilon$-ball
$B_\varepsilon \subset \mathrm{GL}(U_0)$ centered at the neutral
element of $\mathrm{GL}(U_0)$.  This property gets transferred from
$\mathrm{H}_{p,\,q}$ to $M_0 \simeq \mathrm{H}_{p,\,q} /
\mathrm{U}_{p,\,q}$ by the projection $h \mapsto h \,
\sigma(h)^{-1}$. Thus, for $X \in \mathfrak{gl}(U_0)$ the definition
\begin{displaymath}
    (\widehat{X} \Phi)(m_1 , m_0) := \frac{d}{dt} \Phi(m_1 ,
    \mathrm{e}^{-tX} m_0 \, \sigma(\mathrm{e}^{tX})) \Big\vert_{t = 0}
\end{displaymath}
makes sense, and $X \mapsto \widehat{X}$ is a $\mathfrak{gl}
(U_0)$-action on $\mathcal{F}\,$.  Since the assignment $X \mapsto
\widehat{X}$ is complex linear on the left and (via $\sigma$) complex
anti-linear on the right, one gets from it a left action $X \mapsto
\widehat{X}^L$ and a right action $X \mapsto \widehat{X}^R$ by
setting
\begin{equation}
    \widehat{X}^L := {\textstyle{\frac{1}{2}}} (\widehat{X}
    - \mathrm{i} \, \widehat{(\mathrm{i}X)}) \;, \quad
    \widehat{X}^R := {\textstyle{\frac{1}{2}}} (\widehat{X}
    + \mathrm{i} \, \widehat{(\mathrm{i}X)}) \;.
\end{equation}
These two actions commute, i.e., $[\widehat{X}^L , \widehat{Y}^R] =
0$ for all $X , Y \in \mathfrak{gl}(U_0)$.  Altogether we now have
left and right actions of the even part $\mathfrak{g}_0 \simeq
\mathfrak{gl}(U_1) \oplus \mathfrak{gl}(U_0)$ of our complex Lie
superalgebra $\mathfrak{g} = \mathfrak{gl}(U)$.

To describe the action of the odd part, $\mathfrak{g}_1 =
\mathfrak{gl}(U)_1\,$, we first of all adapt our notation. Although
the $G_\mathbb{R}$-principal bundle $P$ was introduced as
\begin{displaymath}
    P = P_1 \times P_0 = (\mathrm{U}_n \times \mathrm{U}_n) \times
    (\mathrm{H}_{p,\,q} \stackrel{1,\sigma}{\hookrightarrow}
    \mathrm{GL}(U_0)\times \mathrm{GL}(U_0) ) \;,
\end{displaymath}
we now rearrange factors to interpret $P$ as a subset of the direct
product of two copies of $G = (\mathrm{GL}(U_1)\times \mathrm{GL}
(U_0)\stackrel{\mathrm{diag}}{\hookrightarrow} \mathrm{GL}(U) )$:
\begin{displaymath}
    P \subset G \times G \;,
\end{displaymath}
and write $p = (x\, ,y)$ for the points of $P$ accordingly. Such an
element $(x\, ,y) \in P$ consists of $x = \mathrm{diag}(u_L\, , h)$
and $y = \mathrm{diag}(u_R\, ,\sigma(h))$ where $h \in \mathrm{H}_{p,
q}$ and $u_L\, , u_R \in \mathrm{U}_n\,$.

Recalling from Sect.\ \ref{sect:3.4} the meaning of the tautological
object
\begin{displaymath}
    \xi = \sum \varphi^i \otimes \mathsf{F}_i \in
    \mathfrak{g}_{\mathbb{R}, 1}^\ast \otimes
    \mathfrak{g}_{\mathbb{R},1}^{\vphantom{\ast}} \;,
\end{displaymath}
where $\{ \mathsf{F}_i \}$ is some basis of $\mathfrak{g}_{
\mathbb{R},1} = \mathfrak{u}(U^s)_1$, and the dual basis $\{
\varphi^i \}$ is viewed as a set of generators of $\wedge
(\mathfrak{g}_{\mathbb{R},1}^\ast)$, we then consider the supermatrix
\begin{equation}\label{super matrix}
    P \ni (x\, ,y) \mapsto \Xi = x\, \mathrm{e}^{\xi} y^{-1} \;.
\end{equation}
{}From the relation $x \, \mathrm{e}^\xi y^{-1} = (xg)\,
\mathrm{e}^{\sum \varphi^i \mathrm{Ad}^\ast (g^{-1})_{i
\phantom{j}}^{\phantom{i}j} \otimes \mathsf{F}_j} (yg)^{-1} $ for $g
\in G_\mathbb{R}\,$, the matrix entries of $\Xi$ are $G_\mathbb{R}
$-equivariant functions $P \to \wedge (\mathfrak{g}_{\mathbb{R}
,1}^\ast)$ and, thus, sections of $\mathcal{F}\,$.

The rest of the development exactly follows Sect.\
\ref{sect:g1-action} and we indicate it only very briefly. Enlarging
the complex Grassmann algebra $\wedge(\mathfrak{g}_1^\ast) =
\mathbb{C} \otimes \wedge( \mathfrak{g}_{ \mathbb{R},1}^\ast)$ by an
anti-commuting parameter $\eta$ one shows that (see Lem.\ \ref{odd
rearrange} and its proof) if $Y$ is an odd generator $Y \in
\mathfrak{g}_1 = \mathfrak{gl}(U)_1$ there exist even-type functions
$y \mapsto \alpha_Y^i(y) \in \wedge^\mathrm{even}
(\mathfrak{g}_1^\ast)$ and odd-type functions $y \mapsto \beta_Y^j(y)
\in \wedge^\mathrm{odd}(\mathfrak{g}_1^\ast)$ so that
\begin{displaymath}
    (x \, \mathrm{e}^\xi y^{-1}) \mathrm{e}^{\eta \otimes Y} = x\,
    \mathrm{e}^{ \sum (\varphi^i + \alpha_Y^i(y) \eta) \otimes
    \mathsf{F}_i} y^{-1} \mathrm{e}^{\sum \beta_Y^j(y) \eta \otimes
    \mathsf{E}_j} \;,
\end{displaymath}
where $\{\mathsf{E}_j\}$ is a basis of $\mathfrak{g}_{\mathbb{R},0}\,
$. By the general principle explained in Prop.\ \ref{prop 3.2p} and
its proof, these functions determine a right action $Y \mapsto
\widehat{Y}^R$ of $\mathfrak{g}_1 = \mathfrak{gl}(U)_1\,$. The left
action $Y \mapsto \widehat{Y}^L$ is constructed in the same fashion.

Let us now summarize the material of this subsection.
\begin{prop}
The complex Lie superalgebra $\mathfrak{g} = \mathfrak{gl}(U)$ acts
on the graded-com\-mutative algebra $\mathcal{F} = \Gamma(M ,
\mathbb{C} \otimes \wedge F^\ast)$ on the left and right by $X
\mapsto \widehat{X}^L$ resp.\ $X \mapsto \widehat{X}^R$.
\end{prop}

\subsection{Extending the character $\chi\,$}
\label{sect:extend-char}

Note that if $\Xi$ is decomposed as $\Xi = \begin{pmatrix} \mathsf{A}
& \mathsf{B} \\ \mathsf{C} &\mathsf{D} \end{pmatrix}$ according to $U
= U_1 \oplus U_0 \,$, the matrix entries of the blocks $\mathsf{A}$
and $\mathsf{D}$ are even sections of $\mathcal{F} \,$, and those of
the blocks $\mathsf{B}$ and $\mathsf{C}$ are odd sections.

For the following definition recall $V = U \otimes \mathbb{C}^N$, and
let $u \in \mathrm{U}_N\,$.
\begin{defn}\label{def 4.2}
By the reciprocal of the superdeterminant of $\mathrm{Id}_V - \Xi
\otimes u$ we mean
\begin{displaymath}
    \mathrm{SDet}(\mathrm{Id}_V - \Xi \otimes u)^{-1} =
    \frac{\mathrm{Det}\big( \mathrm{Id}_{V_1} - \mathsf{A}
    \otimes u + (\mathsf{B} \otimes u)\,(\mathrm{Id}_{V_0} -
    \mathsf{D} \otimes u)^{-1} (\mathsf{C} \otimes u) \big)}
    {\mathrm{Det}(\mathrm{Id}_{V_0} - \mathsf{D} \otimes u)} \;.
\end{displaymath}
\end{defn}
\begin{rem}
This is Def.\ \ref{SDet} (with Prop.\ \ref{prop 1.3}) transcribed to
the present setting.  We will show later, by Lem.\ \ref{lem 4.13},
that $\mathrm{Id}_{V_0} - \mathsf{D} \otimes u$ has an inverse for
all $u \in \mathrm{U}_N$ and $\mathsf{D} = \Xi \vert_{U_0 \to
U_0}\,$, thereby ensuring the \emph{global} existence of $\mathrm
{SDet}(\mathrm{Id}_V - \Xi \otimes u)^{-1}$.
\end{rem}

Our goal is to relate the character $t \mapsto \chi(t)$ to a section
of $\mathcal{F}\,$.  To that end, let us reorganize the notation of
Cor.\ \ref{cor 2.5} so that $t = (t_1 , t_0) \in \mathrm{GL}(U_1)
\times \mathrm{GL}(U_0)$ with
\begin{equation}\label{def t0t1}
    t_1 = t_1^+ + t_1^- =
    \sum_{k=1}^{p+q} \mathrm{e}^{\mathrm{i} \psi_k}
    E_{kk}^{11} \;, \quad t_0 = t_0^+ + t_0^- =
    \sum_{k=1}^{p+q} \mathrm{e}^{\phi_k} E_{kk}^{00} \;,
\end{equation}
where $\{ E_{ij}^{\tau \tau^\prime} \}$ is a standard basis of
$\mathrm{End} (U_1 \oplus U_0)$. Although the parameters $\psi_k$ and
$\phi_k$ previously assumed values in $\mathbb{C}\,$, we now
\emph{restrict their range to the real numbers}, while retaining
(from Cor.\ \ref{cor 2.5}) the condition
\begin{equation}\label{range of phi}
    \phi_j < 0 < \phi_l \quad (1 \le j \le p < l \le p+q) \;.
\end{equation}
Imposing these restrictions, we may regard the diagonal
transformations $t = (t_1, t_0)$ as points of $M = M_1 \times M_0\,$.
Indeed, $t_1$ now lies in $\mathrm {U}_1^n \subset \mathrm{U}_n =
M_1\,$; and it is easy to see that $t_0 \in \mathrm{H}_{p,\,q}$ and
$t_0 = \sigma(t_0)^{-1}$, and therefore we have $t_0 \in M_0\,$.

Thus we now choose to view our character $t \mapsto \chi(t)$ as a
function of the diagonal transformations $t \in M\,$. At the same
time, we restrict the assignment of (\ref{howe factors}), i.e.,
\begin{displaymath}
    t \mapsto \rho(t) = R(t \otimes \mathrm{Id}_N) \;,
\end{displaymath}
to the domain for $t$ as above.  Then from Prop.\ \ref{howe-char} we
still have the expression
\begin{equation}\label{char-det-4.1}
    \chi(t) = \mathrm{STr}_{\mathcal{V}_\lambda}\, \rho(t) =
    \int_{\mathrm{U}_N} \frac{ \mathrm{Det}(\mathrm{Id} - t_1
    \otimes u)} {\mathrm{Det}(\mathrm{Id} - t_0 \otimes u)}\, du \;.
\end{equation}

As a preparatory step for Prop.\ \ref{prop 4.2} below, recall from
Sect.\ \ref{sect:howe} the representation
\begin{displaymath}
    \rho_\ast : \,\, \mathfrak{gl}_{n|n} \to
    \mathfrak{gl}(\mathcal{V}_\lambda) \;,
\end{displaymath}
and from Sect.\ \ref{sect:3.4} that by the character of a
representation $(\mathcal{V}_\lambda, \rho_\ast, \rho)$ of a Lie
supergroup $\mathfrak{G} = (\mathfrak{g}\, , G)$ with supermatrix
$\Xi = x\, \mathrm{e}^\xi y^{-1} = x\, \mathrm{e}^{\sum \varphi^i
\otimes \mathsf{F}_i} y^{-1}$, we mean
\begin{equation}\label{char susyM}
    \mathrm{STr}_{\mathcal{V}_\lambda}\, \rho(x)\, \mathrm{e}^{
    \sum \varphi^i \rho_\ast(\mathsf{F}_i)} \rho(y^{-1}) \equiv
    \mathrm{STr}_{\mathcal{V}_\lambda}\, \rho(\Xi) \;.
\end{equation}
In the present refined setting, where the $G$-principal bundle $G
\times G \to G$, $(g,h) \mapsto gh^{-1}$ has been replaced by the
$G_\mathbb{R}$-principal bundle
\begin{displaymath}
    G \times G \supset P \to M \subset G \;,
    \quad (x\, ,y) \mapsto xy^{-1} \;,
\end{displaymath}
and the associated complex vector bundle $(G \times G) \times_G^{
\vphantom{\dagger}} \mathfrak{g}_1 \to G$ by
\begin{displaymath}
    F = P \times_{G_\mathbb{R}}^{\vphantom{\dagger}}
    \mathfrak{g}_{\mathbb{R},1} \to M \;,
\end{displaymath}
we still mean the same thing.
\begin{prop}\label{prop 4.2}
The assignment $t \mapsto \rho(t)$ extends to a semigroup
representation
\begin{displaymath}
    \rho : \,\, \mathrm{GL}_n \times \mathrm{H}_{p,\,q}
    \,\, \to \,\, \mathrm{GL}(\mathcal{V}_\lambda) \;,
\end{displaymath}
and a corresponding Lie group representation
\begin{displaymath}
    \rho^\prime : \,\, G_\mathbb{R} = \mathrm{U}_n \times
    \mathrm{U}_{p,\,q} \,\, \to \,\, \mathrm{U}(\mathcal{V}_\lambda)
    \;,
\end{displaymath}
which are compatible with the $\mathfrak{gl}_{n|n} $-representation
$\rho_\ast \,$. The character associated with $(\mathcal{V}_\lambda,
\rho_\ast, \rho)$ by equation (\ref{char susyM}) exists globally as
an analytic section of the algebra $\mathcal{F} = \Gamma(M,
\mathbb{C} \otimes \wedge F^\ast)$, and is given by an extension of
the integral formula (\ref{char-det-4.1}):
\begin{equation}\label{needs ancestor}
    \mathrm{STr}_{\mathcal{V}_\lambda}\, \rho(\Xi) = \int_{\mathrm
    {U}_N} \mathrm{SDet}(\mathrm{Id}- \Xi \otimes u)^{-1} du \;.
\end{equation}
\end{prop}
\begin{rem}
$\rho^\prime$ corresponds to $\rho$ in the sense that $\rho(xg) =
\rho(x) \rho^\prime(g)$ and $\rho((yg)^{-1}) = \rho^\prime(g)^{-1}
\rho(y^{-1})$ for $g \in G_\mathbb{R}$ and $x\, , y^{-1} \in
\mathrm{GL}_n \times \mathrm{H}_{p,\,q}\,$. These relations and the
compatibility of representations guarantee that $(x\, ,y) \mapsto
\mathrm {STr}_{\mathcal {V}_\lambda} \, \rho(x)\, \mathrm{e}^{\sum
\varphi^i \rho_\ast( \mathsf{F}_i)} \rho(y^{-1})$ is a
$G_\mathbb{R}$-equi\-variant mapping and hence defines a section
$\chi \in \mathcal{F}$ in the usual way:
\begin{displaymath}
    \chi(xy^{-1}) = \big[(x\, ,y) \,;\,\mathrm{STr}_{
    \mathcal{V}_\lambda} \, \rho(x)\, \mathrm{e}^{\,\sum
    \varphi^i \rho_\ast(\mathsf{F}_i)} \rho(y^{-1}) \big] \;.
\end{displaymath}
\end{rem}
Although it might seem that Prop.\ \ref{prop 4.2} extends the
character formula of Cor.\ \ref{cor 2.12} in a straightforward
manner, we do not know any easy way of proving the existence of the
representations $\rho$ and $\rho^\prime$; and to avoid a lengthy
interruption in the flow of our argument, we must postpone the proof
of Prop. \ref{prop 4.2} until the end of Sect.\ \ref{sect:key}.

\subsection{All Casimir invariants vanish on $\mathcal{V}_\lambda$}
\label{sect:Cas0}

Prop.\ \ref{prop 4.2} achieves the important step of extending the
torus function $t \mapsto \chi(t)$ to a section of the
graded-commutative algebra $\mathcal{F} = \Gamma(M, \mathbb{C}
\otimes \wedge F^\ast)$. Hence, taking Prop.\ \ref{prop 4.2} for
granted, we can apply the powerful machinery reviewed in Sect.\
\ref{sect:3}: the Lie superalgebra $\mathfrak{g} = \mathfrak{gl}(U) =
\mathfrak{gl}_{n|n}$ acts on $\mathcal{F}$ by left- or
right-invariant derivations; this induces two actions of the
universal enveloping algebra $\mathsf{U}(\mathfrak{gl}_{n|n})$ on
$\mathcal{F}$; and the Casimir invariants, i.e.\ the central elements
of $\mathsf{U}(\mathfrak{gl}_{n|n})$, then act on $\mathcal{F}$ by
$\mathfrak{gl}_{n|n}$-invariant differential operators, the
Laplace-Casimir operators. What's most important is that, since the
$\mathfrak{gl}_{n|n}$-representation $( \mathcal{V}_\lambda \, ,
\rho_\ast)$ is irreducible, the character $\chi$ is an eigenfunction
of all of these operators.

The next step is to calculate the Laplace-Casimir eigenvalues. To
that end, recall from Lem.\ \ref{Ber0} the expression for the
degree-$ \ell$ Casimir invariant $I_\ell$ of $\mathfrak{gl}_{n|
n}\,$. There exists a simple heuristic why all of these invariants
must be identically zero in such a representation as $(\mathcal{V}
_\lambda, \rho)$.  (In this subsection we use the simplified notation
$\rho_\ast \equiv \rho$.)

First of all, one argues that each invariant $I_\ell$ can be
expressed as an anti-commutator of two odd elements of the universal
enveloping algebra $\mathsf {U}(\mathfrak{gl}_{n|n})$. Indeed, using
the basic bracket relations (\ref{basic brackets}) of
$\mathfrak{gl}_{n|n}$ it is straightforward to verify that
\begin{displaymath}
    I_\ell = [ Q \, , F^{(\ell)} ] \;,
\end{displaymath}
where $Q\,$, $F^{(\ell)} \in \mathsf{U}( \mathfrak{gl}_{n|n} )$ are
given in terms of a standard basis $\{ E_{ij}^{\tau \tau^\prime} \}$
by
\begin{displaymath}
    Q = \sum_{i=1}^{n} E_{i\,i}^{10} \;, \quad F^{(\ell)} = \sum
    E_{i j_2}^{0\, \tau_2}(-1)^{\tau_2} E_{j_2 j_3}^{\tau_2\tau_3}
    (-1)^{\tau_3} \cdots E_{j_{\ell-1} j_\ell}^{\tau_{\ell-1}
    \tau_\ell} (-1)^{\tau_\ell} E_{j_\ell \, i}^{\tau_\ell 1} \;.
\end{displaymath}

If the representation space $\mathcal{V}_\lambda$ were
finite-dimensional, we could now argue that
\begin{displaymath}
    \mathrm{STr}_{\mathcal{V}_\lambda} \, \rho(I_\ell) =
    \mathrm{STr}_{\mathcal{V}_\lambda} \, [ \rho (Q) ,
    \rho(F^{(\ell)}) ] = 0 \;,
\end{displaymath}
since the supertrace of any bracket vanishes.  On the other hand,
since $I_\ell$ is a Casimir invariant, the operator $\rho (I_\ell
)$ on the irreducible representation space $\mathcal{V}_\lambda$
must be a multiple of unity: $\rho(I_\ell) = \alpha(I_\ell) \times
\mathrm{Id}_{\mathcal{V}_\lambda}\,$. In finite dimension we could
therefore say that
\begin{displaymath}
    \mathrm{STr}_{\mathcal{V}_\lambda}\,\rho(I_\ell)=\alpha(I_\ell)
    \, \mathrm{STr}_{\mathcal{V}_\lambda} \, \mathrm{Id} =
    \alpha(I_{\ell}) (\mathrm{dim}\, \mathcal{V}_{\lambda,0} -
    \mathrm{dim}\, \mathcal{V}_{\lambda,1}) \;.
\end{displaymath}
Inspection of $\mathcal{V}_\lambda$ shows that there is exactly
one vector in its even subspace which has no partner in the odd
subspace -- this vector is the "vacuum". In this sense, the
dimension of $\mathcal{V}_{\lambda,0}$ exceeds that of
$\mathcal{V}_{\lambda,1}$ by one. Hence $\mathrm{STr}_{
\mathcal{V}_\lambda} \, \rho (I_\ell) = 1 \cdot \alpha(I_\ell)$,
and from the previous result $\mathrm{STr}_{\mathcal{V}_\lambda}
\, \rho(I_\ell) = 0$ we would be forced to conclude $\alpha
(I_\ell) = 0$.  Let us record this conclusion for later use.
\begin{lem}\label{finite V Cas}
In an irreducible $\mathfrak{gl}_{n|n}$-representation on a
finite-dimensional vector space $V$ with $\mathrm{dim}\, V_0 \not=
\mathrm{dim}\, V_1\,$, all Casimir invariants $I_\ell$ of
$\mathfrak{gl}_{n|n}$ are identically zero.
\end{lem}
While the argument leading to Lem.\ \ref{finite V Cas} is correct
for finite dimension, it is not sound for the case of our
infinite-dimensional representation space $\mathcal{V}_\lambda\,$,
as the traces $\mathrm{STr}_{\mathcal{V}_ \lambda} \,
\rho(I_\ell)$ and $\mathrm{ STr}_{\mathcal{V}_\lambda}
\mathrm{Id}$ do not converge. Nevertheless, the conclusion still
holds true:
\begin{prop}\label{Cas zero}
In the representation $(\mathcal{V}_\lambda \, ,\rho)$ all Casimir
invariants $I_\ell$ vanish.
\end{prop}
\begin{proof}
The heuristic argument becomes rigorous on regularizing the
traces. This is done with the help of a difference of two sums of
generators,
\begin{displaymath}
    \Lambda = \sum_{\tau = 0}^1 \sum_{j = 1}^p E_{jj}^{\tau\tau} -
    \sum_{\tau = 0}^1 \sum_{l = 1}^q E_{l+p,l+p}^{\tau\tau} \;,
\end{displaymath}
which is not a Casimir invariant of $\mathfrak{gl}_{n|n}\,$, but does
lie in the center of $\mathsf{U}( \mathfrak{gl}_{p|p} \oplus
\mathfrak{gl}_{q|q})$. It commutes with both $Q$ and $F^{(\ell)}$
because these respect the vector space decomposition $\mathbb{C}^n =
\mathbb{C}^p \oplus \mathbb{C}^q$. On $\mathcal{V}_\lambda$ it is
represented by the operator
\begin{eqnarray*}
    \rho(\Lambda) &=& \sum_{j = 1}^{p} \sum_{a = 1}^N \big(
    \varepsilon(e_{1,j}^+ \otimes e_a) \iota(f_+^{1,j} \otimes f^a)
    + \mu(e_{0,j}^+ \otimes e_a)\delta(f_+^{0,j} \otimes f^a) \big)
    \\ &+& \sum_{l = 1}^{q} \sum_{a = 1}^N \big(\varepsilon(f_-^{1,l}
    \otimes f^a) \iota(e_{1,l}^- \otimes e_a) + \mu(f_-^{0,l} \otimes
    f^a) \delta(e_{0,l}^- \otimes e_a) \big) \;,
\end{eqnarray*}
where the notation of Sect.\ \ref{sect:2} is being used. $\Lambda$ is
called the particle number in physics.

Now if $t$ is any positive real parameter, inserting the operator
$\mathrm{e}^{-t\, \rho(\Lambda)}$ cuts off the infinite contribution
from high particle numbers and makes the traces of the heuristic
argument converge. Thus, on the one hand we now rigorously have
\begin{displaymath}
    \mathrm{STr}_{\mathcal{V}_\lambda} \, \mathrm{e}^{- t \,
    \rho(\Lambda)} \rho(I_\ell) = \mathrm{STr}_{\mathcal{V}_\lambda}
    \, [ \rho (Q) , \mathrm{e}^{- t \, \rho(\Lambda)}
    \rho(F^{(\ell)}) ] = 0 \;,
\end{displaymath}
and on the other hand, since $\rho(I_\ell) = \alpha(I_\ell) \times
\mathrm{Id}_{\mathcal{V}_\lambda}\,$,
\begin{displaymath}
    \mathrm{STr}_{\mathcal{V}_\lambda} \, \mathrm{e}^{- t \,
    \rho(\Lambda)} \rho(I_\ell) = \alpha(I_\ell) \, \mathrm{STr}_{
    \mathcal{V}_\lambda}\,\mathrm{e}^{- t \, \rho(\Lambda)} \;.
    \end{displaymath}
To compute the last trace, notice that $\mathrm{e}^{- t \Lambda}$ is
the diagonal transformation
\begin{displaymath}
    \mathrm{e}^{-t \Lambda} = \mathrm{e}^{-t} \sum  E_{jj}^{\tau\tau}
    + \mathrm{e}^{t} \sum E_{l+p,l+p}^{\tau\tau} \;,
\end{displaymath}
and use the formula (\ref{char-det-4.1}) to obtain
\begin{displaymath}
    \mathrm{STr}_{\mathcal{V}_\lambda} \, \mathrm{e}^{- t\, \rho
    (\Lambda)} = \int_{\mathrm{U}_N} \frac{\mathrm{Det}^p(\mathrm{Id}-
    \mathrm{e}^{-t} u) \, \mathrm{Det}^q( \mathrm{Id}- \mathrm{e}^{t}
    u)} {\mathrm{Det}^p( \mathrm{Id} - \mathrm{e}^{-t} u) \,
    \mathrm{Det}^q( \mathrm{Id} - \mathrm{e}^{t} u)} \, du = 1 \;.
\end{displaymath}
Therefore
\begin{displaymath}
    0 = \mathrm{STr}_{\mathcal{V}_\lambda} \, \mathrm{e}^{-t\,
    \rho(\Lambda)} \rho(I_\ell) = \alpha(I_\ell) \;,
\end{displaymath}
and the proposition is proved.
\end{proof}
\begin{rem}
Up to this point, everything would have gone through for the most
general case of ratios (\ref{R for U(N)}) with an arbitrary number of
characteristic polynomials and their complex conjugates in the
numerator and denominator. However, we now had to require these
numbers to be pairwise equal, or else Prop.\ \ref{Cas zero} would
have been false.
\end{rem}
As an immediate consequence of Prop.\ \ref{Cas zero} and the
relationship between Casimir invariants $I_\ell$ and
Laplace-Casimir operators $D(I_\ell)$, we have:
\begin{cor}\label{0 eigfctn}
The irreducible $\mathrm{GL}_{n|n}$-character $\chi(\Xi) = \mathrm
{STr}_{\mathcal{V}_\lambda} \, \rho(\Xi)$ lies in the kernel of
the ring of $\mathfrak{gl}_{n|n}$-invariant differential operators
$D(I_\ell):$
\begin{displaymath}
    D(I_\ell) \chi = 0 \quad (\ell \in \mathbb{N}) \;.
\end{displaymath}
\end{cor}
\begin{rem}
Thus the $\mathfrak {gl}_{n|n}$-representation
$(\mathcal{V}_\lambda, \rho)$ is \emph{degenerate} with the
trivial representation (the Laplace-Casimir eigenvalues are the
same), and is \emph{atypical}. Sadly, the characters of atypical
representations are not covered by the known generalization
\cite{berezin} of the Weyl character formula to the case of Lie
supergroups.  To compute the character $\chi(t) = \mathrm{STr}_{
\mathcal{V}_\lambda}\, \rho(t)$, further effort is required.
\end{rem}

\subsection{Maximal torus}
\label{sect:maxtorus}

In Sect.\ \ref{sect:radialpart} we reviewed some of Berezin's results
for the radial parts of the Laplace-Casimir operators $D(I_\ell)$ for
$\mathfrak{gl}_{m|n}\,$.  To bring these to bear on our problem, we
first need to establish the existence of a kind of maximal torus for
the real-analytic manifold $M = M_1 \times M_0\,$. The main step here
is the following lemma.

Its proof needs two basic facts (established by Cor. \ref{cor 5.4}):
the semigroup $\mathrm{H}_{p,\,q}$ is con\-nected, and the elements
$h \in \mathrm{H}_{p,\,q}$ satisfy the inequality $h s h^\dagger < s$
as well as $h^\dagger s h < s$.
\begin{lem}\label{lem 4.13}
Each element of $M_0 = \{h\sigma(h)^{-1} \mid h \in
\mathrm{H}_{p,\,q} \}$ is brought to diagonal form by a
pseudo-unitary transformation $g \in \mathrm{U}_{p,\,q}\,$. The
eigenvalues lie in $\mathbb{R}_+ \setminus \{ 1 \}$.
\end{lem}
\begin{proof}
On general grounds, every $m\in M_0$ has at least one eigenvector $v
\not= 0\,$. Notice that the eigenvalue $\lambda$ must be non-zero
because $m = h\, \sigma(h) ^{-1} = h s h^\dagger s$ has an inverse.
Now the inequality $h s h^\dagger < s$ for $h \in \mathrm{H}_{p,\,q}$
leads to
\begin{displaymath}
    \lambda \langle sv , v \rangle = \langle sv , mv \rangle =
    \langle sv , (hsh^\dagger) sv \rangle < \langle sv , v \rangle
    \;.
\end{displaymath}
One then infers that $\langle v , s v \rangle = \langle sv , v
\rangle \not= 0$, and that $\lambda$ is real and $\lambda \not= 1$.
If $\langle v , s v \rangle < 0$ then $\lambda > 1$, and if $\langle
v , sv \rangle > 0$ then $\lambda< 1$.  In the former case we fix the
normalization of $v$ by the condition $\langle v , sv \rangle = -1$
and in the latter case by $\langle v , sv \rangle = 1$.

The pseudo-Hermitian form $\langle \, , \, \rangle_s : \, U_0 \times
U_0 \to \mathbb{C}$, $(u,v) \mapsto \langle u , sv \rangle$ is
non-degenerate. Therefore, since $\langle v , v \rangle_s \equiv
\langle v, sv\rangle \not= 0\,$, the vector space $U_0$ decomposes as
a direct sum
\begin{displaymath}
    U_0 = \mathbb{C} v \oplus U^\perp
\end{displaymath}
where $U^\perp$ is the subspace of $U_0$ which is $\langle \, , \,
\rangle_s$-orthogonal to the complex line $\mathbb{C} v\,$.

Now $m = h s h^\dagger s$ is self-adjoint w.r.t.\ the
pseudo-Hermitian form $\langle\, , \, \rangle_s$ and hence leaves the
decomposition $U_0 = \mathbb{C}v \oplus U^\perp$ the same. Thus one
can split off the eigenspace $\mathbb{C}v$ from $U_0$ and repeat the
whole discussion for $m$ restricted to $U^\perp$.  The restriction
$m^ \prime := m \vert_{U^\perp}$ has again at least one eigenvector,
say $v^\prime$, and $\langle v^\prime, v^\prime \rangle_s \not= 0$ by
the same argument as before.  The eigenvalue $\lambda^ \prime$ is
real with $\lambda^\prime \not= 0$ and $\lambda^\prime \not= 1$.
Since $m^\prime$ is $\langle \, , \, \rangle_s$ self-adjoint and
preserves the decomposition of $U^\perp$ by the complex line
$\mathbb{C}v^\prime$ plus its orthogonal complement, one can split
off $\mathbb{C}v^\prime$ from $U^\perp$. By continuing in this way,
one concludes that $m$ is diagonalizable, and that the
$(p+q)$-dimensional vector space $U_0$ decomposes as a $\langle\, ,
\,\rangle_s$-orthogonal direct sum of $m$-eigenspaces $\mathbb{C}v_1,
\ldots, \mathbb{C}v_{p+q} \,$.

It remains to show that the diagonalizing transformation $g$ lies in
the pseudo-unitary group $\mathrm{U}_{p,\,q}\,$.  For this, let $\{
e_i \}_{i=1, \ldots, p+q}$ be an orthonormal basis of $U_0$ such that
$e_1, \ldots, e_p$ span the $s$-positive subspace $U_0^+$, and
$e_{p+1}, \ldots, e_{p+q}$ span $U_0^-$. It is clear that if
$\lambda^\pm$ are two real numbers in the range $0 < \lambda^+ < 1$
and $1 < \lambda^- < \infty \,$, then
\begin{displaymath}
    m = \lambda^+ \, \mathrm{Id}_{U_0^+} \oplus \lambda^-
    \, \mathrm{Id}_{U_0^-}
\end{displaymath}
lies in $M_0 \,$. Indeed, $m$ equals $h\, \sigma(h)^{-1}$ for $h =
\sqrt{\lambda^+}\, \mathrm{Id}_{U_0^+} \oplus \sqrt{\lambda^-}\,
\mathrm{Id}_{U_0^-} \in \mathrm{H}_{p,\,q}\,$. Note that the
signature of $\mathrm{Id}_{U_0} - m$ for this $m$ is the same as the
signature of $s\,$.

Now, since the manifold $\mathrm{H}_{p,\,q}$ is connected, so is $M_0
= \{ h\, \sigma(h)^{-1} \mid h \in \mathrm{H}_{p,\,q} \}$. When $m$
is pushed around in $M_0 \,$, the eigenvalues of $m$ move but, as we
have seen, they are always real and never hit zero or one. Therefore,
they stay in $\mathbb{R}_+ \setminus \{ 1 \}$ and the signature of
$\mathrm{Id}_{U_0} - m$ for $m \in M_0\, $ is an invariant and is
given by $s\,$. By reordering the indices, one can arrange for the
positive signature eigenspace of $\mathrm{Id}_{U_0 }- m$ to be
spanned by the first $p$ eigenvectors $v_1, \ldots , v_p$ of $m\,$. A
linear transformation $g \in \mathrm {GL}(U_0)$ diagonalizes $m$ if
$v_i = g e_i$ for $i = 1, \ldots, p+q\,$. Since the $m$-eigenvectors
$v_i$ have been constructed to form an orthonormal system with
respect to the pseudo-Hermitian form $\langle\,,\,\rangle_s \,$, it
follows that $g$ is pseudo-unitary: $g \in \mathrm{U}_{p,\,q}\,$.
\end{proof}
\begin{rem}
The numerical part of the operator $\mathrm{Id}_{V_0} - \mathsf{D}
\otimes u$ of Def.\ \ref{def 4.2} is $\mathrm{Id}_{V_0} - m \otimes
u$ for $u \in \mathrm{U}_N$ and some $m \in M_0\,$. Since the
eigenvalues of $m$ lie in $\mathbb {R}_+ \setminus \{ 1 \}$, we see
that $\mathrm {Id}_{V_0} - \mathsf{D} \otimes u$ in fact has an
inverse.
\end{rem}
Every element $m_1 \in M_1 = \mathrm{U}_n$ is diagonalizable by a
unitary transformation and has unitary eigenvalues.  In
combination with Lem.\ \ref{lem 4.13} this has the following
consequence for $m = (m_1,m_0) \in M_1 \times M_0 = M$.
\begin{cor}\label{torus exists}
Let $T = T_1 \times T_0$ with
\begin{displaymath}
    T_1 = \mathrm{U}_1^{n} \;, \quad T_0 = (0,1)^p
    \times (1,\infty)^q \quad (p + q = n) \;,
\end{displaymath}
be the set of diagonal transformations $t = (t_1,t_0)$ defined in
Eqs.\ (\ref{def t0t1}) and (\ref{range of phi}) with real-valued
$\psi_k$ and $\phi_k$ restricted by $\phi_j < 0 < \phi_l\,$. This
Abelian semigroup $T$ is a maximal torus for $M = M_1 \times M_0$ in
the sense that for every $m \in M$ there exist $t \in T$ and $g \in
G_\mathbb{R} = \mathrm{U}_n \times \mathrm{U}_{p,\,q}$ such that
\begin{displaymath}
    g^{-1} m g = t \;.
\end{displaymath}
\end{cor}
Recall now from Sect.\ \ref{sect:weights} that we view our variable
parameters $\psi_k$ and $\phi_k$ as complex linear functions on a
complex Cartan subalgebra $\mathfrak{h}\subset \mathfrak{gl}_{n|n}
\,$. Let $\mathfrak{h}_\mathbb{R} \subset \mathfrak{h}$ be the real
form defined by the requirement that if $H \in
\mathfrak{h}_\mathbb{R}$ then $\psi_k(H) \in \mathbb{R}$ and
$\phi_k(H) \in \mathbb{R}$ for $k = 1, \ldots, n\,$. The condition
imposed after Eq.\ (\ref{def t0t1}) can then be rephrased as the
statement that our parameters restrict to linear functions
\begin{equation}
    \psi_k :\,\, \mathfrak{h}_\mathbb{R}\to \mathbb{R} \quad
    \text{and} \quad \phi_k : \,\, \mathfrak {h}_\mathbb{R}
    \to \mathbb{R} \quad (k = 1, \ldots, n) \;.
\end{equation}
The further conditions $\phi_j(H) < 0 < \phi_l(H)$ for $1 \le j \le p
< l \le n$ select an open subspace $\mathfrak{t} \subset
\mathfrak{h}_\mathbb{R}$ -- note that $\mathfrak{t}$ is not a vector
space -- with the property
\begin{equation}
    \exp \mathfrak{t} = T \;.
\end{equation}

\subsection{$W$-invariance}\label{sect:weyl-inv}

From Prop.\ \ref{prop 4.2} the character $\chi(\Xi)$ exists as an
analytic section of $\mathcal{F}$, and this section is radial. Since
the domain $M = M_1\times M_0$ is invariant under conjugation by $g
\in G_\mathbb{R} = \mathrm{U}_n \times \mathrm {U}_{p,\,q}\,$, so is
the character. Therefore, if $g \in G_\mathbb {R}$ is any element
which normalizes $T$, i.e. $\forall t \in T : g t g^{-1} \in T$, the
restriction $\chi: \, T \to \mathbb {C}$ inherits the invariance
property
\begin{equation}\label{eq W-invariance}
    \chi(t) = \chi(g t g^{-1}) \;.
\end{equation}

The transformations $t \mapsto g t g^{-1} =: w \cdot t$ of $T$
arising in this way form the Weyl group $W$. Since the neutral
element $t = e$ is contained in the closure of $T\,$, one can
differentiate the $W$-action $t \mapsto w \cdot t$ at the fixed point
$t = e$ to obtain an induced linear action $H \mapsto \mathrm{Ad}(g)
H =: w(H)$ of $W$ on tangent vectors $H \in \mathfrak{h}_\mathbb{R}
\,$. This action in turn induces an action of $W$ on the linear
functions $f: \, \mathfrak{h}_\mathbb {R} \to \mathbb{C}$ by $w(f) =
f \circ w^{-1}$.
\begin{prop}\label{Weyl inv}
The Weyl group of our problem is $W = \mathrm{S}_{n} \times
(\mathrm{S}_p \times \mathrm{S}_q)$ where the symmetric group
$\mathrm{S}_{n}$ permutes the $n = p+q$ functions
\begin{displaymath}
    (\psi_1\, , \ldots, \psi_n) \;,
\end{displaymath}
while the first and second factor in $\mathrm{S}_p \times
\mathrm{S}_q$ permute the first $p$ resp.\ last $q$ entries of
\begin{displaymath}
    (\phi_1 \, ,\ldots, \phi_p \, , \phi_{p+1}\, ,\ldots, \,
    \phi_{p+q}) \;.
\end{displaymath}
The character $\chi(t) = \mathrm{STr}_{\mathcal{V}_\lambda} \,
\rho(t)$ satisfies $\chi(t) = \chi(w\cdot t)$ for all $w\in W$.
\end{prop}
\begin{proof}
Let $H = (H_1, H_0) \in \mathfrak{t} \subset \mathfrak{h}_\mathbb
{R}$ be a regular point, i.e., $H$ viewed as a diagonal linear
transformation of $\mathbb{C}^{n|n}$ has eigenvalues $\mathrm{i}
\psi_1(H) , \ldots, \phi_n(H)$ which are pairwise distinct and
non-zero. Consider the orbit of such a point $H$ under the adjoint
action of $G_\mathbb{R} = \mathrm{U}_n \times \mathrm{U}_{p,\,q}\,$.
Every distinct intersection of this orbit with $\mathfrak{t}$ amounts
to one element of the Weyl group $W = W_1 \times W_0\,$. Since
conjugation by $g \in G_\mathbb{R}$ leaves the eigenvalues of $H$
unchanged, every point of intersection must correspond to a
permutation of these eigenvalues.  The $n$ eigenvalues $\mathrm{i}
\psi_1(H), \ldots, \mathrm{i}\psi_n(H)$ are imaginary and conjugation
by $(u_1,e) \in G_\mathbb{R}$ with suitably chosen $u_1 \in
\mathrm{U}_n$ allows to arbitrarily permute them; hence $W_1 =
\mathrm{S}_{n}\,$. The $n = p+q$ eigenvalues $\phi_1(H), \ldots,
\phi_n(H)$ are real, with the first $p$ being negative ($\phi_j < 0$)
and the last $q$ positive ($\phi_l > 0$). Conjugation by $(e,u_0)$
with $u_0 \in \mathrm{U}_{p,\,q}$ never mixes these two sets but only
permutes them separately. Thus $W_0 = \mathrm {S}_p \times
\mathrm{S}_q\,$.

The $W$-invariance $\chi(t) = \chi(w \cdot t)$ is a restatement of
Eq.\ (\ref{eq W-invariance}).
\end{proof}

\subsection{Radial differential equations for $\chi\,$}
\label{sect:4.2}

Being the character (or supertrace) of a representation, the analytic
section $\chi \in \mathcal{F}$ is radial. Hence, if $\dot{D}(I_\ell)$
is the radial part of the $\mathfrak{gl}_{n|n}$-invariant
differential operator $D(I_\ell)$, the equation $D(I_\ell) \chi = 0$
reduces to $\dot{D}(I_\ell) \chi(t) = 0$ for the $W$-invariant torus
function $t \mapsto \chi(t)$.

We now write the system $(\ell \in \mathbb{N})$ of differential
equations $\dot{D} (I_\ell) \chi(t) = 0$ in explicit form by drawing
on Berezin's results (Thm.\ \ref{Ber1} of Sect.\ \ref{sect:3}). For
that purpose we fix a set $\Delta^+ = \Delta_0^+ \cup \Delta_1^+$ of
positive roots of the Lie superalgebra $\mathfrak {gl}_{n|n} \,$. By
the same choice that was made in Sect.\ \ref{sect:weights} before, we
take the even resp.\ odd positive roots to be
\begin{eqnarray*}
    &&\Delta_0^+ : \quad \mathrm{i}\psi_k - \mathrm{i}
    \psi_{k^\prime} \;, \quad \phi_k - \phi_{k^\prime}
    \quad (1 \le k^\prime < k \le n) \;, \\ &&\Delta_1^+ :
    \quad \mathrm{i}\psi_k - \phi_j \;, \quad \phi_l -
    \mathrm{i}\psi_k \quad (1 \le j \le p < l \le n) \;.
\end{eqnarray*}
Note that if $H \in \mathfrak{t}\,$, then $\mathfrak{Re}\,
\beta(H) > 0$ for all roots $\beta \in \Delta_1^+$.

While we have chosen to regard our parameters $\psi_k$ and
$\phi_k$ as linear functions on $\mathfrak {h}_\mathbb{R} \supset
\mathfrak{t}\,$, another possibility would have been to use
composition with the logarithm to define them as \emph{local
coordinates} on $T = \exp\, \mathfrak{t}\,$. From the latter
perspective, they determine differential operators $\partial /
\partial \psi_k\,$ and $\partial / \partial\phi_k$ on functions $f
:\, T \to \mathbb{C}$.

Let $D_\ell$ as in Sect.\ \ref{sect:radialpart} (after replacing
$\phi_k \to -\mathrm{i}\phi_k$) be the differential operator
\begin{equation}\label{D_ell formula}
    D_\ell = \sum_{k=1}^n \frac{\partial^\ell}
    {\partial\psi_k^\ell} - (-\mathrm{i})^\ell \sum_{k=1}^n
    \frac{\partial^\ell}{\partial\phi_k^\ell} \;,
\end{equation}
and note that $D_\ell$ is $W$-invariant, i.e., commutes with the
$W$-action on functions $f(t)$.  Notice also that the degree-2
operator $D_2$ is elliptic.
\begin{prop}\label{rad Lapl}
The character $\chi :\, T \to \mathbb{C}$ given by $\chi(t) =
\mathrm{STr}_{\mathcal{V}_\lambda} \, \rho(t)$ satisfies the set of
differential equations (defined on the set of regular points
$T^\prime \subset T$)
\begin{displaymath}
    J^{-1/2} D_\ell \big( J^{1/2} \chi \big) = 0
    \quad (\ell \in \mathbb{N}) \;,
\end{displaymath}
where the function $J^{1/2} : T^\prime \to \mathbb{C}$ is a square
root of
\begin{equation}\label{universal J}
    J(t) = \frac{ \prod_{\alpha \in \Delta_0^+} \sinh^2 \left(
    \frac{1}{2} \alpha(\ln t) \right)} { \prod_{ \beta \in
    \Delta_1^+} \sinh^2 \left(\frac{1}{2} \beta(\ln t)\right)}\;.
\end{equation}
\end{prop}
\begin{proof}
On restricting $\chi \in \mathcal{F}$ to the function $\chi: \,
T\to\mathbb {C}$, it follows from Cor.\ \ref{0 eigfctn} that
$\dot{D}(I_\ell) \chi(t) = 0$ for all $\ell \in \mathbb{N} \,$. By
Thm.\ \ref{Ber1} the radial part $\dot{D} (I_\ell)$ agrees with the
differential operator $J^{-1/2} D_\ell \circ J^{1/2}$ modulo
lower-order terms, $J^{-1/2} Q_{\ell-1} \circ J^{1/2}$. Since the
lower-order operators $Q_{\ell-1}$ are themselves expressed as
polynomials in the $D_k$ and all constant terms $Q_{\ell-1}
(\mathbf{0},\mathbf{0})$ vanish (Cor.\ \ref{no consts}), the system
of equations $\dot{D}(I_\ell) \chi = 0$ is equivalent to the system
$J^{-1/2} D_\ell (J^{1/2} \chi) = 0\,$ ($\ell \in \mathbb{N}$).

The expression for the function $J$ follows from Thm.\ \ref{Ber1} by
analytically continuing from the compact torus $\mathrm{U}_1^n \times
\mathrm{U}_1^n$ to the Abelian semigroup $T\,$.
\end{proof}
The only input we required for Prop.\ \ref{rad Lapl} was the system
of differential equations $D(I_\ell) \chi = 0\,$. Since this system
is available also under the conditions of Cor.\ \ref{finite V Cas},
the same conclusion holds true in that modified context. We record
this fact for later use:
\begin{cor}\label{cor:4.12}
Let $\tilde{\chi}$ be the character of an irreducible $\mathfrak{gl}
_{n|n}$-representation on a finite-dimensional $\mathbb{Z}_2$-graded
vector space $V = V_0 \oplus V_1\,$. If $\mathrm{dim}(V_0) \not=
\mathrm{dim}(V_1)$ then $J^{-1/2} D_\ell (J^{1/2} \tilde{\chi}) = 0$
for all $\ell \in \mathbb{N}\,$.
\end{cor}
\begin{rem}
To obtain the statement of Prop.\ \ref{rad Lapl} by reference to
standard text, it was necessary to go to the expense of extending
$\chi$ to a section of $\mathcal{F} = \Gamma(M, \mathbb{C} \otimes
\wedge F^\ast)$. With the system of differential equations $J^{-1/2}
D_\ell (J^{1/2} \chi) = 0\,$ now established, the objects $M$ and
$\mathcal{F}$ have served their purpose, and we restrict all further
considerations to the torus function $\chi : \, T \to \mathbb{C}\,$.
\end{rem}

\subsection{Proof of the main theorem}\label{sect:proof}

We have now accumulated enough information about the character $t
\mapsto \chi(t)$ to prove Thm.\ \ref{thm0}. Let us summarize what
we know.
\begin{itemize}
\item[A.] $\chi$ is a $W$-invariant analytic function on the
Abelian semigroup $T$ of Cor.\ \ref{torus exists}.
\item[B.] $\chi$ has an expansion in terms of weights (Cor.\ \ref{cor
2.14}) with coefficients in the limited range given by Prop.\
\ref{prop 2.13}.
\item[C.] $\chi$ by Prop.\ \ref{rad Lapl} obeys the differential
equations $J^{-1/2} D_\ell (J^{1/2} \chi) = 0$ $(\ell \in \mathbb
{N})$.
\end{itemize}
The proof of Thm.\ \ref{thm0} is done in two steps: we first show
that the conditions (A-C) admit at most one solution $\chi \,$;
afterwards we will write down the solution and verify that it has the
required properties. Of course, once Thm.\ \ref{thm0} has been proved
for $t \in T$, the result immediately extends to the complex torus by
analytic continuation.

\subsubsection{Uniqueness of the solution}\label{sect:unique}

We shall prove that the coefficients of the weight expansion of
$\chi$ (property B) are completely determined by $W$-invariance
(property A) and the system of differential equations (property C).

Our first step is to establish a good way of representing the square
root function $J^{1/2}$.  Let $\delta: \,\mathfrak{h} \to \mathbb{C}$
be half the sum (in the $\mathbb{Z}_2$-graded sense) of positive
roots:
\begin{equation}
    \delta = {\textstyle{\frac{1}{2}}} \sum\nolimits_{\alpha \in
    \Delta_0^+} \alpha - {\textstyle{\frac{1}{2}}} \sum\nolimits_{
    \beta \in \Delta_1^+} \beta \;,
\end{equation}
and choose a square root of the expression (\ref{universal J}) for
$J :$
\begin{equation}\label{sqrt of J}
    J^{1/2} = \mathrm{e}^{\delta}\,\frac{\prod_{\alpha\in\Delta_0^+}
    \frac{1}{2}(1-\mathrm{e}^{-\alpha})}{\prod_{\beta\in\Delta_1^+}
    \frac{1}{2} (1 - \mathrm{e}^{-\beta})} \circ \ln \;.
\end{equation}
Notice that while the function $t \mapsto \mathrm{e}^{\delta(\ln t)}$
is double-valued, the square root ambiguity cancels in the
differential operator $J^{-1/2} D_\ell \circ J^{1/2}$.

The following subsystem of $r = 2(p+q)-1$ positive roots $\{\sigma_1,
\ldots, \sigma_r \} \subset \Delta^+$:
\begin{eqnarray}
    &&\phi_j - \phi_{j-1} \,\, (j = 2, \ldots, p) \;, \quad
    \mathrm{i}\psi_1 - \phi_p \;, \quad \mathrm{i}\psi_k -
    \mathrm{i}\psi_{k-1} \,\, (k = 2, \ldots, p+q) \;, \nonumber
    \\ &&\phi_{p+1} - \mathrm{i}\psi_{p+q} \;, \quad \phi_l -
    \phi_{l-1} \,\, (l = p+2, \ldots, p+q) \label{simple roots} \;,
\end{eqnarray}
is a system of simple roots, i.e., every positive root $\alpha$ is
uniquely expressed as a linear combination $\alpha = \sum_i a_i
\sigma_i$ with non-negative integers $a_i\,$.
\begin{lem}\label{expand sqrtJ}
The function $\mathrm{e}^{-\delta \circ \ln} J^{1/2} : T \to
\mathbb{C}$ has an absolutely convergent expansion in terms of
simple roots $\sigma_1 ,\ldots, \sigma_r :$
\begin{displaymath}
    \mathrm{e}^{-\delta \circ \ln} J^{1/2} = \sum A_b \,
    \mathrm{e}^{- \sum_{i=1}^r b_i \sigma_i \circ \ln} \quad
    (A_b \in \mathbb{C}\, , \,\, A_{b= \mathbf{0}} \not= 0) \;,
\end{displaymath}
where the sum runs over sets $b \equiv \{ b_1, \ldots, b_r \}$
composed of non-negative integers $b_i \,$.
\end{lem}
\begin{proof}
Expand the factors $1-\mathrm{e}^{-\beta \circ \ln}$ in the
denominator of (\ref{sqrt of J}) as geometric series. If $\ln t
\in \mathfrak {t}$ one has $| \mathrm{e}^{-\beta(\ln t)} | =
\mathrm{e}^{- \mathfrak{Re}\, \beta(\ln t)} < 1$ for all $\beta
\in \Delta_1^+$. Therefore, these series converge absolutely on
$T\,$, and after expressing all of the roots $\alpha, \beta \in
\Delta^+$ in terms of the simple roots $\sigma_i \,$, the series
can be reordered to the stated form, with the resulting series
still being absolutely convergent.
\end{proof}
To draw a conclusion from the differential equations $J^{-1/2}
D_\ell (J^{1/2} \chi) = 0$, we need some understanding of the
joint kernel of the set of differential operators $D_\ell \,$.
Consider the exponential function $f = \mathrm{e}^{\sum_k (
\mathrm{i}m_k \psi_k - n_k \phi_k)}$ for some set of numbers
$m_k\,$, $n_k \,$.
\begin{lem}\label{permute only}
The function $f = \mathrm{e}^{\sum_k (\mathrm{i} m_k \psi_k - n_k
\phi_k) \circ \ln}$ is annihilated by all of the operators
$D_\ell$ $(\ell \in \mathbb{N}$) if and only if $\{ m_1, \ldots,
m_{p+q} \}$ is the same set as $\{ n_1, \ldots, n_{p+q} \}$.
\end{lem}
\begin{proof}
$f$ is a joint eigenfunction:
\begin{displaymath}
    \mathrm{e}^{-\sum_k (\mathrm{i}m_k \psi_k - n_k \phi_k)
    \circ \ln} \, D_\ell \big( \mathrm{e}^{\sum_k (\mathrm{i}m_k
    \psi_k - n_k \phi_k)\circ \ln} \big) = \mathrm{i}^\ell
    \sum\nolimits_{k=1}^{p+q} (m_k^\ell - n_k^\ell) \;,
\end{displaymath}
and the eigenvalue vanishes for all $\ell \in \mathbb{N}$ if and
only if $n_k = \pi(m_k)$ $(k = 1, \ldots, p+q)$ for some
permutation $\pi \in \mathrm{S}_{p+q}\,$.
\end{proof}
If we were dealing with a non-degenerate representation, the joint
eigenspace of the operators $D_\ell$ (corresponding to the given set
of Casimir eigenvalues) would have a dimension no larger than the
order of the Weyl group. In contrast, the joint kernel of the
operators $D_\ell$ is seen to be huge! One might therefore think that
our character $\chi$ is hopelessly underdetermined by the system of
equations $J^{-1/2} D_\ell( J^{1/2} \chi ) = 0$. However, the
situation will be saved by Prop.\ \ref{prop 2.13}, which says that
$\gamma = \sum (\mathrm{i} m_k \psi_k - n_k \phi_k)$ is a weight of
$\mathcal{V}_\lambda$ only if the coefficients $m_k$ and $n_k$ are
integers in the range
\begin{equation}\label{restr range}
    n_j \le 0 \le m_k \le N \le n_l \quad
    (1 \le j \le p < l \le p+q) \;.
\end{equation}

Recall now from Cor.\ \ref{cor 2.14} the weight expansion of the
character $\chi\,$, which involved unknown multiplicities for the
weights $\gamma \in \Gamma_\lambda$. For present purposes, let
this expansion be written as
\begin{equation}\label{weight expansion}
    \chi(t) = \sum\nolimits_{\gamma \in \Gamma_\lambda}
    B_\gamma \,\, \mathrm{e}^{\gamma(\ln t)} \;,
\end{equation}
where $B_\gamma = (-1)^{|\gamma|} \mathrm{dim}(V_\gamma)$ are the
unknowns. Note $B_{\lambda_N} = 1$, since the highest-weight space is
one-dimensional. $W$-invariance then implies $B_{w(\lambda_N)} = 1$
for all $w \in W$.

Every weight $\gamma \in \Gamma_\lambda$ is uniquely represented
in terms of the highest weight $\lambda_N$ and the simple roots
(\ref{simple roots}) as $\gamma = \lambda_N - \sum_{i=1}^r c_i
\sigma_i$ with non-negative integers $c_i\,$. We refer to the
uniquely determined number $\sum_i c_i \ge 0$ as the \emph{degree}
of the weight $\gamma\,$.
\begin{lem}\label{unique}
If the system of differential equations $J^{-1/2} D_\ell (J^{1/2}
\chi) = 0$ ($\ell \in \mathbb{N}$) admits a $W$-invariant solution
of the form (\ref{weight expansion}) with $B_{\lambda_N} = 1$ and
weights $\gamma$ that have coefficients in the range (\ref{restr
range}), then this solution is unique.
\end{lem}
\begin{proof}
By multiplication with the function $\mathrm{e}^{-\delta \circ
\ln} J^{1/2}$, write the system of differential equations in the
equivalent form
\begin{displaymath}
    \forall \ell \in \mathbb{N}: \quad 0 =
    \mathrm{e}^{-\delta \circ \ln} D_\ell(J^{1/2} \chi) \;,
\end{displaymath}
and use for $J^{1/2}$ the absolutely convergent expansion given by
Lem.\ \ref{expand sqrtJ}. By Cor.\ \ref{cor 2.14} the domain of
absolute convergence of the weight expansion (\ref{weight
expansion}) of $\chi$ contains $T\,$. Thus one may insert
(\ref{weight expansion}) to obtain
\begin{equation}\label{eq 4.26}
    \forall \ell \in \mathbb{N}: \quad 0 = \sum A_b
    \sum\nolimits_{\gamma} B_\gamma \, \mathrm{e}^{-\delta
    \circ \ln} D_\ell \left(\mathrm{e}^{(\delta + \gamma
    - \sum b_i \sigma_i) \circ \ln } \right) \;,
\end{equation}
where the $b$-sum is over $b_i \ge 0\,$.  The statement to be
proved now is that the solution of these equations for the
unknowns $B_\gamma$ is unique (assuming that a solution exists).

Interchanging the sum over $\gamma \in \Gamma_\lambda$ with the sum
over the $b_i$'s and making the substitution $\gamma \to \gamma +
\sum b_i \sigma_i\,$, one reorganizes the system (\ref{eq 4.26}) as
\begin{equation}\label{eq 4.30p}
    \forall \ell \in \mathbb{N}: \quad 0 = \sum\nolimits_{\gamma}
    \left( \sum\nolimits_{\sum b_i \ge 0} A_b B_{\gamma + \sum b_i
    \sigma_i} \right) \mathrm{e}^{- \delta \circ \ln} D_\ell
    \left(\mathrm{e}^{(\delta + \gamma) \circ \ln } \right) \;.
\end{equation}
By standard reasoning the functions $\mathrm{e}^{(\delta + \gamma)
\circ \ln}$ which appear in the outer sum, all are linearly
independent of each other. They are eigenfunctions of the operators
$D_\ell\,$. The inner sum is a finite sum because the degree of the
summation variable $\gamma^{\,\prime} \equiv \gamma + \sum b_i
\sigma_i$ decreases with increasing values of the coefficients $b_i$
and the multiplicity $B_{\gamma^{\,\prime}}$ vanishes when this
degree is less than the degree of the highest weight $\lambda_N\,$.

The plan then is to solve the system of equations (\ref{eq 4.26}) for
the unknowns $B_\gamma$ recursively, by starting from $B_{\lambda_N}
=1$ and proceeding in ascending order of the total degree of $\gamma
\in \Gamma_\lambda\,$. An unknown multiplicity $B_\gamma$ will be
uniquely determined by the recursive scheme (\ref{eq 4.30p}) if the
function $\mathrm {e}^{\,(\delta + \gamma)\circ\ln}$ does not lie in
the joint kernel of the differential operators $D_\ell\,$. Indeed, in
that case one deduces from (\ref{eq 4.30p}) the equation
\begin{equation}\label{determine B}
    0 = A_\mathbf{0} B_\gamma + \sum\nolimits_{\sum b_i \ge 1}
    A_b \, B_{\gamma + \sum b_i \sigma_i}\;,
\end{equation}
and since $A_{b = \mathbf{0}} \not= 0\,$, this determines $B_\gamma$
from the already known $B_{\tilde\gamma}$ for the weights
$\tilde{\gamma} = \gamma + \sum b_i \sigma_i$ with $\sum b_i \ge 1$,
which are of lower degree than the weight $\gamma$ in question.

It remains to show that there is no indeterminacy from solutions of
$D_\ell (\mathrm{e}^{ (\delta + \gamma)\circ \ln}) = 0$ for $\gamma
\in \Gamma_\lambda\,$. The half sum of positive roots is readily
computed to be
\begin{displaymath}
    \delta = \sum\nolimits_{k=1}^{p+q} (k - p -
    {\textstyle{\frac{1}{2}}}) \mathrm{i} \psi_k +
    \sum\nolimits_{j=1}^p (j - {\textstyle{\frac{1}{2}}})
    \phi_j + \sum\nolimits_{l=p+1}^{p+q} (l - p - q -
    {\textstyle{\frac{1}{2}}}) \phi_l \;,
\end{displaymath}
and adding $\gamma \in \Gamma_\lambda$ one gets from (\ref{restr
range}) that $\delta + \gamma = \sum_k (\mathrm{i} \tilde{m}_k
\psi_k - \tilde{n}_k \phi_k)$ with
\begin{eqnarray*}
    \tilde{n}_{p-j+1} &\le& j - p - {\textstyle{\frac{1}{2}}}
    \quad (j = 1,\ldots, p) \;,\\ k - p - {\textstyle{\frac{1}{2}}}
    &\le& \tilde{m}_k \,\,\, \le \,\,\, N+k-p-{\textstyle{
    \frac{1}{2}}} \quad (k = 1, \ldots, p+q) \;, \\
    N + l - p - {\textstyle{\frac{1}{2}}} &\le&
    \tilde{n}_{2p+q-l+1} \quad (l = p+1, \ldots, p+q) \;.
\end{eqnarray*}
By a simple induction one proves that, given these inequalities,
there exists just one way to arrange for the $\tilde{m}_k$ to be a
permutation of the $\tilde{n}_k\,$. This unique arrangement occurs
when the inequalities are saturated as $\tilde{m}_j = \tilde{n}_{p
-j+1} = j-p+\frac{1}{2}$ (for $j = 1, \ldots, p$) and $\tilde {m}_l =
\tilde{n}_{2p+q-l+1} = N + l - p - \frac{1}{2}$ (for $l = p+1,
\ldots, p+q$). Returning from $\gamma + \delta$ to $\gamma$ we see
that these values of the coefficients $\tilde{m}_k$ and $\tilde{n}_k$
correspond to the coefficients $m_k$ and $n_k$ of the highest weight
$\lambda_N\,$.

By Lem.\ \ref{unique} it follows that $\gamma = \lambda_N$ is the
only solution (for $\gamma \in \Gamma_\lambda$) of the system of
equations $D_\ell (\mathrm{e}^{ (\delta + \gamma)\circ \ln}) = 0$ for
all $\ell \in \mathbb{N}$. Thus there is no indeterminacy and the
weight expansion (\ref{weight expansion}) for $\chi$ has now been
shown to be uniquely determined.
\end{proof}

\subsubsection{Explicit solution}

As a preparation for writing down and establishing a solution of
the conditions (A-C), we trisect the system of positive roots
$\Delta^+$.

Recall from Sect.\ \ref{sect:weights} the decomposition
\begin{displaymath}
    \mathrm{Hom}(U^+ , U^-) \oplus \mathfrak{gl}(U^+) \oplus
    \mathfrak{gl}(U^-) \oplus \mathrm{Hom}(U^- , U^+) \hookrightarrow
    \mathfrak{gl}(U) \;.
\end{displaymath}
$\Delta_\lambda^+$ was defined in (\ref{eq 2.40}) as the set of
positive roots $\alpha$ with the property that the root space
$\mathfrak{g}^{- \alpha}$ lies in the space of degree-increasing
operators $\mathfrak{g}^{(2)} = \mathrm{Hom}(U^-,U^+) \subset
\mathfrak{n}^{-}$; these are the roots listed in (\ref{eq 2.41}). The
complementary set $\Delta^+ \setminus \Delta_\lambda^+$ splits into
two subsets associated with $\mathfrak {gl}(U^+) =
\mathfrak{gl}_{p|p}$ and $\mathfrak{gl}(U^-) = \mathfrak{gl}_{q|q}:$
\begin{equation}
    \Delta_p^+ = \{ \alpha \in \Delta^+ \mid
    \mathfrak{g}^\alpha \subset \mathfrak{gl}(U^+) \} \;,
    \quad \Delta_q^+ = \{ \alpha \in \Delta^+ \mid
    \mathfrak{g}^\alpha \subset \mathfrak{gl}(U^-) \} \;.
\end{equation}
The first set $\Delta_p^+$ is made from the sequence
\begin{displaymath}
    \phi_1 \, , \ldots, \phi_p \, , \mathrm{i}\psi_1 \, ,
    \ldots, \mathrm{i} \psi_p \;,
\end{displaymath}
by taking differences $x - y$ with $x$ occurring later in the
sequence than $y\,$, just like in (\ref{order coords}). The other
set, $\Delta_q^+ \,$, is obtained by doing the same with the
sequence
\begin{displaymath}
    \mathrm{i}\psi_{p+1} \, , \ldots, \mathrm{i}\psi_{p+q} \, ,
    \phi_{p+1} \, , \ldots, \phi_{p+q} \;.
\end{displaymath}
Thus we can express the system of positive roots $\Delta^+$ as a
disjoint union of three sets:
\begin{equation}
    \Delta^+ = \Delta_p^+ \cup \Delta_q^+ \cup \Delta_\lambda^+
    \;.
\end{equation}

Let the corresponding factorization of the function $J$ of
(\ref{universal J}) be written as
\begin{equation}
    J = J_p \, J_q \, Z^{-2} \;.
\end{equation}
If $\Delta_p^+ = \Delta_{p,0}^+ \cup \Delta_{p,1}^+$ is the
decomposition into even and odd roots, the first factor is
\begin{equation}
    J_p = \frac{ \prod_{\alpha \in \Delta_{p,0}^+}
    \sinh^2(\frac{1}{2} \alpha)} {\prod_{\beta \in \Delta_{p,1}^+}
    \sinh^2(\frac{1}{2} \beta)} \circ \ln \;,
\end{equation}
and the second factor, $J_q\,$, is analogous. The third factor is the
inverse square of
\begin{equation}\label{define Z(t)}
    Z = \frac{ \prod_{\beta \in \Delta_{\lambda,1}^+}
    \sinh(\frac{1}{2} \beta)} {\prod_{\alpha \in \Delta_{
    \lambda,0}^+} \sinh(\frac{1}{2} \alpha)} \circ \ln =
    \prod_{j=1}^p \prod_{l=p+1}^{p+q} \frac{(1 -
    \mathrm{e}^{\mathrm{i}\psi_j - \phi_l})
    (1 - \mathrm{e}^{\phi_j - \mathrm{i}\psi_l})}{(1 -
    \mathrm{e}^{\mathrm{i}\psi_j - \mathrm{i}\psi_l})
    (1 - \mathrm{e}^{\phi_j - \phi_l})} \circ \ln \;.
\end{equation}
All of these functions exist on the subset of \emph{regular
elements} $T^\prime \subset T$.

Recall from Sect.\ \ref{sect:weyl-inv} that Weyl group elements $w
\in W$ act primarily on the torus $T$ by $t \mapsto w \cdot t \,$;
this induces an action on functions $f :\, T \to \mathbb{C}$ by
$w(f)(t) = f(w^{-1} \cdot t)$. We also have induced actions on the
tangent vector space $\mathfrak{h}_\mathbb{R}\supset \mathfrak {t}$
and on linear functions on $\mathfrak{h}_\mathbb{R}\,$. Let
$W_\lambda \subset W$ be the subgroup that stabilizes the highest
weight $\lambda_N\,$.

If $f: \, T \to \mathbb{C}$ is $W_\lambda$-invariant, define the
$W$-symmetrized function $S_W f$ by
\begin{displaymath}
    S_W f(t) := \sum_{[w] \in W / W_\lambda} f(w^{-1} \cdot t) \;.
\end{displaymath}
Note that since the set $\Delta_\lambda^+$ is invariant under the
induced action of $W_\lambda\,$, so is the function $Z$
constructed from it.

\begin{lem}\label{property C}
The function $\chi : T^\prime \to \mathbb{C}$ defined by
\begin{displaymath}
    \chi(t) = S_W \chi_\lambda(t) \;, \quad \chi_\lambda(t)
    = \mathrm{e}^{\lambda_N (\ln t)} Z(t) \;,
\end{displaymath}
is a solution of the system of differential equations $J^{-1/2}
D_\ell( J^{1/2} \chi) = 0$ (property C).
\end{lem}
\begin{proof}
Take the square root $J^{1/2} = J_p^{1/2} J_q^{1/2} Z^{-1}$. Then
notice that the differential operators $D_\ell\,$, which are sums of
powers of partial derivatives, split according to $\mathfrak{g}^{(0)}
= \mathfrak{gl}(U^+) \oplus \mathfrak{gl}(U^-) = \mathfrak{gl}_{p|p}
\oplus \mathfrak{gl}_{q|q}$ as $D_\ell = D_{\ell}^+ + D_{\ell}^-$, so
that
\begin{displaymath}
    J_p^{-1/2} J_q^{-1/2} D_\ell \circ J_p^{1/2} J_q^{1/2} =
    J_p^{-1/2} D_{\ell}^+ \circ J_p^{1/2} +
    J_q^{-1/2} D_{\ell}^- \circ J_q^{1/2} \;.
\end{displaymath}
Since the $D_\ell$ are $W$-invariant and commute with the
symmetrizer $S_W$, one has
\begin{displaymath}
    J^{-1/2} D_\ell (J^{1/2} \chi) = J^{-1/2} S_W D_\ell (J_p^{1/2}
    J_q^{1/2} \mathrm{e}^{\lambda_N \circ \ln} ) \;.
\end{displaymath}
With these identities in place, in order to establish property C it
is sufficient to show that the function $t \mapsto \mathrm{e}^{
\lambda_N(\ln t)}$ obeys the system of differential equations
\begin{displaymath}
    J_p^{-1/2} D_\ell^+ (J_p^{1/2} \mathrm{e}^{\lambda_N \circ \ln}) =
    J_q^{-1/2} D_\ell^- (J_q^{1/2} \mathrm{e}^{\lambda_N \circ \ln})
    = 0 \quad (\ell \in \mathbb{N}) \;.
\end{displaymath}
But these follow in turn from Cor.\ \ref{cor:4.12}. Indeed,
$\mathrm{e}^{ \lambda_N \circ \ln}$ with $\lambda_N = N \sum_{l =
p+1}^{p+q} (\mathrm{i}\psi_l - \phi_l)$ is the primitive character of
$\mathrm{GL}_{p|p} \times \mathrm{GL}_{q|q}$ which is given by the
trivial representation of $\mathrm{GL}_{p|p}$ and the $\mathrm
{SDet}^{-N}$-representation of $\mathrm{GL}_{q|q}\,$. Both of these
representations are one-dimensional and thus of unequal even and odd
dimension ($\mathrm{dim}\, V_0 = 1$ and $\mathrm{dim}\, V_1 = 0$).
Hence Cor.\ \ref{cor:4.12} applies, and property C is now
established.
\end{proof}
Next, we address the question whether the function $\chi$ defined
in Lem.\ \ref{property C} possesses the property A of our list.
For that, we write the function $Z$ of (\ref{define Z(t)}) in the
form
\begin{displaymath}
    Z = \prod_{j=1}^p \prod_{l=p+1}^{p+q} \frac{\sinh\big(
    \frac{1}{2} (\phi_l-\mathrm{i}\psi_j)\big)\,\sinh\big(
    \frac{1}{2} (\mathrm{i}\psi_l-\phi_j)\big)}{\sinh\big(
    \frac{1}{2} (\mathrm{i}\psi_l-\mathrm{i}\psi_j)\big)\,
    \sinh\big( \frac{1}{2} (\phi_l-\phi_j)\big)}\circ \ln \;.
\end{displaymath}
Clearly, in the domain $T$, where $\phi_j < 0 < \phi_l\,$, the
function $\mathrm{e}^{\lambda_N \circ \ln}\, Z$ and all of its
$W$-translates are analytic in the $\phi$-variables. $Z$, however,
has poles at $\psi_l = \psi_j$ (mod $2\pi\mathrm{i}$).

The issue at stake now is what happens with these singularities under
sym\-metrization by the Weyl group $W$. To answer that, let us adopt
the convention that $j,j^\prime, j^{\prime\prime} \in \{ 1, \ldots, p
\}$ and $l,l^\prime, l^{\prime \prime} \in \{ p+1, \ldots, p+q \}$.
Then, fixing any pair $j,l\,$, let $T_{jl} \subset T$ be the zero
locus of the factor $\sin\big( \frac{1}{2}(\psi_l - \psi_j)\big)$ of
the denominator of $Z\,$, excluding the zero loci of other factors.
If $w \in W$, there exist two types of outcome for the action of $w$
on the corresponding functions $\psi_j\, , \psi_l\,$. Either we have
one of the two situations
\begin{displaymath}
    w(\psi_j) = \psi_{j^\prime}\;, \,\, w(\psi_l) = \psi_{l^\prime}
    \;, \quad \text{or} \quad w(\psi_j) = \psi_{l^\prime} \;, \,\,
    w(\psi_l) = \psi_{j^\prime} \;,
\end{displaymath}
for some pair $j^\prime, l^\prime\,$, in which case the singular
factor $\sin\big(\frac{1}{2}(\psi_l - \psi_j)\big)$ still occurs
in the denominator of the transformed function $w(Z)$; or else we
have
\begin{displaymath}
    w(\psi_j) = \psi_{j^\prime} \;, \,\, w(\psi_l) = \psi_{j^{
    \prime\prime}}\;,\quad\text{or}\quad w(\psi_j) = \psi_{l^\prime}
    \;, \,\, w(\psi_l) = \psi_{l^{\prime\prime}} \;,
\end{displaymath}
in which case the singular factor is absent and $w(Z)$ is obviously
well-behaved on $T_{jl}\,$. Thus we may organize the sum over
$W$-translates for $\chi$ in two parts:
\begin{equation}\label{reg and sing}
    \chi = \frac{A^{(jl)}}{\sin\big(\frac{1}{2} (\psi_l - \psi_j)
    \circ \ln \big)} + B^{(jl)} \;,
\end{equation}
where $B^{(jl)}$ is the sum of all terms that are non-singular on
$T_{jl}\,$.

Now observe that there exists an element $w_{jl} \in W$ with the
property that its action interchanges $\psi_j$ with $\psi_l$ and
leaves all other $\psi_k$ the same.  This $w_{jl}$ clearly
preserves the organization of $\chi$ into singular and
non-singular terms. Since $\chi$ is $W$-invariant, we have $w_{jl}
( B^{(jl)}) = B^{(jl)}$ and, since $\sin\big( \frac{1}{2} (\psi_l
- \psi_j)\big)$ is $w_{jl}$-odd, it follows that
\begin{displaymath}
    w_{jl} \, (A^{(jl)}) = - A^{(jl)} \;.
\end{displaymath}
Given the expression for $Z$, this means that $A^{(jl)}$ vanishes
at least linearly on $T_{jl}\,$. Thus the apparent pole is
cancelled and both summands in (\ref{reg and sing}) remain finite
there.

Simultaneous poles from several factors of the denominator $\prod
\sin\big(\frac{1}{2}(\psi_l - \psi_j)\big)$ are handled in the
same way. Thus we conclude that the $W$-invariant function $\chi
:\, T^\prime \to \mathbb{C}$ defined in Lem.\ \ref{property C}
continues to an analytic function $\chi :\, T \to \mathbb{C}$.
\begin{prop}\label{soln exists}
The analytic function $\chi :\, T \to \mathbb{C}$ which is given
by $\chi = S_W \chi_\lambda$ and $\chi_\lambda(t) = \mathrm{e}^{
\lambda_N (\ln t)} Z(t)$ is a solution to the problem posed by the
conditions (A-C).
\end{prop}
\begin{proof}
It remains to prove that the proposed solution $\chi$ has property
B, i.e., if $\chi$ is expanded as $\chi(t) = \sum_\gamma B_\gamma
\, \mathrm{e}^{\gamma(\ln t)}$, then all of the weights $\gamma =
\sum_k (\mathrm{i}m_k \psi_k - n_k \phi_k)$ occurring in this
expansion have integer coefficients in the range
\begin{displaymath}
    n_j \le 0 \le m_k \le N \le n_l \;,
\end{displaymath}
and $B_{\lambda_N} = 1$. We still refer to the linear functions
$\gamma$ as "weights", even though in the present context we must
not use the fact that they are the weights of a representation.

The first step is to expand the function $Z$ by exponentials. For
that purpose, we add some real constants to the imaginary
parameters $\mathrm{i} \psi_k$ and define a positive chamber
$\tilde{\mathfrak{t}}^+ \subset \mathfrak{h}$ of real dimension
$2(p+q)$ by
\begin{displaymath}
    \tilde{\mathfrak{t}}^+ : \quad \phi_1 < \ldots < \phi_p <
    \mathfrak{Re}\, (\mathrm{i}\psi_1) < \ldots < \mathfrak{Re}\,
    (\mathrm{i}\psi_{p+q}) < \phi_{p+1} < \ldots < \phi_{p+q} \;.
\end{displaymath}
Note that if $H \in \tilde{\mathfrak{t}}^+$ then $\mathfrak{Re}\,
\alpha(H) > 0$ for all $\alpha \in \Delta^+$. Put $\tilde{T}_+ :=
\exp \tilde{\mathfrak{t}}^+\,$.

Now rewrite $Z$ from (\ref{define Z(t)}) as
\begin{displaymath}
   Z = \frac{ \prod_{\beta \in \Delta_{\lambda,1}^+}
   (1 - \mathrm{e}^{-\beta})} {\prod_{\alpha \in \Delta_{
    \lambda,0}^+} (1 - \mathrm{e}^{-\alpha})} \circ \ln \;.
\end{displaymath}
Since $\Delta_{\lambda,0}^+ \subset \Delta^+$, and $|\mathrm{e}^{-
\alpha(H)}| < 1$ for $H \in \tilde{\mathfrak{t}}^+$ and $\alpha \in
\Delta^+$, each factor $(1 - \mathrm{e}^{-\alpha \circ \ln} )^{-1}$
of $\chi_\lambda = \mathrm{e}^{\lambda_N \circ \ln}\, Z$ can be
expanded on $\tilde{T}_+$ as a convergent geometric series.

A slightly modified expansion procedure works for all of the
$W$-translates $w(\chi_\lambda)$: if a root $w(\alpha)$ for $\alpha
\in \Delta_{\lambda ,0}^+$ is positive, we expand as before; if it is
not, we first rearrange
\begin{displaymath}
    (1 - \mathrm{e}^{-w(\alpha)})^{-1} = - \mathrm{e}^{w(\alpha)}
    (1 - \mathrm{e}^{w(\alpha)})^{-1} \;,
\end{displaymath}
and then expand. In this way, we produce for $\chi = S_W
\chi_\lambda$ an expansion
\begin{equation}\label{eq 4.45}
    \chi(t) = \sum\nolimits_\gamma B_\gamma \,
    \mathrm{e}^{\gamma(\ln t)} \;, \quad \gamma =
    \sum\nolimits_k (\mathrm{i}m_k \psi_k - n_k \phi_k) \;,
\end{equation}
with integers $m_k$ and $n_k\,$, which is absolutely convergent on
$\tilde{T}_+\,$.

In the case of the coefficients $n_k\,$, since the action of the Weyl
group $W$ preserves the inequalities $\phi_j < 0 < \phi_l\,$, the
desired restriction on the range is obvious from $\lambda_N = N \sum
(\mathrm{i}\psi_l - \phi_l)$ and the system of $\lambda _N$-positive
roots $\Delta_{\lambda, 0}^+ \,$. What needs to be proved, though, is
the inequality for the coefficients $m_k :$
\begin{equation}\label{range of m_k}
    0 \le m_k \le N \quad (1 \le k \le p+q) \;.
\end{equation}

To that end, notice first of all that the set of $(p+q)$-tuples
$(m_1, \ldots, m_{p+q})$ occurring in (\ref{eq 4.45}) certainly is a
\emph{bounded} subset, say $D_{\lambda_N}\,$, of $\mathbb{Z}^{p+q}$.
Indeed, if this was not so, then there would be an immediate
contradiction with the known analyticity of $\chi$ when any one of
the factors $\mathrm{e}^{\mathrm{i}\psi_j - \mathrm{i} \psi_l}$
passes through unity. (In other words, our expansion by exponentials
must remain convergent when we let $\tilde{t} \in \tilde{T}_+$ tend
to $t \in T$.)

Although sharp bounds on $D_{\lambda_N}$ are not easy to establish
directly, the following statement is immediate: $D_{\lambda_N}$
lies in the intersection of the sets of $m_1 \ge 0$ and $m_{p+q}
\le N\,$. Indeed, writing the proposed solution $\chi$ as
\begin{equation}\label{propose soln}
    \chi = \sum_{[w] \in W/W_\lambda} \mathrm{e}^{w(\lambda_N)}
    \prod_{j = 1}^p \, \prod_{l = p+1}^{p+q} \frac{ (1 -
    \mathrm{e}^{\phi_j - \mathrm{i} w(\psi_l)})\, (1 -
    \mathrm{e}^{\mathrm{i} w(\psi_j) - \phi_l})} {(1 -
    \mathrm{e}^{\mathrm{i} w(\psi_j) - \mathrm{i} w(\psi_l)})
    \, (1 - \mathrm{e}^{\phi_j - \phi_l})} \circ \ln \;,
\end{equation}
and inspecting the terms in the sum over $W$-translates where
$w(\psi_l) \not= \psi_1$ for all $l = p+1, \ldots, p+q\,$, it is
clear that negative powers of $\mathrm{e}^{\mathrm{i}\psi_1}$ never
arise from such terms in our expansion.  For the summands where
$w(\psi_l) = \psi_1$ does occur for some $l\,$, we write
\begin{displaymath}
    \prod_j \frac{1 - \mathrm{e}^{\phi_j - \mathrm{i}w(\psi_l)}}
    {1-\mathrm{e}^{\mathrm{i}w(\psi_j) - \mathrm{i} w(\psi_l)}}
    = \prod_j \mathrm{e}^{- \mathrm{i} w(\psi_j)} \frac{
    \mathrm{e}^{\phi_j} - \mathrm{e}^{\mathrm{i}\psi_1}}
    {1 - \mathrm{e}^{\mathrm{i}\psi_1 - \mathrm{i}w(\psi_j)}} \;,
\end{displaymath}
and since the roots $\mathrm{i}w(\psi_j) - \mathrm{i}\psi_1$ are in
$\Delta^+$, the new denominators are ready for expansion, and we see
that the dependence on $\psi_1$ is always $\mathrm{e}^{\mathrm{i}m_1
\psi_1}$ with $m_1 \ge 0\,$.

Similarly, the inequality $m_{p+q} \le N$ is obvious, with the
exception of the terms of the $[w]$-sum where $w(\psi_j) = \psi_{p
+ q}$ for some $j \in \{ 1, \ldots, p\}$. But in those cases we
write
\begin{displaymath}
    \prod_l \frac{1 - \mathrm{e}^{\mathrm{i}w(\psi_j) - \phi_l}}
    {1-\mathrm{e}^{\mathrm{i}w(\psi_j) - \mathrm{i}w(\psi_l)}} =
    \prod_l \mathrm{e}^{\mathrm{i}w(\psi_l)} \frac{\mathrm{e}^{-
    \phi_l}- \mathrm{e}^{-\mathrm{i}\psi_{p+q}}}{1 - \mathrm{e}^{
    \mathrm{i}w(\psi_l) - \mathrm{i}\psi_{p+q}}} \;,
\end{displaymath}
where all denominators are again of the form $(1 - \mathrm {e}^{-
\alpha})^{-1}$ with $\alpha \in \Delta^+$.  Because $\mathrm{e}^{
\lambda_N}$ contains a factor $\mathrm{e}^{ \mathrm{i} N \psi_{p+q}}$
we cannot infer $m_{p+q} \le 0$, but it does follow that $m_{p+q} \le
N\,$.

Thus we have established that $D_{\lambda_N}$ does not intersect
either of the sets $m_1 < 0$ or $m_{p+q} > N\,$. This of course still
leaves a set much bigger than the desired one of (\ref{range of
m_k}).

We now \emph{repeat}, however, the whole procedure for \emph{another}
choice of highest weight $w(\lambda_N) \not= \lambda_N$ ($w \in W$).
Thus, replacing the system of positive roots $\Delta^+$ by a system
of positive roots $w(\Delta^+)$, and the domain $\tilde{T}_+$ by a
\emph{transformed} domain $w(\tilde{T}_+)$, we produce another
expansion for $\chi$ by exponentials where the $(p+q)$-tupels of
integers $(m_1 , \ldots, m_{p+q})$ in (\ref{eq 4.45}) are another set
$D_{w(\lambda_N)} \subset \mathbb{Z}^{p+q}$. There exists such a
choice of highest weight $w_{w(\lambda_N)}$ that the roles formerly
played by $\psi_1$ and $\psi_{p+q}$ are now played by $\psi_2$ resp.\
$\psi_{p+q-1}\,$. With this particular choice, by the same argument
as before, the set $D_{w(\lambda_N)}$ has zero intersection with $m_2
< 0$ and with $m_{p+q-1}
> N\,$.

A priori, the second expansion could be different from the first one.
However, since the two expansions represent the same analytic
function and both sets $D_{\lambda_N}$ and $D_{w(\lambda_N)}$ are
bounded, they must actually coincide: $D_{\lambda_N} = D_{w
(\lambda_N)}$. It follows that the integers $m_k$ of our two
identical expansions have zero intersection with each of the sets
$m_1 < 0\,$, $m_2 < 0\,$, $m_{p+q-1} > N$, and $m_{p+q}
> N\,$.

By continuing in this fashion, one eventually cuts down the range of
the integers $m_k$ to that of (\ref{range of m_k}).  It is easy to
see that $B_{\lambda_N} = 1$.  This completes the proof.
\end{proof}
{}From (\ref{propose soln}) our solution to the problem posed by
the conditions (A-C) coincides with the one stated in Thm.\
\ref{thm0}. Since the solution is unique by Lem.\ \ref{unique},
the proof of that theorem is now finished but for the need to
prove Prop.\ \ref{prop 4.2}.

\subsection{Proof of Cor.\ \ref{cor 1.2}}
\label{sect:cor1.2}

The claim made in Cor.\ \ref{cor 1.2} is that
\begin{eqnarray}
    &&\int\limits_{\mathrm{U}_N} \frac{\prod_{j=1}^p
    \mathrm{Det}(\mathrm{Id}_N - \mathrm{e}^{\mathrm{i}
    \psi_j}\,u)\, \prod_{l=p+1}^{p+q} \mathrm{Det}
    (\mathrm{Id}_N - \mathrm{e}^{-\mathrm{i}\psi_l}\,
    \bar{u})}{\prod_{j^\prime = 1}^{p^\prime}\mathrm{Det}
    (\mathrm{Id}_N - \mathrm{e}^{\phi_{j^\prime}} u)\,
    \prod_{l^\prime = p^\prime +1}^{p^\prime + q^\prime}
    \mathrm{Det}(\mathrm{Id}_N-\mathrm{e}^{-\phi_{l^\prime}}
    \,\bar{u})} \, du \label{eq 4.41} \\ &=& \frac{1}{p! \, q!}
    \sum_{w \in \mathrm{S}_{p+q}}\,\,\prod_{k=p+1}^{p+q}
    \frac{\mathrm{e}^{N\,\mathrm{i} w(\psi_k)}}
    {\mathrm{e}^{N\,\mathrm{i} \psi_k}} \times
    \, \frac{\prod_{j^\prime,l}(1 - \mathrm{e}^{\phi_{j^\prime}
    - \mathrm{i} w(\psi_l)}) \, \prod_{j,l^\prime}
    (1 - \mathrm{e}^{\mathrm{i} w(\psi_j) - \phi_{l^\prime}})}
    {\prod_{j,l} (1-\mathrm{e}^{\mathrm{i} w(\psi_j) -
    \mathrm{i}w(\psi_l)}) \, \prod_{j^\prime , l^\prime}
    (1 - \mathrm{e}^{\phi_{j^\prime} - \phi_{l^\prime}})}\nonumber
\end{eqnarray}
holds as long as $p^\prime \le p + N$ and $q^\prime \le q + N\,$.

This formula is deduced from (\ref{propose soln}) by the method of
induction as follows. To start the induction process, note that if $p
= p^\prime$ and $q = q^\prime$, then (\ref{eq 4.41}) is just
(\ref{propose soln}) restated by dividing both sides by
$\mathrm{e}^{\lambda_N}$ and replacing the sum over cosets $\sum_{[w]
\in W / W_\lambda}$ by the normalized sum over group elements $(p! \,
q!)^{-1} \sum_{w \in \mathrm{S}_{p+q}}$.

Assume now that (\ref{eq 4.41}) holds true for some set of
non-negative integers $p$, $p^\prime$, $q$, and $q^\prime$ in the
specified range. Then by taking $\mathrm{e}^{-\mathrm{i}
\psi_{p+q}}$ to zero, and assuming $q \ge 1$ and $q^\prime \le
(q-1) + N\,$, let us show that (\ref{eq 4.41}) still holds true
when $q$ is lowered to $q - 1\,$.

For this purpose, divide the sum over permutations  $w \in
\mathrm{S}_{p+q}$ into two partial sums which are set apart by
whether $w^{-1}(\psi_{p + q}) = \psi_j$ for $j \in \{ 1, \ldots, p
\}$ or $w^{-1}(\psi_{p + q}) = \psi_l$ for $l \in \{ p+1 , \ldots,
p+q \} \,$. In the former case the corresponding summand on the
right-hand side of (\ref{eq 4.41}) goes to zero in the limit of
$\mathrm{e}^{- \mathrm{i} \psi_{p+q}} \to 0\,$. Indeed, collecting
all factors containing $w(\psi_j) \equiv \psi_{p+q}$ one gets
\begin{displaymath}
    \frac{1}{\mathrm{e}^{N\, \mathrm{i} \psi_{p+q}}} \times
    \frac{\prod_{l^\prime}(1 - \mathrm{e}^{\mathrm{i} \psi_{p+q}
    - \phi_{l^\prime}})}{\prod_l (1 - \mathrm{e}^{\mathrm{i}
    \psi_{p+q} -w(\psi_l)})} \;,
\end{displaymath}
which scales as $\mathrm{e}^{-(N + q - q^\prime) \mathrm{i}
\psi_{p + q}}$ when $\mathrm{e}^{- \mathrm{i} \psi_{p+q}}$ is
small, and since $N + q - q^\prime \ge 1$ by assumption, the
contributions from such terms vanish when $\mathrm{e}^{-
\mathrm{i}\psi_{p+q}}$ is set to zero.

Now turn to the second case, where $w^{-1}(\psi_{p+q}) = \psi_l$ for
some $l \in \{ p+1 , \ldots, p+q \}$.  There are $q \times (p+q-1)!$
such permutations. By explicit invariance under the second factor of
$\mathrm{S}_p \times \mathrm{S}_q$ one may assume that $w(\psi_{p+q})
= \psi_{p+q}$ and make the replacement
\begin{displaymath}
    q!^{-1} \sum\nolimits_{w \in \mathrm{S}_{p+q}} \to
    (q-1)!^{-1} \sum\nolimits_{w \in \mathrm{S}_{p+q-1}}
\end{displaymath}
where $\mathrm{S}_{p+q-1} \subset \mathrm{S}_{p+q}$ is the subgroup
of permutations that fix $\psi_{p+q}\,$. In all of these terms the
product of factors containing the singular variable $\psi_{p + q}$ is
\begin{displaymath}
    \frac{\mathrm{e}^{N\, \mathrm{i} \psi_{p+q}}}{\mathrm{e}^{
    N\, \mathrm{i}\psi_{p+q}}} \times \frac{\prod_{j^\prime}
    (1 - \mathrm{e}^{\phi_{j^\prime} - \mathrm{i}\psi_{p+q}})}
    {\prod_j (1 - \mathrm{e}^{\mathrm{i} w(\psi_j) - \mathrm{i}
    \psi_{p+q}})} \;,
\end{displaymath}
which goes to unity when $\mathrm{e}^{-\mathrm{i} \psi_{p+q}}$ is set
to zero.  The remaining factors on the right-hand side of (\ref{eq
4.41}) reproduce the desired answer for the lowered value $q-1\,$. On
the left-hand side, setting $\mathrm{e}^{- \mathrm{i} \psi_{p+q}}$ to
zero just removes the factor $\mathrm{Det}(\mathrm{Id}_N - \mathrm{e}
^{-\mathrm{i} \psi_{p+q}}\, \bar{u} )$. Altogether this shows that
the induction step $q \to q-1$ is valid in the specified range.

In very much the same manner one establishes the validity of the
induction step of $p \to p-1$ for $p \ge 1 $ and $p^\prime \le
(p-1) + N\,$.

Finally, it is clear that one can always lower $p^\prime$ and
$q^\prime$ by sending $\mathrm{e}^{\phi_{j^\prime}} \to 0$ resp.\
$\mathrm{e}^{-\phi_{l^\prime}} \to 0$.  In this way (\ref{eq 4.41})
follows in the full range $p^\prime \le p + N$ and $q^\prime \le q +
N$.

\section{Proof of the extended character formula}
\label{sect:key}\setcounter{equation}{0}

Key to our determination of the character $\chi$ was property C (the
system of differential equations) which in turn resulted from the
statement that $\chi :\, T \to \mathbb{C}$ extends to a radial
section $\chi \in \mathcal{F}$ by Prop.\ \ref{prop 4.2}. The proof of
that proposition had to be postponed because of length; we return to
it in the current section.

To prepare for the situation of our Howe dual pair $(\mathfrak{gl}
_{n|n}\,, \mathrm{U}_N)$, we go back again to the basic setting
(Sect.\ \ref{sect:susyframe}) of a $\mathbb{Z}_2$-graded vector space
$V = V_1 \oplus V_0$ with Hermitian subspaces $V_0$ and $V_1$ and
orthogonal decompositions $V_\tau^{\vphantom{+}} = V_\tau^+ \oplus
V_\tau^-$ $(\tau = 0,1)$. Introducing $s := \mathrm{Id}_{ V_0^+}
\oplus (- \mathrm{Id}_{V_0^-})$ we associate with the pseudo-unitary
vector space $V_0^s \equiv (V_0\, , s)$ a non-compact Lie group
$\mathrm{U} (V_0^s)$ and a semigroup $\mathrm{H}(V_0^s)$ by
\begin{equation}\label{def H(V0s)}
    \mathrm{U}(V_0^s) = \{g \in \mathrm{GL}(V_0) \mid g^\dagger s g
    = s \} \;, \quad \mathrm{H}(V_0^s) = \{ g \in \mathrm{GL}(V_0)
    \mid g^\dagger s g < s \} \;.
\end{equation}
Note that $\mathrm{H}(V_0^s)$ is open in $\mathrm{GL}(V_0)$, and that
$\mathrm{U}(V_0^s) \subset \overline{\mathrm{H}(V_0^s)}$.

{}From Def.\ \ref{def 2.8} recall the definition of the
spinor-oscillator module $\mathcal{A}_V$ of $V$. We are now going to
construct representations $R$ and $R^\prime$ of the semigroup
$\mathrm{GL}(V_1) \times \mathrm{H}(V_0^s)$ and the Lie group
$\mathrm{U}(V_1) \times \mathrm{U}(V_0^s)$, which integrate the
infinitesimal representation $R_\ast : \,\mathfrak{gl}(V) \to
\mathfrak{gl} (\mathcal{A}_V)$ of Sect.\ \ref{sect:susyframe}. This,
in combination with some superanalysis, will eventually lead to a
proof of the character formula (\ref{needs ancestor}) of Prop.\
\ref{prop 4.2}.

Our first step is to construct $R : \, \mathrm{GL}(V_1) \times
\mathrm{H}(V_0^s) \to \mathrm{GL}(\mathcal{A}_V)$. For the first
factor this is straightforward. We start from the canonical
representation $\sigma : \, \mathrm{GL}(V_1) \to \mathrm{GL}(\wedge
V_1)$ by $\sigma (g) (v \wedge v^\prime \wedge \ldots) = (gv) \wedge
(gv^\prime) \wedge \ldots$. Then, choosing for $d = \mathrm{dim}
(V_1^-)$ a unitary generator $\Omega \in \wedge^d({V_1^-}^\ast)$  we
have a non-degenerate pairing $\wedge^\bullet (V_1^-)\otimes \wedge^{
d- \bullet} (V_1^-) \to \mathbb{C}$, $(a,b) \mapsto \Omega (a\wedge
b)$ and hence an isomorphism $\tau : \, \wedge(V_1^-) \to \wedge
({V_1^-}^\ast)$. This gives a $\mathrm{GL} (V_1)$-representation on
$\wedge (V_1^+) \otimes \wedge({V_1^-}^\ast)$ and hence on
$\mathcal{A}_V$. We denote it by $R_1 : \, \mathrm{GL}(V_1) \to
\mathrm{GL}(\mathcal{A}_V )$. Since $\sigma$ is unitary on
$\mathrm{U}(V_1)$, so is $R_1^\prime := R_1 \vert_{\mathrm{U}(V_1)}$.

Recall that $g \in \mathrm{GL}(V_1)$ acts on $v + \varphi \in V_1^
{\vphantom{\ast}} \oplus V_1^\ast$ by $g (v + \varphi) = gv + \varphi
\circ g^{-1}$. Recall also that in Def.\ \ref{def 2.8} a Clifford
action $w \mapsto \mathbf{c}(w)$ of $W_1^{\vphantom{\ast}} = V_1^{
\vphantom{\ast}} \oplus V_1^\ast$ on $\mathcal{A}_V$ was specified.
\begin{lem}\label{cor 4.6p}
The Clifford action $\mathbf{c}: \, W_1 \to \mathrm{End}(\mathcal
{A}_V)$ is $\mathrm{GL}(V_1)$-equivariant:
\begin{displaymath}
    R_1(g)^{-1} \mathbf{c}(w) R_1(g) = \mathbf{c}(g^{-1} w)\quad
    (\text{for all~} g \in \mathrm{GL}(V_1) \text{~and~} w \in W_1)\;.
\end{displaymath}
\end{lem}
\begin{proof}
This is an immediate consequence of the formula $\mathrm{Ad} \circ
\exp = \exp \circ \mathrm{ad}$ and the fact that $[ R_\ast(X) ,
\mathbf{c}(w) ] = \mathbf{c}(X \cdot w)$ for $X \in \mathfrak{gl}
(V_1) \hookrightarrow \mathfrak{gl}(V)$.
\end{proof}

\subsection{More about the semigroup $\mathrm{H}(V_0^s)\,$}

We now turn to the sector of the vector space $V_0^{\vphantom{\ast}}
= V_0^+ \oplus V_0^-$ with pseudo-unitary structure given by the
involution $s\,$. To construct a representation of the semigroup
$\mathrm{H}(V_0^s)$ on the oscillator module $\mathrm{S}(V_0^+ \oplus
{V_0^-}^\ast)$ some serious analytic issues have to be addressed. The
first prerequisite for that is Lem.\ \ref{lem 4.2p}, which provides
some needed information about $\mathrm {H}(V_0^s)$. In the statement
of that lemma and its proof, we succumb to the convenience of
denoting the unit operator on $V_0$ simply by $\mathrm{Id}_{V_0}
\equiv 1$. As usual, if $X \in \mathrm{End} (V_0)$, let $X \mapsto
X^\dagger$ be the operation of taking the adjoint with respect to the
Hermitian structure of $V_0\,$, and define $\mathfrak{Re}\, X =
\frac{1}{2}(X + X^\dagger)$.
\begin{lem}\label{lem 4.2p}
If $\zeta_{\,s}^+(V_0)$ denotes the set of complex linear
transformations
\begin{displaymath}
    \zeta_{\,s}^+(V_0) := \{ X \in \mathrm{End}(V_0) \mid
    \mathfrak{Re}\, X > 0 \;,\, \mathrm{Det}(X + s) \not= 0\} \;,
\end{displaymath}
the (Cayley-type) rational mapping
\begin{displaymath}
    a : \,\, \mathrm{H}(V_0^s) \to \mathrm{End}(V_0) \;,
    \quad h \mapsto s \, \frac{1+h}{1-h} = a_h
\end{displaymath}
is a bijection from the semigroup $\mathrm{H}(V_0^s)$ to
$\zeta_{\,s}^+(V_0)$.
\end{lem}
\begin{proof}
$h \in \mathrm{H}(V_0^s)$ cannot have an eigenvalue at unity, for
if there existed an eigenvector $\psi = h \psi$ then the defining
inequality $h^\dagger s h < s$ of $\mathrm{H}(V_0^s)$ would imply
\begin{displaymath}
    \langle \psi , s \psi \rangle = \langle h \psi , s h \psi
    \rangle = \langle \psi , h^\dagger s h \psi \rangle < \langle
    \psi , s \psi \rangle \;,
\end{displaymath}
which is a contradiction.  Thus $1-h$ is regular, and the map $a$
exists on $\mathrm{H}(V_0^s)$.

Adding and subtracting terms gives
\begin{displaymath}
    0 < s - h^\dagger s h = {\textstyle{\frac{1}{2}}}(1 - h^\dagger)
    s (1 + h) + {\textstyle{\frac{1}{2}}}(1 + h^\dagger)s(1 - h)\;.
\end{displaymath}
Multiplication by $(1-h^\dagger)^{-1}$ and $(1-h)^{-1}$ from the
left resp.\ right then results in
\begin{displaymath}
    0 < {\textstyle{\frac{1}{2}}}(a_h^{\vphantom{\dagger}} +
    a_h^\dagger)\;,
\end{displaymath}
which is the statement $\mathfrak{Re}\, a_h > 0$. If $X = a_h$ then
$X + s = 2s(1-h)^{-1}$ has an inverse and hence a non-vanishing
determinant. Thus $h \mapsto a_h$ maps $\mathrm{H} (V_0^s)$ into
$\zeta_{\,s}^+(V_0)$.

By the same token, the converse is also true: if $X \in
\zeta_{\,s}^+(V_0)$ then
\begin{displaymath}
    h = a^{-1}(X) = 1 - (X+s)^{-1} 2s = (X+s)^{-1} (X-s)
\end{displaymath}
satisfies $h^\dagger s h < s$ and lies in $\mathrm{H}(V_0^s)$. Thus
$a :\, \mathrm{H}(V_0^s) \to \zeta_{\,s}^+(V_0)$ is a bijection.
\end{proof}
\begin{cor}\label{cor 5.4}
The manifold $\mathrm{H}(V_0^s)$ is connected, and is closed under
$h \mapsto s h^\dagger s\,$.
\end{cor}
\begin{proof}
The space of all $X \in \mathrm{End}(V_0)$ with $\mathfrak{Re}\, X
> 0$ is convex and hence connected.  The connected property cannot
be lost by removing the solution set of $\mathrm{Det}(X + s) = 0$;
thus the space $\zeta_{\,s}^+(V_0)$ is connected. Clearly, the
bijection $a^{-1} :\, \zeta_{\,s}^+(V_0) \to \mathrm{H} (V_0^s)$ is a
continuous map, and it follows that $\mathrm {H}(V_0^s)$ is
connected.

By Lem.\ \ref{lem 4.2p}, if $h \in \mathrm{H}(V_0^s)$, there exists
$X \in \zeta_{\,s}^+(V_0)$ so that $h = 1 - (X + s)^{-1} 2s\,$. Take
the adjoint and conjugate by $s$ to get $s h^\dagger s = 1 -
(X^\dagger + s)^{-1} 2s\,$. Since $\zeta_{\,s}^+(V_0)$ is obviously
closed under $X \mapsto X^\dagger$, so is $\mathrm{H} (V_0^s)$ under
$h \mapsto s h^\dagger s \,$.
\end{proof}
\begin{rem}
It follows that if $h \in \mathrm{GL}(V_0)$ and $h^\dagger s h < s$
then one also has $h s h^\dagger < s\,$. In terms of the automorphism
$\sigma : \, \mathrm{GL}(V_0) \to \mathrm{GL}(V_0)$ of (\ref{def
sigma}), one can say that $\mathrm{H}( V_0^s)$ is closed under $h
\mapsto \sigma(h)^{-1}$. By setting $V_0^+ = \mathbb{C}^p$ and $V_0^-
= \mathbb{C}^q$ one gets the properties of the semigroup
$\mathrm{H}_{p,\,q}$ which were stated without proof in Sects.\
\ref{sect:P-to-M} and \ref{sect:maxtorus}.
\end{rem}

\subsection{Representation of $\mathrm{H}(V_0^s)$ and
$\mathrm{U}(V_0^s)$ on $\mathcal{A}_V$}\label{sect:G0-rep}

Our approach is inspired by Howe's construction \cite{howe2} of
the Shale-Weil representation of the metaplectic group via the
oscillator semigroup: we first construct a representation of the
semigroup $\mathrm{H} (V_0^s)$, and then pass to its closure to
obtain a representation of $\mathrm{U}(V_0^s) \subset
\overline{\mathrm{H}(V_0^s)}$.

To begin, if a vector $v \in V_0$ is written according to the
orthogonal decomposition $V_0^{\vphantom{+}} = V_0^+ \oplus V_0^-$ as
$v = v_+ + v_- \,$, we assign to it a linear operator $\mathsf{T}_v$
on $\mathcal{A}_V$ by
\begin{equation}\label{normal order}
    \mathsf{T}_v =
    \mathrm{e}^{\mathrm{i} \mu(v_+) + \mathrm{i}\delta(c\,v_+)} \,
    \mathrm{e}^{\mathrm{i}\delta(v_-) + \mathrm{i}\mu(c\,v_-)}\;.
\end{equation}
Since $\mathcal{A}_V$ is equipped with its canonical Hermitian
structure -- cf.\ Sect.\ \ref{sect:2.3} and (\ref{eq eps-iota}) -- in
which the relations $\mu(v_+)^\dagger = \delta(c\, v_+)$ and
$\delta(v_-)^\dagger = \mu(c\, v_-)$ hold, such operators are
unitary: ${\mathsf{T}_v}^\dagger = \mathsf{T}_{-v} =
{\mathsf{T}_v}^{-1}$. By a straightforward computation using the
canonical commutation relations (\ref{CCR}), one verifies the
composition law
\begin{equation}\label{heisenberg}
    \mathsf{T}_u \mathsf{T}_{v} = \mathsf{T}_{u + v} \, \mathrm{e}^{
    - \mathrm{i} \mathfrak{Im} \langle u ,\, s v \rangle} \;,
\end{equation}
where $\langle \, , \, \rangle$ is the Hermitian scalar product of
the Hermitian vector space $V_0\,$. (The $\mathsf {T}_v$ with this
law define a representation of the so-called Heisenberg group).

Now, with a rapidly decreasing function $F$ on $V_0$ associate an
operator $\mathrm{Op}(F)$ by
\begin{equation}
  \mathrm{Op}(F) = \int_{V_0} F(v) \mathsf{T}_v \,\, d\mathrm{vol}(v)
\end{equation}
where $d\mathrm{vol}(v)$ is Lebesgue measure on $V_0$ normalized by
$\int_{V_0} \mathrm{e}^{-\frac{1}{2} \langle v , v \rangle}
d\mathrm{vol}(v) = 1$. The multiplication law for such operators is
readily seen to be
\begin{equation}\label{op mult}
    \mathrm{Op}(F_1)\mathrm{Op}(F_2) =\mathrm{Op}(F_1 \sharp F_2)
\end{equation}
where $F_1 \sharp F_2$ is the convolution product twisted by the
multiplier in (\ref{heisenberg}):
\begin{equation}
    (F_1 \sharp F_2)(u) = \int_{V_0} F_1(u - v) F_2(v) \,
    \mathrm{e}^{-\mathrm{i} \mathfrak{Im} \langle u ,\, s v
    \rangle} \, d\mathrm{vol}(v) \;.
\end{equation}

The next step is to specialize this convolution product to a good
space of Gaussian functions on $V_0\,$.  Recall the definition of
the set $\zeta_{\,s}^+(V_0) \subset \mathrm{End} (V_0)$ from Lem.\
\ref{lem 4.2p}, and with every element $X \in \zeta_{\,s}^+(V_0)$
associate a Gaussian function $\gamma_X : \, V_0 \to \mathbb{C}$
by
\begin{equation}
    \gamma_X (v) = \mathrm{Det}(X + s) \, \mathrm{e}^{- \frac{1}{2}
    \langle v, \, X v \rangle} \;.
\end{equation}
If $X$ and $Y$ lie in $\zeta_{\,s}^+(V_0)$, then $\mathfrak{Re}(X+Y)
> 0\,$, so that the density
\begin{displaymath}
    v \mapsto | \gamma_X (u-v) \gamma_Y(v)|\, d\mathrm{vol}(v)
\end{displaymath}
decreases rapidly at infinity and the twisted convolution product
$\gamma_X \sharp \gamma_Y$ exists.

Doing the Gaussian convolution integral by completing the square,
we easily find
\begin{equation}\label{convolution}
  \gamma_X \sharp \gamma_Y = \gamma_{X \circ Y} \;,
\end{equation}
with the composition law $(X,Y) \mapsto X \circ Y$ given by
\begin{equation}\label{AcB}
    X \circ Y := X - (X-s) (X+Y)^{-1} (X+s)
    = (Y+s) (X+Y)^{-1} (X+s) - s\;.
\end{equation}

{}From Lem.\ \ref{lem 4.2p} recall now the bijective map (we again
abbreviate $\mathrm{Id}_{V_0} \equiv 1$)
\begin{displaymath}
    a : \,\, \mathrm{H}(V_0^s) \to \zeta_{\,s}^+(V_0) \;,
    \quad x \mapsto s \, \frac{1+x}{1-x} = a_x \;.
\end{displaymath}
\begin{lem}\label{lem 4.6p}
The mapping $\mathsf{R}:\, \mathrm{H}(V_0^s) \to \mathrm{End}
(\mathcal{A}_V)$ defined by
\begin{displaymath}
    x \mapsto \mathsf{R}(x) = \mathrm{Det}(a_x + s) \int_{V_0}
    \mathrm{e}^{-\frac{1}{2} \langle v , \, a_x v \rangle} \,
    \mathsf{T}_v \, d\mathrm{vol}(v)
\end{displaymath}
is a representation of the semigroup $\mathrm{H}(V_0^s)$ by
contractions, i.e., $\parallel \mathsf{R}(x)
\parallel_\mathrm{op}\, \le 1$.
\end{lem}
\begin{proof}
With respect to multiplication in $\mathrm{H}(V_0^s)$ the function $x
\mapsto a_x$ behaves as
\begin{eqnarray*}
    a_{xy} + s &=& 2s (1-xy)^{-1} = 2s \big( (1-x) + (1-y) -
    (1-x)(1-y) \big)^{-1} \\ &=& \big( (a_x + s)^{-1} +
    (a_y + s)^{-1} - (a_x + s)^{-1} 2 s (a_y + s)^{-1}
    \big)^{-1} \\ &=& (a_y + s) \, (a_x + a_y)^{-1} (a_x + s)
    = a_x \circ a_y + s \;,
\end{eqnarray*}
which is exactly the composition law (\ref{AcB}).  Thus the bijection
$a^{-1} :\, \zeta_{\,s}^+(V_0) \to \mathrm{H}(V_0^s)$ transforms the
complicated product (\ref{AcB}) into plain composition of linear
operators:
\begin{displaymath}
    a^{-1}(X \circ Y) = a^{-1}(X) \, a^{-1}(Y) \;.
\end{displaymath}

Setting $\mathsf{R}(x) := \mathrm{Op}(\gamma_{a_x})$ and using
(\ref{convolution}, \ref{op mult}) one gets
\begin{displaymath}
    \mathsf{R}(xy) = \mathrm{Op}(\gamma_{a_x \circ a_y}) =
    \mathrm{Op}(\gamma_{a_x} \sharp \gamma_{a_y}) =
    \mathrm{Op}(\gamma_{a_x}) \mathrm{Op}(\gamma_{a_y})
    = \mathsf{R}(x) \mathsf{R}(y) \;,
\end{displaymath}
which says that $x \mapsto \mathsf{R}(x)$ is a semigroup
representation.

For the proof of the contraction property we refer to \cite{howe2},
Chapter 15.
\end{proof}
Now recall that the group of pseudo-unitary transformations
$\mathrm{U}(V_0^s)$ defined in (\ref{def U(p,q)}) is found in the
closure $\overline{\mathrm{H}(V_0^s)}$.  Hence we can define
$\mathsf{R}^\prime(g)$ for $g \in \mathrm{U}(V_0^s)$ by taking the
limit of $\mathsf{R}(x_j)$ for any sequence $\{ x_j \}_{j \in \mathbb
{N}}$ in $\mathrm{H}(V_0^s)$ that converges to $g$.  (Convergence
here means convergence in a strong sense, namely w.r.t.\ the bounded
strong$^{\ast}$ topology. For the details of this argument we refer
to \cite{howe2}, Chapter 16.)

In particular, if a sequence $\{ x_j \}_{j \in \mathbb{N}}$ in
$\mathrm{H}(V_0^s)$ converges to an element $g$ of the open and
dense subset of $\mathrm{U}(V_0^s)$ where $1-g \in \mathrm{End}
(V_0)$ is regular, then using $\mathsf{T}_v^\dagger = \mathsf{T}
_{-v}^{\vphantom{\dagger}}$ and
\begin{displaymath}
    a_x^\dagger = (1-x^\dagger)^{-1} 2s - s =
    2s(1-sx^\dagger s)^{-1} - s = a_{s x^\dagger s} \;,
\end{displaymath}
one sees that the limit operator $\mathsf{R}^\prime(g) := \lim_{j
\to \infty} \mathsf{R}(x_j)$ has an adjoint which is
\begin{displaymath}
  \mathsf{R}^\prime(g)^\dagger = \lim_{j \to \infty}
  \mathsf{R}(s x_j^\dagger s) = \mathsf{R}^\prime(g^{-1}) \;,
\end{displaymath}
since $s g^\dagger s = g^{-1}$ for $g \in \mathrm{U}(V_0^s)$.
Moreover, if $\{ x_j \}$ and $\{ y_j \}$ are two sequences in
$\mathrm{H}(V_0^s)$ approaching $g$ resp.\ $h$ in $\mathrm{U}
(V_0^s)$, then the semigroup law $\mathsf{R}(x_j) \mathsf{R}(y_j)
= \mathsf{R}(x_j y_j)$ delivers the group law $\mathsf{R}^\prime
(g) \mathsf{R}^\prime(h) = \mathsf{R}^\prime(gh)$ by continuity of
the limit.

We thus arrive at the following statement.
\begin{lem}\label{lem 4.7p}
Taking the limit (where still $\epsilon \cdot 1 \equiv \epsilon \,
\mathrm{Id}_{V_0}$)
\begin{displaymath}
  \mathsf{R}^\prime(g) = \lim_{\epsilon\to 0+} \mathrm{Det}
  (a_g + s + \epsilon \cdot 1) \int_{V_0} \mathrm{e}^{-
  \frac{1}{2} \langle v , (a_g + \epsilon \cdot 1) v \rangle}
  \mathsf{T}_v \,\, d\mathrm{vol}(v)
\end{displaymath}
yields a unitary representation $\mathsf{R}^\prime :\, \mathrm{U}
(V_0^s) \to \mathrm {U}(\mathcal{A}_V)$.  This representation is
compatible with the semigroup representation $\mathsf{R} : \,
\mathrm{H}(V_0^s) \to \mathrm{GL}(\mathcal{A}_V)$ in the sense
that $\mathsf{R}(xg) = \mathsf{R}(x) \mathsf{R}^\prime(g)$ and
$\mathsf{R}(gx) = \mathsf{R}^\prime(g) \mathsf{R}(x)$ for all $x
\in \mathrm{H}(V_0^s)$ and $g \in \mathrm{U}(V_0^s)$.
\end{lem}
Since the construction of $\mathsf{R}^\prime$ makes no reference
to the re\-presentation of the Lie algebra $\mathfrak{u}(V_0^s)$
by the restriction $R_\ast \vert_{\mathfrak{u} (V_0^s)}\,$, their
relationship is as yet an open question.

To answer it, pick any $1 - X \equiv x \in \mathrm{H}(V_0^s)$ subject
to the condition $X^\dagger s + s X > 0$. For such $X = 1 - x \,$,
the curve $t \mapsto x_t := 1 - tX$ for $0 < t \le 1$ is a curve in
$\mathrm{H}(V_0^s)$.  Indeed, the con\-dition for $x_t$ to lie in
$\mathrm{H}(V_0^s)$ is $x_t^\dagger s x_t^{\vphantom{\ast}} < s \,$,
which translates to
\begin{displaymath}
     t^2 X^\dagger s X < t (X^\dagger s + s X) \;,
\end{displaymath}
and since $0 < X^\dagger s + s X$ and this inequality holds for $t =
1$, it holds for all $0 < t \le 1$.

Note that $X^{-1}$ exists, and that $\mathfrak{Re}\, Xs > 0\,$. Now
insert $a_{x_t} = 2s(1 - x_t)^{-1} - s = 2s (tX)^{-1} -s$ into the
expression for $\mathsf{R}(x_t)$ from Lem.\ \ref{lem 4.6p}, and
rescale $v \to \sqrt{t} \, v\, $:
\begin{eqnarray*}
    \mathsf{R}(1-tX) &=& \mathrm{Det}(t X s / 2)^{-1}
    \int_{V_0} \mathrm{e}^{-\langle v , (X s)^{-1} v \rangle}
    \mathrm{e}^{\mathrm{i}\sqrt{t}\, \mu(v_+) + \mathrm{i}
    \sqrt{t}\, \delta(c\, v_+)} \\ &&\hspace{3cm} \times \,\,
    \mathrm{e}^{\mathrm{i}\sqrt{t}\, \delta(v_-) + \mathrm{i}
    \sqrt{t} \, \mu(c\, v_-)} \, \mathrm{e}^{\frac{t}{2}
    \langle v ,\, s v \rangle} \, d\mathrm{vol}(\sqrt{t} v) \;.
\end{eqnarray*}
Using $\mathrm{Det}(tXs/2)^{-1} d \mathrm{vol}(\sqrt{t} v) =
\mathrm{Det}(Xs/2)^{-1} d\mathrm{vol}(v)$, take the one-sided
de\-rivative at $t = 0+ :$
\begin{eqnarray*}
    (d \mathsf{R})_e (-X) &=& - \mathrm{Det}(Xs/2)^{-1} \int_{V_0}
    \mathrm{e}^{-\langle v , (Xs)^{-1} v \rangle} \big( \mu(v_+) +
    \delta(v_-) \big) \\ &&\hspace{3cm} \times \,\, \big( \delta(c\,
    v_+) + \mu(c\, v_-) \big) \, d\mathrm{vol}(v) \;.
\end{eqnarray*}
To do the Gaussian integral over $v\,$, expand the integration vector
$v$ in orthonormal bases $\{ e_i^+ \} $ and $\{ e_j^- \}$ of $V_0^+$
resp.\ $V_0^-$, and use $c\, e_i^\pm = \langle e_i^\pm , \cdot
\rangle = f_i^\pm$. If $X$ is decomposed into blocks according to
$V_0^{\vphantom{\ast}} = V_0^+ \oplus V_0^-$, the final result
assumes the form
\begin{displaymath}
    (d\mathsf{R})_e \begin{pmatrix} \mathsf{A}&\mathsf{B}\\
    \mathsf{C} &\mathsf{D} \end{pmatrix} = \sum \big(\mu(\mathsf{A}
    e_i^+) + \delta(\mathsf{C} e_i^+) \big) \delta(f_i^+) - \sum \big(
    \mu(\mathsf{B} e_j^-) + \delta(\mathsf{D} e_j^-) \big) \mu(f_j^-) \;.
\end{displaymath}
This is exactly the operator which is assigned to $X = \sum (X e_i^+)
\otimes f_i^+ + \sum (X e_j^-)\otimes f_j^-$ by the spinor-oscillator
representation of Def.\ \ref{def 2.8}.

Since $\mathsf{R}^\prime :\, \mathrm{U}(V_0^s) \to \mathrm{U}
(\mathcal{A}_V)$ is constructed from $\mathsf{R} : \, \mathrm{H}
(V_0^s) \to \mathrm{GL}(\mathcal{A}_V)$ by a continuous limit
procedure, the same formula holds for $X \in \mathfrak{u}(V_0^s)$.
We thus have:
\begin{lem}\label{lem 4.10}
The infinitesimal of the representation $\mathsf{R}^\prime : \,
\mathrm{U}(V_0^s) \to \mathrm{U}(\mathcal{A}_V)$ agrees with the
Lie superalgebra representation $R_\ast : \, \mathfrak{gl}(V) \to
\mathfrak{gl}(\mathcal{A}_V)$ restricted to $\mathfrak{u}(V_0^s):$
\begin{displaymath}
    (d \mathsf{R}^\prime)_e = R_\ast \vert_{\mathfrak{u}(V_0^s)} \;.
\end{displaymath}
\end{lem}
Another question to be addressed concerns the relation between the
semigroup representation $\mathsf{R}$ and the Weyl action $\mathbf{w}
: \, W_0 \to \mathrm{End}(\mathcal{A}_V)$ of Def.\ \ref{def 2.8}. To
answer it in a good way, let us introduce for every $v + \varphi \in
W_0$ the operator
\begin{equation}
    \widetilde{\mathsf{T}}_{v + \varphi} = \mathrm{e}^{\mathrm{i}
    \mu(v_+) + \mathrm{i}\delta(v_-)} \, \mathrm{e}^{\mathrm{i}
    \delta(\varphi_+) - \mathrm{i}\mu(\varphi_-)} \;,
\end{equation}
where $w = v + \varphi = v_+ + v_- + \varphi_+ + \varphi_-$ is the
orthogonal decomposition according to $W_0^{\vphantom{\ast}} =
V_0^{\vphantom{\ast}} \oplus V_0^\ast = V_0^+ \oplus V_0^- \oplus
{V_0^+}^\ast \oplus {V_0^-}^\ast$.  For $x \in \mathrm{GL}(V_0)$
recall $x \cdot w = xv + \varphi \, x^{-1}$.
\begin{lem}\label{lem 4.10p}
For all $w \in W_0$ and $x \in \mathrm{H}(V_0^s) \subset
\mathrm{GL}(V_0)$ the following relation holds:
\begin{displaymath}
    \mathsf{R}(x) \widetilde{\mathsf{T}}_w =
    \widetilde{\mathsf{T}}_{x \, w} \, \mathsf{R}(x) \;.
\end{displaymath}
\end{lem}
\begin{rem}
Being unbounded, $\widetilde{\mathsf{T}}_{v + \varphi}$ is not an
operator on the $L^2$-space of $\mathcal{A}_V$. Still, it does exist
as an operator on the algebra of holomorphic functions on $V_0^+
\oplus (V_0^-)^\ast$.
\end{rem}
\begin{proof}
By the canonical commutation relations (\ref{CCR}), the operators
$\widetilde{\mathsf{T}}_{v + \varphi}$ satisfy
\begin{displaymath}
    \widetilde{\mathsf{T}}_{v + \varphi} \widetilde{\mathsf{T}}_{
    v^\prime + \varphi^\prime} = \widetilde{\mathsf{T}}_{v + v^\prime
    + \varphi + \varphi^\prime}\, \mathrm{e}^{- \varphi(v^\prime)} \;.
\end{displaymath}
Using this composition law and the identity $\mathsf{T}_u =
\mathrm{e}^{- \frac{1}{2} \langle u ,\, s u \rangle} \widetilde{
\mathsf{T}}_{u + csu} $ in the defining expression for
$\mathsf{R}(x)$ in Lem.\ \ref{lem 4.6p}, one writes the product
$\mathsf{R}(x) \widetilde{\mathsf{T}}_{v + \varphi}$ in the form
\begin{displaymath}
    \mathsf{R}(x) \widetilde{\mathsf{T}}_{v + \varphi} =
    \mathrm{Det}(a_x + s) \int_{V_0} \mathrm{e}^{- \frac{1}{2}
    \langle u , (a_x + s) u \rangle + \varphi((1-x)^{-1}u)}
    \widetilde{\mathsf{T}}_{u + xv + csu} \,\, d\mathrm{vol}(u) \;.
\end{displaymath}
Here a shift of integration variables $u \to u - (1-x) v$ and $csu
\to csu - \varphi$ has been made, and the relation $\frac{1}{2} (a_x
+ s)(1-x) = s$ was used.
%
%
Similarly, one evaluates the product $\widetilde{\mathsf{T}}_{xv +
\varphi \, x^{-1}} \mathsf{R}(x)$ by shifting $c s u \to c s u -
\varphi \, x^{-1}$. The resulting expression is the same as for
$\mathsf{R}(x) \widetilde{ \mathsf{T}}_{v + \varphi}\,$.  Since $xv +
\varphi \, x^{-1} = x \cdot (v + \varphi)$, this already proves the
lemma.
\end{proof}
\begin{cor}\label{cor 4.11p}
The semigroup representation $\mathsf{R}$ is compatible with the
Weyl action $\mathbf{w} : \, W_0 \to \mathrm{End}(\mathcal{A}_V)$
of Def.\ \ref{def 2.8} in the sense that
\begin{displaymath}
    \mathsf{R}(x) \mathbf{w}(w) = \mathbf{w}(x \, w) \mathsf{R}(x)
\end{displaymath}
for all $w \in W_0$ and $x \in \mathrm{H}(V_0^s)$.
\end{cor}
\begin{proof}
Differentiating the curve of operators $t \mapsto \widetilde
{\mathsf{T}}_{tw}$ at $t = 0$ gives the Weyl action:
\begin{displaymath}
    \frac{d}{dt}\Big\vert_{t=0} \widetilde{\mathsf{T}}_{tw} =
    \mathrm{i}\mu(v_+) + \mathrm{i}\delta(v_-) + \mathrm{i} \delta(
    \varphi_+) - \mathrm{i}\mu(\varphi_-) = \mathrm{i}
    \mathbf{w}(w) \;.
\end{displaymath}
Therefore the assertion follows by differentiating the statement of
Lem.\ \ref{lem 4.10p}.
\end{proof}
One final statement is needed to clarify the relation of $\mathsf{R}$
with earlier structure.
\begin{lem}\label{lem 5.11}
For every $x \in \mathrm{H}(V_0^s)$ and $Y \in \mathfrak{gl}(V_0)$
there exists some $\epsilon > 0$ so that $\mathrm{e}^{tY} x \in
\mathrm{H}(V_0^s)$ for all $t \in [-\epsilon, \epsilon]$. The
derivative of $\mathsf{R}$ along the curve $t \mapsto \mathrm{e}^{tY}
x$ in $x$ is
\begin{displaymath}
    \frac{d}{dt} \mathsf{R}(\mathrm{e}^{tY} x)
    \Big\vert_{t = 0} = R_\ast(Y) \mathsf{R}(x) \;,
\end{displaymath}
where $R_\ast \equiv R_\ast \vert_{\mathfrak{gl}(V_0)}$ is the
restriction of the representation $R_\ast : \, \mathfrak{gl}(V) \to
\mathfrak{gl}(\mathcal{A}_V)$.
\end{lem}
\begin{proof} The first statement just says that the semigroup
$\mathrm{H}(V_0^s)$ is open in $\mathrm{GL}(V_0)$.

For the second statement use the formula
\begin{displaymath}
    \frac{d}{dt} a_{\mathrm{e}^{tY} x} \Big\vert_{t = 0} = 2s
    \frac{d}{dt}\, (1 - \mathrm{e}^{tY} x)^{-1}\Big\vert_{t = 0}
    = (a_x + s) Y x \, (1-x)^{-1}
\end{displaymath}
to compute the derivative as
\begin{eqnarray*}
    \frac{d}{dt} \mathsf{R}(\mathrm{e}^{tY}x) \Big\vert_{t = 0}
    &=& \mathrm{Det}(a_x + s) \int_{V_0} \mathrm{e}^{- \frac{1}{2}
    \langle u ,\, a_x u \rangle} f_{Y,x}(u) \mathsf{T}_u
    \,\, d\mathrm{vol}(u) \;,\\ f_{Y,x}(u) &=& \mathrm{Tr}_{V_0}\,
    Y x \,(1-x)^{-1} - {\textstyle{\frac{1}{2}}} \big\langle u ,
    (a_x + s) Y x\,(1-x)^{-1} u \big\rangle \;.
\end{eqnarray*}

On the other hand, identifying $\mathfrak{gl}(V_0) \simeq
V_0^{\vphantom {\ast}}\otimes V_0^\ast$ let $Y = v \otimes \varphi$
and notice
\begin{displaymath}
    R_\ast(v \otimes \varphi) = \mathbf{w}(v) \mathbf{w}(\varphi)
    = - \frac{\partial^2}{\partial t \partial t^\prime} \widetilde{
    \mathsf{T}}_{tv + t^\prime \varphi}\Big\vert_{t=t^\prime =0}\;.
\end{displaymath}
In order to exploit this relation, consider the product
\begin{displaymath}
    \widetilde{\mathsf{T}}_{v+\varphi} \, \mathsf{R}(x) =
    \mathrm{Det}(a_x + s) \int_{V_0} \mathrm{e}^{- \frac{1}{2}
    \langle u , (a_x + s)u \rangle - \varphi(u)} \,
    \widetilde{\mathsf{T}}_{ u + v + csu + \varphi} \,\,
    d\mathrm{vol}(u)\;.
\end{displaymath}
Now make a shift of integration variables $u \to u - v$ and $c s u
\to c s u - \varphi\,$, and use the identity $\frac{s}{2} (a_x + s) -
1 = x\,(1-x)^{-1}$ to obtain
\begin{displaymath}
    \widetilde{\mathsf{T}}_{v+\varphi} \, \mathsf{R}(x) =
    \mathrm{Det}(a_x + s) \int_{V_0} \mathrm{e}^{- \frac{1}{2}
    \langle u ,\, a_x u \rangle + \frac{1}{2}\langle u , (a_x + s)
    v \rangle + \varphi(x\,(1-x)^{-1}(u-v) )} \mathsf{T}_u \,\,
    d\mathrm{vol}(u) \;.
\end{displaymath}
Substitute $v \to tv$ and $\varphi \to t^\prime \varphi$ and
differentiate at $t = t^\prime = 0$ to arrive at a formula for
$R_\ast(v \otimes \varphi) \mathsf{R}(x)$.  The result is the same
as our expression for the derivative of $\mathsf{R}( \mathrm{e}^
{tY}x)$ at $t = 0$ specialized to the case $Y = v \otimes
\varphi$. The formula of the lemma now follows because every $Y
\in \mathfrak{gl}(V_0)$ can be expressed as a linear combination
$Y = \sum v_i \otimes \varphi_i \,$.
\end{proof}

\begin{lem}\label{lem convergence}
Let $P$ be any element of the Clifford-Weyl algebra of $W = V \oplus
V^\ast$ acting on $\mathcal{A}_V$ by the Clifford-Weyl action
$\mathbf{q}$ of Def.\ \ref{def 2.8}. Then the operator $
\mathbf{q}(P) \mathsf{R}(x)$ for $x \in \mathrm{H}(V_0^s)$ is
trace-class and $\mathrm{STr}_{\mathcal{A}_V}\, \mathbf{q}(P)
\mathsf{R}(x)$ depends analytically on $x\,$.
\end{lem}
\begin{proof}
The representation space $\mathcal{A}_V$ is isomorphic to the tensor
product of $\mathcal{A}_{V_1} := \wedge(V_1^+ \oplus {V_1^-}^\ast)$
with $\mathcal{A}_{V_0} := \mathrm{S}(V_0^+ \oplus {V_0^-}^\ast)$,
and the operator $\mathsf{R}(x)$ acts non-trivially only on the
second factor. Since the first factor $\mathcal{A} _{V_1}$ is a
finite-dimensional vector space, the assertion is true if
$\mathrm{Tr}_{ \mathcal{A}_{V_0}} \mathbf{w}(P) \mathsf{R}(x)$ exists
and depends analytically on $x$ for every Weyl algebra element $P \in
\mathfrak{w} (W_0)$, operating by the Weyl action $\mathbf {w}$ on
$\mathcal{A}_{V_0}\,$.

Applying the unitary operator $\mathsf{T}_v$ to $\phi_0 \equiv 1 \in
\mathbb{C} \subset \mathcal {A}_{V_0}$ one gets a unit vector $\phi_v
:= \mathsf{T}_v \phi_0 \in \mathcal{A}_{V_0} \,$, which is called a
\emph{coherent state}. From the relation (\ref{heisenberg}) one has
\begin{displaymath}
    (\phi_v \, , \mathsf{T}_u \, \phi_v) = (\phi_0 \, ,
    \mathsf{T}_{-v} \mathsf{T}_u \mathsf{T}_v \phi_0) =
    \mathrm{e}^{-2\mathrm{i}\mathfrak{Im} \langle u ,\, sv
    \rangle} (\phi_0\, , \mathsf{T}_u \phi_0) = \mathrm{e}^{-2
    \mathrm{i}\mathfrak{Im} \langle u ,\, sv \rangle - \frac{1}{2}
    \langle u , u \rangle} \;,
\end{displaymath}
where $( \, , \, )$ denotes the Hermitian scalar product of
$\mathcal{A}_{V_0}\,$.

We will show that $\mathrm{Tr}_{\mathcal{A}_{V_0}} \mathbf{w}(P)
\mathsf{R}(x) < \infty$ by integrating over coherent states. First,
consider $( \phi_v\, , \mathsf{R}(x) \phi_v)$, using for
$\mathsf{R}(x)$ the formula of Lem.\ \ref{lem 4.6p}.  Since $|
\mathrm{e}^{-2\mathrm{i} \mathfrak {Im}\langle u , \, sv \rangle}| =
1$ and the Gaussian density $u \mapsto \mathrm{e}^{-\frac{1}{2}
\langle u ,\, \mathfrak{Re} (a_x) u \rangle} \mathrm{e}^{-\frac{1}{2}
\langle u , u \rangle} d\mathrm{vol}(u)$ has finite integral over
$V_0\,$, one may take the coherent-state expectation inside the
integral to obtain
\begin{displaymath}
    (\phi_v \, , \mathsf{R}(x) \phi_v) = \mathrm{Det}(a_x + s)
    \int_{V_0} \mathrm{e}^{-\frac{1}{2} \langle u , (a_x +1) u
    \rangle - 2 \mathrm{i} \mathfrak{Im} \langle u , \, s v
    \rangle } \,\, d\mathrm{vol}(u) \;.
\end{displaymath}
The Gaussian integral over $u \in V_0$ converges absolutely, and
performing it by completing the square one gets
\begin{displaymath}
    (\phi_v \, , \mathsf{R}(x) \phi_v) = \mathrm{Det}(a_x + s)
    \, \mathrm{Det}(a_x + 1)^{-1} \, \mathrm{e}^{-\frac{1}{2}
    \langle s v, \, (a_x + 1)^{-1} s v \rangle} \;,
\end{displaymath}
which is a rapidly decreasing function of $v\,$.

Now the Heisenberg group action
%
%
$\mathrm{U}_1 \times V_0 \to \mathrm{U}(\mathcal{A}_{V_0})$, $(z\,
,u) \mapsto z \, \mathsf{T}_u\,$, is well known (as part of the
Stone-von Neumann Theorem) to be irreducible on $\mathcal{A}_{V_0}
\,$. Therefore, if $L \in \mathrm{End} (\mathcal{A}_{V_0})$ is a
trace-class operator, its trace is computed by the integral
\begin{displaymath}
    \mathrm{Tr}_{\mathcal{A}_{V_0}} L = \int_{V_0}
    (\phi_v \, , L \, \phi_v) \, d\mathrm{vol}(v) \Big/ \int_{V_0}
    \mathrm{e}^{-\langle v , v \rangle} d\mathrm{vol}(v) \;.
\end{displaymath}
Indeed, the function $L \mapsto \int_{V_0} (\phi_v \, , L \, \phi_v)
\, d\mathrm{vol}(v)$ does not change when $L$ is conjugated by
$\mathsf{T}_u\,$, and by the irreducibility of $u \mapsto
\mathsf{T}_u$ any such function must be proportional to the operation
of taking the trace. The constant of proportionality is determined by
inserting some rank-one projector, say $L = \phi_0 (\phi_0 \, ,
\cdot)$, and using $|(\phi_0 \, , \phi_v)|^2 = \mathrm{e}^{-\langle v
, v \rangle}$.

Since $v \mapsto (\phi_v \, , \mathsf{R}(x) \phi_v)$ is a decreasing
Gaussian function and $x \mapsto a_x$ depends analytically on $x \in
\mathrm{H}(V_0^s)$, it is now clear that the trace of $\mathsf{R}(x)$
exists and is analytic in $x\,$. An easy computation gives
\begin{displaymath}
    \mathrm{Tr}_{\mathcal{A}_{V_0}} \mathsf{R}(x) = \mathrm{Det}
    \big( {\textstyle{\frac{1}{2}}} (a_x + s) \big) = \mathrm{Det}(s)
    \, \mathrm{Det}(1 - x)^{-1} \;.
\end{displaymath}

Essentially the same argument goes through for the case of
$\mathbf{w}(P) \mathsf{R}(x)$ with a poly\-nomial $P \in \mathfrak{w}
(W_0)$. Indeed, since the expectation $(\phi_v \, , \mathbf{w}(P)
\mathsf{T}_u \, \phi_v)$ is easily verified to be $\mathrm{e}^{-
2\mathrm {i}\mathfrak{Im} \langle u , \, sv \rangle}$ times a
polynomial in $u$ and $v$ of finite degree, $(\phi_v \, , \mathbf{w}
(P) \mathsf{R}(x) \phi_v)$ still is a rapidly decreasing function of
$v\,$ and therefore its integral -- the trace of $\mathbf{w}(P)
\mathsf{R}(x)$ -- exists and depends analytically on $x\,$.
\end{proof}

\subsection{Combining the representations}

Let us piece together the various building stones we have
accumulated. {}From Lem.\ \ref{lem 4.6p} we have a representation
$\mathsf{R}$ of the semigroup $\mathrm{H}(V_0^s)$ which we now rename
to
\begin{displaymath}
    R_0 : \,\, \mathrm{H}(V_0^s) \to \mathrm{End}(\mathcal{A}_V) \;.
\end{displaymath}
Notice that $R_0$ coincides with the representation $\omega \otimes
\tilde{\omega}$ of Sect.\ \ref{sect:2.4} if $\mathrm{H}(V_0^s)$ is
restricted to $\mathrm{H}^<(V_0^+) \times \mathrm{H}^>(V_0^-)
\hookrightarrow \mathrm{H}(V_0^s)$. We also recall that we have a
$\mathrm{GL}(V_1)$-representation
\begin{displaymath}
    R_1 : \,\, \mathrm{GL}(V_1) \to \mathrm{GL}(\mathcal{A}_V) \;,
\end{displaymath}
which agrees with $\sigma \otimes \tilde{\sigma}$ of Sect.\
\ref{sect:2.1} upon restriction to $\mathrm{GL}(V_1^+) \times
\mathrm{GL}(V_1^-) \hookrightarrow \mathrm{GL}(V_1)$.

Combining $R_1$ with $R_0$ we have
\begin{equation}
    R : \,\, \mathrm{GL}(V_1) \times \mathrm{H}(V_0^s) \to
    \mathrm{End}(\mathcal{A}_V) \;, \quad (g,h) \mapsto
    R_1(g) R_0(h) \;.
\end{equation}
As a corollary to Lem.\ \ref{lem 4.7p} we also have a unitary
representation of the direct product of real Lie groups
$\mathrm{U} (V_1) \times \mathrm{U}(V_0^s)$.  We denote this by
\begin{equation}
    R^\prime :\,\, \mathrm{U}(V_1) \times \mathrm{U}(V_0^s) \,
    \to \, \mathrm{U}(\mathcal{A}_V) \;.
\end{equation}

In the next statement, the adjoint action $\mathrm{Ad}(x) : \,
\mathfrak{gl}(V) \to \mathfrak{gl}(V)$ for $V = V_1 \oplus V_0$ is
meant to be conjugation by $x \in \mathrm{GL}(V_1) \times
\mathrm{H}(V_0^s) \hookrightarrow \mathrm{GL}(V)$.
\begin{prop}\label{lem intertwine}
The semigroup representation $R : \, \mathrm{GL}(V_1) \times
\mathrm{H}(V_0^s) \to \mathrm{End}(\mathcal{A}_V)$ and the Lie
superalgebra representation $R_\ast :\, \mathfrak{gl}(V) \to
\mathfrak{gl}(\mathcal{A}_V)$ are compatible in that
\begin{displaymath}
    R(x)\, R_\ast(Y) = R_\ast(\mathrm{Ad}(x)Y)\, R(x)
\end{displaymath}
for all $Y \in \mathfrak{gl}(V)$ and $x \in \mathrm{GL}(V_1) \times
\mathrm{H}(V_0^s)$.
\end{prop}
\begin{proof}
Let $\mathbf{c} + \mathbf{w} :\, W_1 \oplus W_0 \to \mathrm{End}
(\mathcal{A}_V)$ be the Clifford-Weyl action of Def.\ \ref{def 2.8}.
If $x = (x_1,x_0) \in \mathrm{GL}(V_1) \times \mathrm{H} (V_0^s)$,
then by combining Lem.\ \ref{cor 4.6p} and Cor.\ \ref{cor 4.11p} we
have
\begin{equation}\label{intertwine}
    R(x) \big( \mathbf{c}(w_1) + \mathbf{w}(w_0) \big) = \big(
    \mathbf{c}(x_1 \cdot w_1)+\mathbf{w}(x_0 \cdot w_0)\big) R(x)
    \;.
\end{equation}
By definition, the representation $R_\ast$ is induced from the
spinor-oscillator representation of the Clifford-Weyl algebra
$\mathfrak{q}(W)$, which in turn is determined by the Clifford-Weyl
action $\mathbf{c} + \mathbf{w}$. The statement therefore follows
from the property (\ref{intertwine}) of $\mathbf{c} + \mathbf{w}$ in
combination with the fact that by the identification $\mathfrak{gl}
(V) \simeq V \otimes V^\ast$ the action of $\mathrm{GL}(V_1) \times
\mathrm{GL}(V_0) \hookrightarrow \mathrm{GL}(V)$ on $W = V \oplus
V^\ast$ by $x \, (v + \varphi) = xv + \varphi \, x^{-1}$ corresponds
to the adjoint action $\mathrm {Ad}(x) (v \otimes \varphi) = x\, (v
\otimes \varphi) x^{-1}$.
\end{proof}
\begin{cor}
The group representation $R^\prime :\, \mathrm{U}(V_1) \times
\mathrm{U}(V_0^s) \to \mathrm{U}(\mathcal{A}_V)$ and the
superalgebra representation $R_\ast :\, \mathfrak{gl}(V) \to
\mathfrak{gl}(\mathcal{A}_V)$ are compatible in the sense that
\begin{displaymath}
    R^\prime(g) R_\ast(X) R^\prime(g)^{-1} = R_\ast(\mathrm{Ad}(g)X)
\end{displaymath}
for all $X \in \mathfrak{gl}(V)$ and $g \in \mathrm{GL}(V_1) \times
\mathrm{U}(V_0^s)\,$.
\end{cor}
\begin{proof}
$R^\prime(g)$ for $g \in \mathrm{U}(V_1) \times \mathrm{U}(V_0^s)$
has an inverse. Thus the assertion follows from Prop.\ \ref{lem
intertwine} by sending $x \to g$ and multiplying with
$R^\prime(g)^{-1}$ on the right.
\end{proof}
The next result is an easy consequence of Lem.\ \ref{lem 5.11} in
the present, combined setting.
\begin{prop}\label{prop 5.14}
If $x \in \mathrm{GL}(V_1) \times \mathrm{H}(V_0^s)$ and $Y \in
\mathfrak{g}_0 \equiv \mathfrak{gl}(V)_0\,$, the directional
deri\-vative of $R$ at $x$ along $Y$ exists and is given by
\begin{displaymath}
    \frac{d}{dt} R(\mathrm{e}^{tY} x)
    \Big\vert_{t = 0} = R_\ast(Y) R(x) \;.
\end{displaymath}
\end{prop}

\subsection{Proof of Prop.\ \ref{prop 4.2}}

We finally tackle the true goal of the current section and return to
the Howe pair setting based on the decomposition $V = U \otimes
\mathbb{C}^N$ and its refinement $V_\tau^\pm = U_\tau^\pm \otimes
\mathbb{C}^N$ with $U_0^+ = U_1^+ = \mathbb{C}^p$ and $U_0^- = U_1^-
= \mathbb{C}^q$, and $p+q = n\,$.

From the tensor-product decompositions $V_1 = \mathbb{C}^n \otimes
\mathbb{C}^N$ and $V_0 = \mathbb{C}^{p+q} \otimes \mathbb{C}^N$ we
have the embeddings
\begin{displaymath}
    \mathrm{GL}_n \times \mathrm{U}_N \hookrightarrow
    \mathrm{GL}(V_1) \;, \quad \mathrm{H}_{p,\,q} \times
    \mathrm{U}_N \hookrightarrow \mathrm{H}(V_0^s) \;,
\end{displaymath}
which by $V_1 \oplus V_0 = (U_1 \oplus U_0)\otimes \mathbb{C}^N$
combine to a semigroup embedding
\begin{displaymath}
    (\mathrm{GL}_n \times \mathrm{H}_{p,\,q}) \times \mathrm{U}_N
    \hookrightarrow \mathrm{GL}(V_1) \times \mathrm{H}(V_0^s) \;.
\end{displaymath}
Similarly, we have a Lie group embedding
\begin{displaymath}
    (\mathrm{U}_n \times \mathrm{U}_{p,\,q}) \times \mathrm{U}_N
    \hookrightarrow \mathrm{U}(V_1) \times \mathrm{U}(V_0^s) \;.
\end{displaymath}

Using these, we now restrict and project the $\mathfrak{gl}(V)
$-representation $(\mathcal{A}_V , R_\ast , R, R^\prime)$ to a
$\mathfrak{gl}(U)$-representation $(\mathcal {V}_\lambda ,
\rho_\ast , \rho, \rho^\prime)$ on the subalgebra of
$\mathrm{U}_N$-invariants $\mathcal{V}_\lambda \subset
\mathcal{A}_V$.  The semigroup representation $\rho$ is defined by
\begin{displaymath}
    \rho : \,\, \mathrm{GL}_n \times \mathrm{H}_{p,\,q} \to
    \mathrm{GL}(\mathcal{V}_\lambda) \;, \quad x \mapsto
    R(x,\mathrm{Id}_N) \vert_{\mathcal{V}_\lambda} \;,
\end{displaymath}
and the representation $\rho^\prime$ of the Lie group $G_\mathbb {R}
= \mathrm{U}_n \times \mathrm{U}_{p,\,q}$ is defined in the same way:
\begin{displaymath}
    \rho^\prime : \,\, \mathrm{U}_n \times \mathrm{U}_{p,\,q}
    \to \mathrm{U}(\mathcal{V}_\lambda) \;, \quad g \mapsto
    R^\prime(g,\mathrm{Id}_N) \vert_{\mathcal{V}_\lambda} \,.
\end{displaymath}
This gives the existence statement of Prop.\ \ref{prop 4.2}. The
stated compatibility of $\rho$ and $\rho^\prime$ with $\rho_\ast$
is a consequence of the compatibility of $R$ and $R^\prime$ with
$R_\ast\,$.

To go further and prove the key formula (\ref{needs ancestor}),
recall the setting of Sect.\ \ref{sect:csm}: we are given a
real-analytic manifold $M = M_1 \times M_0 \subset \mathrm {GL} (U_1)
\times \mathrm{GL}(U_0)$, a $G_\mathbb{R} $-principal bundle $P \to
M$, an associated vector bundle $F \to M$ with standard fibre
$\mathfrak{g}_{\mathbb{R},1} \equiv \mathfrak{u}(U^s)_1\,$, and a
sheaf of graded-commutative algebras $\mathcal{F} = \Gamma(M,
\mathbb{C} \otimes \wedge F^\ast)$. The next goal is to describe the
cha\-racter of the representation $(\mathcal{V}_\lambda , \rho_\ast ,
\rho, \rho^\prime)$ as a section of $\mathcal{F}$.

Our first observation is that, if $\{ \mathsf{F}_i \}$ is a basis of
$\mathfrak{g}_{\mathbb{R},1}$ and $\{ \varphi^i \}$ is the
corresponding basis of generators of $\wedge (\mathfrak{g}_{
\mathbb{R},1}^\ast)$, the operator
\begin{equation}
    \rho(\Xi) \equiv \rho(x \, \mathrm{e}^\xi y^{-1}) := \rho(x)
    \, \mathrm{e}^{\,\sum \varphi^i \rho_\ast(\mathsf{F}_i)}
    \rho(y^{-1}) \;,
\end{equation}
which is associated to the supermatrix $\Xi = x \, \mathrm{e}^\xi
y^{-1}$ by the representation $(\mathcal{V}_\lambda , \rho_\ast \, ,
\rho, \rho^\prime)$, makes good sense. Indeed, if we replace $(x,y)
\in P$ by $(xg,yg)$ with $g \in G_\mathbb{R}$ and the generators
$\varphi^i$ of $\wedge (\mathfrak{g}_{\mathbb{R},1}^\ast)$ by the
transformed generators $\sum \varphi^j \mathrm {Ad}^\ast(g^{-1}
)_{j\phantom{i}}^{\phantom {j} i}\,$, then
\begin{displaymath}
    \rho(xg) \mathrm{e}^{\,\sum \varphi^j \mathrm{Ad}^\ast(
    g^{-1})_{j\phantom{i}}^{\phantom{j}i} \rho_\ast(\mathsf{F}_i)}
    \rho((yg)^{-1}) = \rho(x)\, \mathrm{e}^{\,\sum \varphi^i
    \rho_\ast(\mathsf{F}_i)} \rho(y^{-1}) \;,
\end{displaymath}
since $\rho(xg) = \rho(x) \rho^\prime(g)$ and $\rho^\prime(g)
\rho_\ast(\mathsf {F}_i) \rho^\prime(g)^{-1} = \rho_\ast
(\mathrm{Ad}(g) \mathsf{F}_i) = \sum \rho_\ast(\mathsf{F}_j)
\mathrm{Ad}(g) _{\phantom{j}i}^{j \phantom{i}} \,$.

As a result, the character $\mathrm{STr}_{\mathcal{V}_\lambda}
\rho(\Xi)$ is a $G_\mathbb{R}$-equivariant mapping from $P$ to the
exterior algebra $\mathbb{C} \otimes \wedge (\mathfrak{g}_{\mathbb{R}
,1}^\ast)$, and thus defines a section $\chi \in \mathcal{F}$ via the
usual isomorphism:
\begin{equation}\label{eq varphi}
    \chi(xy^{-1}) = \big[(x,y);\,\mathrm{STr}_{\mathcal{V}_\lambda}
    \rho(x)\, \mathrm{e}^{\,\sum \varphi^i \rho_\ast(\mathsf{F}_i)}
    \rho(y^{-1}) \big] \;.
\end{equation}
By its definition as a character, this section $\chi$ is radial,
i.e., one has $\chi(gmg^{-1}) = \chi(m)$ for all $g \in G_\mathbb
{R}\,$, and $(\widehat{Y}^L + \widehat{Y}^R) \chi = 0$ for all $Y \in
\mathfrak{g}_{\mathbb{R},1}\,$.
\begin{lem} The section $\chi \in \mathcal{F}$ defined by Eq.\
(\ref{eq varphi}) is radial and analytic.
\end{lem}
\begin{proof}
It remains to establish the analyticity of $\chi\,$. For that
purpose, fixing some pair $(x,y) \in P$ in the fibre over the
point $x y^{-1} = m = (m_1,m_0) \in M$, use the compatibility
statement of Prop.\ \ref{lem intertwine} to write
\begin{displaymath}
    \mathrm{STr}_{\mathcal{V}_\lambda} \rho(\Xi) =
    \mathrm{STr}_{\mathcal{V}_\lambda} \rho(x)\, \mathrm{e}^{\,
    \sum \varphi^i \rho_\ast( \mathsf{F}_i)}\rho(y^{-1})
    = \mathrm{STr}_{\mathcal{V}_\lambda} \mathrm{e}^{\,\sum
    \varphi^i \rho_\ast(\mathrm{Ad}(x)\mathsf{F}_i)} \rho(m) \;.
\end{displaymath}
By the nilpotency of the generators $\varphi^i$, the power series
expansion of the exponential function produces only a finite
number of terms
\begin{displaymath}
    \sum \varphi^{i_1} \cdots \varphi^{i_k} \,
    \mathrm{STr}_{\mathcal{V}_\lambda}
    \rho_\ast(\mathrm{Ad}(x)\mathsf{F}_{i_1}) \cdots
    \rho_\ast(\mathrm{Ad}(x)\mathsf{F}_{i_k}) \, \rho(m) \;.
\end{displaymath}
The numerical coefficients of these terms are of the form
$\mathrm{STr}_{\mathcal{V}_\lambda} \mathbf{q}(P) \rho(m)$ where $P$
is a polynomial in the Clifford-Weyl algebra of $W = V \oplus
V^\ast$. By straightforward adaptation of Lem.\ \ref{lem convergence}
to the present situation, they depend analytically on $m \in M\,$
through $\rho(m)$. They also depend analytically on the factor $x$ of
$m = xy^{-1}$ through $\mathrm{Ad}(x)$.
\end{proof}
The next statement is the main step of completing the proof of Prop.\
\ref{prop 4.2}, as it will allow us to lift the equality of functions
of Cor.\ \ref{cor 2.12} to an equality of sections of $\mathcal{F}$.
\begin{prop}\label{prop 5.16}
An analytic radial section $\vartheta \in \mathcal{F}$ is already
determined by its numerical part, $\mathrm{num}(\vartheta) \in
\Gamma(M,\mathbb{C})$.
\end{prop}
\begin{proof}
Again, for $x y^{-1} = m = (m_1,m_0) \in M$ write
\begin{displaymath}
    \Xi = x \, \mathrm{e}^{\, \sum \varphi^i \otimes
    \mathsf{F}_i} y^{-1} = \mathrm{e}^{\,\sum \varphi^i
    \otimes \mathrm{Ad}(x) \mathsf{F}_i} \, m \;.
\end{displaymath}
Note that $\mathrm{Ad}(x)$ in general sends $\mathsf{F}_i \in
\mathfrak{g}_{\mathbb{R},1}$ into the complex space $\mathfrak{g}_1 =
\mathfrak{g}_{\mathbb{R},1} + \mathrm{i}
\mathfrak{g}_{\mathbb{R},1}$.

{}From Lem.\ \ref{lem 4.13} we know that the spectrum of $m_0 \in
M_0$ lies on the positive real axis in $\mathbb{C}$ while avoiding
unity; whereas the spectrum of $m_1 \in M_1$ lies on the unit circle.
Thus the spectral sets of $m_0$ and $m_1$ are disjoint. As a
consequence, $m$ is regular in the sense of Prop.\ \ref{block diag}
and the supermatrix $\mathrm{e}^{\,\sum \varphi^i \otimes
\mathrm{Ad}(x) \mathsf{F}_i} m$ can be block diagonalized.

This means that if $\mathfrak{g}_0 = \mathfrak{gl}(U)_0$ and
$\mathfrak{g}_1 = \mathfrak{gl}(U)_1\,$, there exist complex
Grassmann envelope elements $\Xi_1 \in \wedge^\mathrm{odd}
(\mathfrak{g}_1^\ast) \otimes \mathfrak{g}_1$ and $\Xi_2 \in
\wedge^\mathrm{even}(\mathfrak{g}_1^\ast) \otimes \mathfrak{g}_0$
such that
\begin{displaymath}
    \mathrm{e}^{\,\sum \varphi^i \otimes \mathrm{Ad}(x)
    \mathsf{F}_i} m = \mathrm{e}^{\Xi_1} \, \mathrm{e}^{\Xi_2}
    m \, \mathrm{e}^{- \Xi_1} \;.
\end{displaymath}
{}From the proof of Prop.\ \ref{block diag} we also know how to
construct these nilpotent elements: we pass to the equivalent
equation
\begin{displaymath}
    \sum \varphi^i \otimes \mathrm{Ad}(x)\mathsf{F}_i =
    \ln \big( \mathrm{e}^{\Xi_1} \mathrm{e}^{\Xi_2}
    \mathrm{e}^{- \mathrm{Ad}(m) \Xi_1} \big) \;,
\end{displaymath}
and, after expanding the right-hand side by the
Baker-Campbell-Hausdorff formula, solve this equation for $\Xi_1$ and
$\Xi_2$ in terms of $\sum \varphi^i \otimes \mathrm{Ad}(x)
\mathsf{F}_i\,$.  Given the basis $\{ \mathsf{F}_i \}$ of
$\mathfrak{g}_{\mathbb{R},1}$ and fixing some basis $\{ \mathsf{E}_a
\}$ of $\mathfrak{g}_{\mathbb{R},0}\,$, we write
\begin{displaymath}
    \Xi_1 = \sum \xi_1^i \otimes \mathsf{F}_i \;, \quad
    \Xi_2 = \sum \xi_2^a \otimes \mathsf{E}_a \;,
\end{displaymath}
where $\xi_2^a$ are even sections of $\mathcal{F}$, and the functions
$\xi_1^i$ are odd.

These preparations do not yet depend on the analytic and radial
property of $\vartheta \in \mathcal{F}$. The rest of the argument,
for clarity, will first be given for the case of $\vartheta =
\chi$ of (\ref{eq varphi}), which is one of the two cases of
relevance below.

The algorithm for constructing $\Xi_1$ and $\Xi_2$ uses the
bracket of $\mathfrak{gl}(U)$, and nothing but that bracket.
Therefore, exactly the same procedure goes through at the
representation level, and we have
\begin{displaymath}
    \sum \varphi^i \rho_\ast( \mathrm{Ad}(x)\mathsf{F}_i ) =
    \ln \big( \mathrm{e}^{\,\sum \xi_1^i \,\rho_\ast(\mathsf{F}_i)}
    \mathrm{e}^{\,\sum \xi_2^a \rho_\ast( \mathsf{E}_a )}\mathrm{e}^{
    -\sum \xi_1^i \,\rho_\ast(\mathrm{Ad}(m)\mathsf{F}_i)} \big) \;.
\end{displaymath}
Exponentiate both sides, multiply by $\rho(m)$ on the right, and
move $\rho(m)$ past the rightmost exponential by Prop.\ \ref{lem
intertwine}. Then take the supertrace,
\begin{displaymath}
    \mathrm{STr}_{\mathcal{V}_\lambda} \mathrm{e}^{\,\sum
    \varphi^i \rho_\ast(\mathrm{Ad}(x) \mathsf{F}_i)} \rho(m) =
    \mathrm{STr}_{\mathcal{V}_\lambda}
    \mathrm{e}^{\,\sum \xi_1^i \, \rho_\ast(\mathsf{F}_i)}
    \mathrm{e}^{\,\sum \xi_2^a \rho_\ast(\mathsf{E}_a)} \rho(m)
    \, \mathrm{e}^{-\sum \xi_1^i \, \rho_\ast(\mathsf{F}_i)}\;,
\end{displaymath}
and use the cyclic property of $\mathrm{STr}$ (or, equivalently,
the radial property of $\chi$) to obtain
\begin{displaymath}
    \mathrm{STr}_{\mathcal{V}_\lambda} \rho(\Xi) =
    \mathrm{STr}_{\mathcal{V}_\lambda} \mathrm{e}^{\,\sum
    \varphi^i \rho_\ast(\mathrm{Ad}(x) \mathsf{F}_i)} \rho(m) =
    \mathrm{STr}_{\mathcal{V}_\lambda} \mathrm{e}^{\,\sum
    \xi_2^a \rho_\ast(\mathsf{E}_a)} \rho(m) \;.
\end{displaymath}
By Prop.\ \ref{prop 5.14} the last expression can be expressed as
a finite sum of derivatives via the $\mathfrak{g}_0$-action on the
left, $\mathsf{E}_a \mapsto \widehat{\mathsf{E}}_a^L :$
\begin{displaymath}
    \mathrm{STr}_{\mathcal{V}_\lambda} \mathrm{e}^{\,\sum \xi_2^a
    \rho_\ast(\mathsf{E}_a)} \rho(m) = \mathrm{e}^{- \sum \xi_2^a
    \widehat{\mathsf{E}}_a^L}\, \mathrm{STr}_{\mathcal{V}_\lambda}
    \rho(m)\;.
\end{displaymath}
Since the function $m \mapsto \mathrm{STr}_{\mathcal{V}_\lambda} \rho
(m)$ is the numerical part $\mathrm{num}(\chi)$ of (\ref{eq varphi}),
we have proved for this case that
\begin{displaymath}
    \chi = \mathrm{e}^{- \sum \xi_2^a \widehat{\mathsf{E}}_a^L}\,
    \mathrm{num}(\chi) \;,
\end{displaymath}
i.e., given the even sections $\xi_2^a$ and derivations
$\widehat{\mathsf{E}}_a^L$ (where of course the sum $\sum \xi_2^a
\widehat{\mathsf{E}}_a^L$ does not depend on the choice of basis but
is invariantly defined), $\chi$ is determined by its numerical values
and a finite number of derivatives thereof.

In order for this argument to go through, all that is needed is
radiality of $\chi$, to cancel the first factor against the last
one in $\mathrm{e}^{\Xi_1} \mathrm{e}^{\Xi_2} m \, \mathrm{e}^{
-\Xi_1}$, and analyticity of $\chi$, to convert $\Xi_2$ in
$\mathrm{e}^{\Xi_2} m$ into a differential operator. It is
therefore clear that the same reasoning applies to any section
$\vartheta \in \mathcal{F}$ which is radial and analytic.
\end{proof}
Finally, recall from Def.\ \ref{def 4.2} exactly what is meant by
the reciprocal of the superdeterminant of $\mathrm{Id}_V - \Xi
\otimes u\,$, and let $\vartheta \in \mathcal{F}$ be the section
which is given by the integral on the right-hand side of
(\ref{needs ancestor}):
\begin{displaymath}
    \vartheta(xy^{-1}) := \Big[ (x,y); \int_{\mathrm {U}_N}
    \mathrm{SDet}\big( \mathrm{Id}_V - x\, \mathrm{e}^{\sum
    \varphi^i \otimes \mathsf{F}_i} y^{-1} \otimes u \big)^{-1}
    \, du \Big] \;.
\end{displaymath}
This section $\vartheta$ is radial, because the superdeterminant has
the property of being radial; and since $\mathrm{SDet}(\mathrm{Id}_V
- \Xi \otimes u)^{-1}$ is analytic in the supermatrix $\Xi = x\,
\mathrm{e}^{\, \sum \varphi^i \otimes \mathsf{F}_i} y^{-1}$ for every
fixed unitary matrix $u\,$, so is $\vartheta$, the result of
integrating over $u \in \mathrm{U}_N\,$.

It now follows that $\vartheta$ equals the character $\chi$ of
(\ref{eq varphi}). Indeed, both $\vartheta$ and $\chi$ are radial
and analytic, and their numerical parts agree by Cor.\ \ref{cor
2.12}. Therefore, by Prop.\ \ref{prop 5.16} they agree as sections
of $\mathcal{F}$. Since the equality $\chi = \vartheta$ is none
other than our formula (\ref{needs ancestor}), the proof of Prop.\
\ref{prop 4.2} is now complete.

\bigskip\noindent{\small \emph{Acknowledgments}. %
The first draft of this paper was written while the authors were
participating in the programme on \emph{Random matrix approaches
in number theory} held at the Newton Institute for Mathematical
Sciences (Cambridge, UK) during the first half of 2004.

\noindent The research of M.R.Z.\ is supported by the Deutsche
Forschungsgemeinschaft via SFB/TR 12.}

\end{document}